\documentclass[floatfix,twocolumn,aps,rmp,showpacs,amsmath,amssymb,eqsecnum]{revtex4}
\usepackage{epsfig}
\begin{document}


\title{The Unruh effect and its applications}

\author{Lu\'\i s C.\ B.\ Crispino}
\email{crispino@ufpa.br}
\affiliation{Faculdade de F\'\i sica, Universidade Federal do Par\'a,
Campus Universit\'ario do Guam\'a, 66075-900, Bel\'em, Par\'a, Brazil}

\author{Atsushi Higuchi}
\email{ah28@york.ac.uk}
\affiliation{Department of Mathematics, University of York, Heslington,
York YO10 5DD, United Kingdom}

\author{George E.\ A.\ Matsas}
\email{matsas@ift.unesp.br}
\affiliation{Instituto de F\'\i sica Te\'orica, Universidade Estadual Paulista,
Rua Pamplona 145, 01405-900, S\~ao Paulo, SP, Brazil}

\date{\today}



\begin{abstract}
It has been thirty years since the discovery of the Unruh effect. 
It has played a crucial role in our understanding that the particle 
content of a field theory is observer dependent. This effect is important 
in its own right and as a way 
to understand the phenomenon of particle emission from black holes 
and cosmological horizons. Here, we review the Unruh effect with 
particular emphasis to its applications. We also comment on a number 
of recent developments and discuss some controversies. Effort is also
made  to clarify what seems to be common misconceptions.
\end{abstract}

\maketitle

\tableofcontents


\section{Introduction}
\label{section:Introduction}

\begin{figure}[t]
\epsfig{file=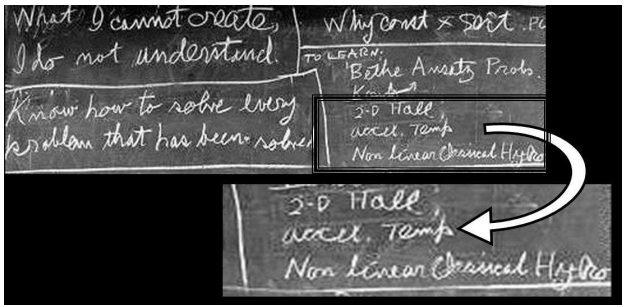,angle=0,width=0.95\linewidth,clip=}
\caption{\label{feyn1} Part of Feynman's blackboard at California
Institute of Technology 
at the time of his death in 1988. At the right-hand side 
one can find ``accel. temp." as one of the issues ``to learn".}
\end{figure}

It has been thirty years since the discovery of the {\em Unruh
effect} \cite{Unruh76} which can be also found under the name of 
{\em Davies-Unruh, Fulling-Davies-Unruh, {\rm and} 
Unruh-Davies-DeWitt-Fulling effect}. This is a conceptually subtle 
quantum field theory result, which has played a crucial role 
in our understanding that the particle content of a field 
theory is observer dependent in a sense to be made precise 
later \cite{Fulling73} [see also \textcite{Unruh77}]. This effect
is important 
in its own right and as a tool to investigate other phenomena 
such as the thermal emission of particles from black holes 
\cite{Hawking74, Hawking75} and cosmological horizons 
\cite{Gibbonsetal77}. 
It is interesting that the Unruh effect was on Feynman's list of things
to learn in his later years (see Fig.~\ref{feyn1}).
In short, the Unruh effect expresses the fact that {\em uniformly 
accelerated observers} in Minkowski spacetime, i.e.~linearly 
accelerated observers with constant proper acceleration (also called 
{\em Rindler observers}), associate a thermal bath of {\em Rindler 
particles} (also called {\em Fulling-Rindler particles}) 
to the {\em no-particle state} of inertial observers
(also called {\em Minkowski vacuum}). 
Rindler particles are associated with positive-energy modes as 
defined by {\em Rindler observers} in contrast to Minkowski particles, 
which are associated with positive-energy modes as defined by inertial 
observers. \textcite{Unruh76} also provides an explanation for 
the conclusion obtained by \textcite{Davies75} 
that an observer undergoing uniform acceleration $a = {\rm const}$ 
in Minkowski spacetime would see a fixed inertial mirror to emit 
thermal radiation with temperature $a \hbar/(2 \pi k c )$, and the reason 
why this is not in contradiction with energy conservation. 
Although there are some accounts in the literature discussing the 
Unruh effect,\footnote{See, e.g.,
\textcite{Birrelletal82,Fullingetal87,Takagi86,
          Ginzburgetal87,WaldQFTCS,Sciamaetal81}.
                       }
we believe that this review will be a useful contribution 
for the reasons we list below. 

Firstly, some authors have recently questioned the existence of the
Unruh effect~\cite{Belinskiietal02, Belinskiietal04}. 
We believe there are two main sources of confusion,
which need to be clarified in order to address these objections.
One is that it has not been 
made clear that the heuristic expression of the Minkowski
vacuum as a superposition of Rindler states makes sense outside as well
as inside the two Rindler wedges.
Although this point is not central to the Unruh
effect~\cite{rebuttal}, it will be useful to point out that this
heuristic expression in fact makes sense in the whole of Minkowski
spacetime. 
Another common source of confusion is that the 
Unruh effect is sometimes tacitly assumed to be the equivalence 
of the {\em excitation rate} of a detector when it is 
(i)~uniformly accelerated in the Minkowski vacuum and 
(ii)~static in a thermal bath of Minkowski particles 
(rather than of Rindler particles). There is no such equivalence 
in general, although this showed up by coincidence in some early 
examples in the literature 
(see discussion in Sec.~\ref{subsubsection:staticdetectors}). 
We emphasize that this point does not contradict 
the fact that the {\em detailed balance relation} 
satisfied by static detectors in a thermal bath of Minkowski particles 
is in general also valid for uniformly accelerated ones in the Minkowski 
vacuum \cite{Unruh76}. The identification of 
the Unruh effect with the behavior of accelerated detectors 
seems to have generated sometimes unnecessary confusion. It is 
worthwhile to emphasize that the Unruh effect is a quantum 
field theory result, which does not depend on the introduction 
of the detector concept. In this sense, it is better to see the 
detailed balance relation satisfied by uniformly accelerated 
detectors as {\em a natural consequence} or {\em application}
rather than a {\em definition} of the Unruh effect. In order to 
exemplify the meaning of the Unruh effect  as the equivalence 
between the Minkowski vacuum and a thermal bath of Rindler 
particles, we collect and discuss a number of illustrative 
physical applications.

The Unruh effect has also been connected with the long-standing 
discussion about whether or not 
sources\footnote{Throughout this review we will use the word ``sources" 
to mean {\em scalar sources, particle detectors} or 
{\em electric charges} depending on the context where it appears.} 
{\em uniformly accelerated} 
from the null past infinity to the null future infinity radiate 
with respect to inertial observers. Although some aspects of this 
issue are surely worthwhile to be investigated and the corresponding 
discussion can be enriched by considering the Unruh 
effect, it is useful to keep in mind that the Unruh effect 
does not depend on a consensus on this issue which seems to be mostly 
semantic~[see \textcite{Fulling05} for a brief discussion on it 
and related references]. We comment briefly on this issue in 
Sec.~\ref{subsubsection:controversy1}. 

Secondly, there have been several recent proposals to detect the
Unruh effect in the laboratory and it is useful to review and assess 
them. We shall emphasize that, although it is certainly fine to interpret 
laboratory observables from the point of view of uniformly accelerated 
observers, the Unruh effect 
itself does not need experimental confirmation any more than free 
quantum field theory does.\footnote{This statement should be understood
in the sense that we are dealing with mathematical constructions that 
stand on their own. The assertions follow from the definitions and so  
do not need to be experimentally verified. The fact that ``Rindler and 
Minkowski perspectives" give consistent physical predictions is a 
consequence of the validity of these constructions.}

\begin{figure}[t]
\epsfig{file=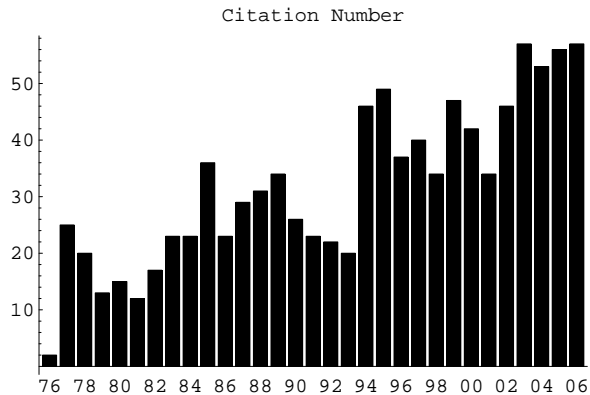,angle=0,width=0.9\linewidth,clip=}
\caption{\label{citationunruh} Histogram depicting the number 
of citations of \textcite{Unruh76} over the years.}
\end{figure}

Finally, there has been an increasing interest in the Unruh effect 
(see Fig.~\ref{citationunruh}) because of its connection with a number of 
contemporary research topics.\footnote{The data in 
Fig.~\ref{citationunruh} should 
not be used to infer any relative or absolute measure of the importance 
of the Unruh effect. It has been introduced only as an illustration on the
increasing interest in this issue.} The thermodynamics of black holes 
and the corresponding information puzzle is one of them. It will be beneficial, 
therefore, to review the literature on the 
generalized second law,\footnote{See, e.g., 
                                 \textcite{UnruhWald82,UnruhWald83,Wald01}.
                                 }
quantum information,\footnote{See, e.g., 
                              \textcite{Bousso02,Peresetal04}.
                              }
and related topics with the Unruh effect as the central theme. 

The review is organized as follows. In Section~\ref{section:Unruheffect} 
we review the derivation of the Unruh effect, 
emphasizing the fact that
the quantum field expanded by the Rindler modes can be used in the whole
of Minkowski spacetime, partly to respond to 
the recent criticisms mentioned above.
We also touch upon more rigorous 
approaches 
such as the Bisognano-Wichmann theorem 
in algebraic field theory and
general theorems on field theories in spacetimes 
with bifurcate Killing horizons.
A brief discussion of the Unruh effect in interacting field theories is
also included. 
In Section~\ref{section:Applications} we present in detail some 
typical examples which illustrate the physical content of the Unruh 
effect. We start by reviewing the behavior of accelerated detectors
which can be also used to describe the physics of accelerated atomic systems. 
Then, we analyze the weak decay of accelerated protons 
and the bremsstrahlung from accelerated charged particles.
Section~\ref{section:Experimentalproposals} is dedicated to the 
discussion of some experimental proposals for laboratory signatures
of the Unruh effect in particle accelerators,
in microwave cavities, in the presence of ultra-intense lasers, 
in the vicinity of
accelerated boundaries, and in hadronic processes. 
In Section~\ref{section:Recentdevelopments} we comment on some recent
developments concerning  the Unruh effect, which include
the possible reduction in fidelity of teleportation when one 
party is accelerated, the decoherence of accelerated systems
and the possible observer-dependence of the entropy concept. 
We conclude the review with a summary in Section~\ref{section:Summary}. 
Throughout this review 
we use natural units $\hbar=c=G=k=1$ and signature $(+ - - -)$ 
unless stated otherwise.

It would be impossible to give a completely balanced account of so much 
work in the literature concerning the Unruh effect.  This review heavily
reflects our own experience with the Unruh effect, 
and we fear that we may have overlooked some important related papers.
However, we hope at least to have included 
a sufficient number of papers to allow the 
readers to trace back to most of the important related results. 

\section{The Unruh effect}
\label{section:Unruheffect}

\subsection{Free scalar field in curved spacetime}
\label{section2:curvedspace}

Even though the Unruh effect is about quantum field theory (QFT) 
in flat spacetime, 
it is useful to review briefly the general framework of non-interacting 
QFT in curved spacetime.  We treat only the simplest
theory, i.e.~the theory of a Hermitian scalar field satisfying the
Klein-Gordon equation.
We present it in a schematic and heuristic way as is done in
\textcite{Birrelletal82}.  
A mathematically more satisfactory treatment can be found, e.g., 
in \textcite{WaldQFTCS}. 

Let us first remind the reader of some important features of QFT in
Minkowski spacetime.
In this spacetime the scalar field is expanded in terms of the
energy-momentum eigenfunctions, and the vacuum state is defined as the
state annihilated by all {\em annihilation operators}, 
i.e.~the coefficient operators of the {\em positive-frequency}
eigenfunctions defined to be those proportional 
to $e^{-ik_0 t}$ with $k_0 > 0$, where $t$ is the time parameter.  The
coefficient operators of the {\em negative-frequency} 
modes proportional to
$e^{ik_0 t}$ are the {\em creation operators}, 
and the states obtained by
applying creation operators 
on the vacuum state are identified with states containing
particles.  Notice that the time-translation symmetry, 
which enables one to define positive- and negative-frequency
solutions to the Klein-Gordon equation, 
plays a crucial role 
in the definition of the vacuum state and the Fock space of particles.
Therefore, in a general curved spacetime with no isometries, there is no
reason to expect the existence of a preferred vacuum state or 
a useful concept of particles.

For simplicity we specialize to $(D+1)$-dimensional 
spacetimes whose metric takes the form
\begin{equation}
ds^2 = [N(x)]^2 dt^2 - G_{ab}(x)dx^a dx^b, \label{sec2:met}
\end{equation}
where $x = (t,{\bf x})$.  The coefficient $N(x)$ 
is called the {\em lapse function}~\cite{Arnowittetal62} 
and $G_{ab}$ is the metric on the spacelike hypersurface of constant $t$.
(All spacetimes we consider in this review have a metric of this form.) 
In this spacetime the minimally-coupled\footnote{It is customary to
allow the field to couple to the scalar curvature.  Thus, 
the general Klein-Gordon equation takes the form 
$(\nabla_\mu\nabla^\mu +\xi R + m^2)\phi = 0$.  
The minimally-coupled scalar field has $\xi=0$ by definition.} 
massive Klein-Gordon equation 
$(\nabla_\mu \nabla^\mu + m^2)\phi = 0$, 
which arises as the Euler-Lagrange equation from the 
Lagrangian density,
\begin{equation}
{\cal L} = \sqrt{-g}\,
           (\nabla_\mu\phi\nabla^\mu\phi - m^2\phi^2 )/2,
\label{sec2:lag}
\end{equation}
takes the form 
\begin{equation}
\partial_t ( N^{-1}\sqrt{G}\partial_t\phi) 
- \partial_a (N \sqrt{G}G^{ab}\partial_b\phi)
+ N\sqrt{G}m^2\phi = 0, \label{sec2:KG}
\end{equation}
where the space indices $a$ and $b$ run from $1$ to $D$.

Given two complex solutions $f_A(x)$ and $f_B(x)$ to the Klein-Gordon 
equation, we define the Klein-Gordon current 
\begin{equation}
J^\mu_{(f_A,f_B)}(x) \equiv f^*_A(x)\nabla^\mu f_B(x)- 
f_B(x)\nabla^\mu f^*_A(x).
\end{equation}  
Then, one can readily show that $\nabla_\mu J^\mu_{(f_A,f_B)}(x) = 0$.  
Hence, the quantity
\begin{equation}
(f_A,f_B)_{\rm KG} \equiv i\int d^D{\bf x}\sqrt{G}n_\mu 
J_{(f_A,f_B)}^\mu \label{sec2:KGinner}
\end{equation}
is independent of $t$, 
where $n_\mu$ is the future-directed unit vector normal to 
the hypersurface $\Sigma$ of constant $t$. (The integral here and
throughout this subsection (Sec.~\ref{section2:curvedspace})
is over the hypersurface $\Sigma$.)
We call this quantity the 
Klein-Gordon inner product of $f_A$ and $f_B$.  
For the metric (\ref{sec2:met}) it takes the following form:
\begin{equation}
(f_A,f_B)_{\rm KG} = i \int d^D{\bf x}\sqrt{G}N^{-1}(f^*_A\partial_t f_B -
f_B\partial_t f^*_A).\label{sec2:KGprod}
\end{equation}
The conjugate momentum density $\pi(x)$ is defined as
$
\pi \equiv {\partial {\cal L}}/{\partial \dot{\phi}}
$,
where $\dot{\phi} \equiv \partial_t\phi$.  
For the metric (\ref{sec2:met}) one finds
\begin{equation}
\pi(x) = N^{-1}\sqrt{G}\dot{\phi}(x). \label{sec2:defpi}
\end{equation}
Note that, if we let $p_A(x)$ and $p_B(x)$ be the conjugate momentum density for the
solutions $f_A(x)$ and $f_B(x)$, respectively,
then the Klein-Gordon inner product can be expressed as
\begin{equation}
(f_A,f_B)_{\rm KG} = i\int d^D{\bf x} [
 f_A^*(x) p_B(x)
- p_A^*(x)f_B(x) ].
\end{equation}
We assume that the Klein-Gordon equation determines the classical field
$\phi(x)$ uniquely given a (well-behaved) initial data $(\phi,\pi)$ on a
hypersurface of constant $t$.  This property is known to hold if
the spacetime is globally hyperbolic with $t= {\rm const}$ hypersurfaces
as the spacelike 
Cauchy surfaces.\footnote{A {\em Cauchy surface} is a closed
hypersurface which is intersected by each inextendible timelike curve once
and only once. A spacetime is said to be {\em globally hyperbolic} if it
possesses a Cauchy surface.  See, e.g., \textcite{Waldbook84}.\label{foot6}}

The quantization of the field $\phi$ proceeds as follows.  
We denote the field operators corresponding to $\phi$ and $\pi$ by
$\hat{\phi}$ and $\hat{\pi}$, respectively. 
One imposes the following equal-time canonical commutation relations:
\begin{eqnarray}
\left[ \hat{\phi}(t,{\bf x}),\hat{\phi}(t,{\bf x}^\prime) \right] 
& = & 
\left[ \hat{\pi}(t,{\bf x}),\hat{\pi}(t,{\bf x}^\prime) \right] 
= 0,
\label{sec2:canonical1}
\\
\left[ \hat{\phi} (t,{\bf x}), \hat{\pi} (t, {\bf x}^{\prime} ) \right]
& = & 
i\delta^D ({\bf x}, {\bf x}^\prime ),
\label{sec2:canonical2}
\end{eqnarray}
where the delta-function 
$\delta^D({\bf x},{\bf x}^\prime)$ is defined by 
\begin{equation}
\int d^D{\bf x}f({\bf x})\delta^D({\bf x},{\bf x}^\prime) 
= f({\bf x}^\prime).
\end{equation}
Note here that there is no density factor $\sqrt{G}$ on the left-hand
side.
For arbitrary complex-valued solutions $f_A(x)$ and $f_B(x)$ 
to the Klein-Gordon equation (\ref{sec2:KG})
(with a suitable integrability conditions) one finds
\begin{equation}
[(f_A,\hat{\phi})_{\rm KG},(\hat{\phi},f_B)_{\rm KG} ] 
= (f_A,f_B)_{\rm KG},   \label{sec2:KGfg}
\end{equation}
from the equal-time canonical commutation 
relations (\ref{sec2:canonical1}) and (\ref{sec2:canonical2}) by using 
Eq.~(\ref{sec2:defpi}).  

Now, assume that there is a complete set of solutions, $\{f_i, f^*_i\}$,
to the 
Klein-Gordon equation (\ref{sec2:KG}) satisfying 
\begin{eqnarray}
(f_i, f_j)_{\rm KG} & = & - (f_i^*,f_j^*)_{\rm KG} = \delta_{ij}, 
\label{sec2:posfreq1}\\
(f_i^*,f_j)_{\rm KG} & = & (f_i,f_j^*)_{\rm KG}
= 0. \label{sec2:posfreq2}
\end{eqnarray}
We assume here that the indices labeling the solutions are discrete for
simplicity of the discussion but its extension to the cases with
continuous labels is straightforward.
In Minkowski spacetime one chooses the positive-frequency modes as
$f_i$'s and, consequently, the negative-frequency modes as $f_i^*$'s. In
a general globally-hyperbolic curved spacetime without isometries, there
are infinitely many ways of choosing the functions $f_i$'s.

Expanding the quantum field $\hat{\phi}(x)$ as
\begin{equation}
\phi(x) = \sum_{i}\left[ \hat{a}_i f_i(x) + 
\hat{a}_i^\dagger f_i^*(x)\right],
\end{equation}
one finds
\begin{equation}
\hat{a}_i =  (f_i,\phi)_{\rm KG}, \;\;\;
\hat{a}_i^\dagger =  (\phi,f_i)_{\rm KG}.
\end{equation}
One can readily show, by using Eqs.~(\ref{sec2:KGfg}),
(\ref{sec2:posfreq1}) and (\ref{sec2:posfreq2}), that
\begin{equation}
[ \hat{a}_i ,\hat{a}_j ]  
= 
[ \hat{a}_i^\dagger, \hat{a}_j^\dagger ] = 0,\;\;\;
[ \hat{a}_i , \hat{a}_j^\dagger ] 
=  
\delta_{i j}. 
\label{sec2:crean}
\end{equation}
Conversely, if these commutation relations 
are satisfied, then the equal-time
canonical commutation relations (\ref{sec2:canonical1}) and
(\ref{sec2:canonical2}) follow.  To prove this, one first shows that any 
two complex-valued solutions $f_A(x)$ and $f_B(x)$ to the Klein-Gordon
equation (\ref{sec2:KG}) satisfy Eq.~(\ref{sec2:KGfg}) by expanding them
in terms of $f_i(x)$ and $f_i^*(x)$ and using the commutators
(\ref{sec2:crean}).  Then, for example, by
letting $f_A(t,{\bf x}) = f_B^*(t,{\bf x})$ and
$p_A(t,{\bf x}) = -p_B^*(t,{\bf x})$, at a given time $t$ 
and evaluating the Klein-Gordon
inner products in Eq.~(\ref{sec2:KGfg}) at time $t$,
one obtains
\begin{eqnarray}
&& \int d^D{\bf x}\, d^D{\bf x}'
f_B(t,{\bf x})p_B(t,{\bf x}')\left[\hat{\phi}(t,{\bf x}),\hat{\pi}(t,{\bf
x}')\right]\nonumber \\
&& = i\int d^D{\bf x}f_B(t,{\bf x})p_B(t,{\bf x}).
\end{eqnarray}
Since $f_B(t,{\bf x})$ and $p_B(t,{\bf x})$ are arbitrary, one can
conclude that Eq.~(\ref{sec2:canonical2}) holds at time $t$. 
Eq.~(\ref{sec2:canonical1}) can be derived in a similar manner.

The operators $\hat{a}_i$ 
and $\hat{a}^\dagger_i$ are called the {\em 
annihilation and creation 
operators}, respectively.  The vacuum state $|0\rangle$ 
is defined by requiring 
$\hat{a}_i|0\rangle=0$.  The Fock space of states is obtained by applying
the creation operators $\hat{a}_i^\dagger$ on the vacuum state 
$|0\rangle$.  We call the operator 
$\hat{N}_i = \hat{a}_i^\dagger \hat{a}_i$ 
(with no summation on the right-hand side) the {\em number operator}
in the mode `$i$'.  
However, note that, since it is not always 
easy to construct a (theoretical) detector model which is excited when 
the eigenvalue of $\hat{N}_i$ 
changes from 1 to 0, say, the operator $\hat{N}_i$ 
does not necessarily lead to a useful particle concept.

Since the coefficient operators $\hat{a}_i$ of the functions $f_i$ 
annihilate the vacuum state $|0\rangle$, the choice of the functions 
$f_i$ satisfying Eqs.~(\ref{sec2:posfreq1}) and (\ref{sec2:posfreq2}) 
determines the vacuum state. For this reason we call the functions 
$f_i$ the {\em positive-frequency modes} and their complex conjugates 
$f_i^*$ the {\em negative-frequency modes} in analogy with the case in
Minkowski spacetime.  Thus, the choice of the 
positive-frequency modes determines the vacuum state.  In a general 
curved spacetime there is no privileged choice of the positive-frequency
modes, and consequently, there is no privileged vacuum state 
unlike in Minkowski spacetime, as we mentioned before.

Now, suppose that two complete sets of positive-frequency modes 
$\{ f_i^{(1)}\}$ and $\{ f_I^{(2)}\}$ satisfy 
the Klein-Gordon inner-product relations (\ref{sec2:posfreq1}) 
and (\ref{sec2:posfreq2}), where the lower-case letters $i$, $j$
are replaced by the upper-case equivalents $I$, $J$ for $f_I^{(2)}$.
Since both sets are complete, the modes 
$f^{(2)}_I$ can be expressed as linear combinations 
of $f_i^{(1)}$ and $f_i^{(1)*}$ and vice versa.  Thus,
\begin{eqnarray}
f_I^{(2)} & = & \sum_i \left[\alpha_{Ii}f_i^{(1)} + 
\beta_{Ii}f_i^{(1)*}\right],\label{sec2:exp1}\\
f_I^{(2)*} & = & \sum_i \left[\alpha_{Ii}^* f_i^{(1)*} + 
\beta_{Ii}^* f_i^{(1)}\right].
\label{sec2:exp2}
\end{eqnarray}
By noting that 
\begin{eqnarray}
\alpha_{Ii} & = & 
( f_i^{(1)},f_I^{(2)})_{\rm KG} = (f_I^{(2)},f_i^{(1)})_{\rm KG}^*,\\
\beta_{Ii} & = & - (f_i^{(1)*}, f_I^{(2)})_{\rm KG} 
= (f_I^{(2)*}, f_i^{(1)})_{\rm KG},
\end{eqnarray}
one can express $f_i^{(1)}$ as a linear combination of 
$f_I^{(2)}$ and $f_I^{(2)*}$ as
\begin{eqnarray}
f_i^{(1)} & = & \sum_I \left[\alpha_{Ii}^* f_I^{(2)} 
- \beta_{Ii}f_I^{(2)*}\right], \label{sec2:expp1}\\
f_i^{(1)*} & = & \sum_I\left[\alpha_{Ii}f_I^{(2)*} - 
\beta_{Ii}^* f_I^{(2)}\right]. \label{sec2:expp2}
\end{eqnarray}

The scalar field $\hat{\phi}(x)$ 
can be expanded using either of the two sets $\{f_i^{(1)}\}$ and 
$\{f_I^{(2)}\}$:
\begin{eqnarray}
\hat{\phi}(x) & = & 
\sum_i \left[ \hat{a}_i^{(1)}f_i^{(1)} + \hat{a}_i^{(1)\dagger}
f_i^{(1)*}\right] \nonumber \\
& = & \sum_I\left[ \hat{a}_I^{(2)}f_I^{(2)} + \hat{a}_I^{(2)\dagger}
f_I^{(2)*}\right].
\end{eqnarray}
Using the expansion given by Eqs.~(\ref{sec2:exp1}) and (\ref{sec2:exp2}),
and comparing the coefficients of $f_i^{(1)}$ and $f_i^{(1)*}$, we find
\begin{equation}
\hat{a}_i^{(1)} = \sum_I (\alpha_{Ii}\hat{a}_I^{(2)} 
+ \beta_{Ii}^* \hat{a}_I^{(2)\dagger}), \label{sec2:bog1}
\end{equation} 
and similarly by using 
Eqs.~(\ref{sec2:expp1}) and (\ref{sec2:expp2}) we have
\begin{equation}
\hat{a}_I^{(2)} = \sum_i (\alpha_{Ii}^*\hat{a}_i^{(1)} 
- \beta_{Ii}^* \hat{a}_i^{(1)\dagger}). \label{sec2:bog2}
\end{equation}
This transformation, which mixes annihilation and creation operators, 
is called a
{\em Bogolubov transformation}, 
and the coefficients $\alpha_{Ii}$ and $\beta_{Ii}$
are called the {\em Bogolubov coefficients}.  
The Bogolubov transformation found its
first major application to QFT in curved spacetime in
the derivation of particle creation in expanding
universes~\cite{Parker68,Sexletal69}. 

The vacuum states $|0_{(1)}\rangle$ and $|0_{(2)}\rangle$ 
corresponding to the two sets of positive-frequency modes 
$\{f_i^{(1)}\}$ and $\{f_I^{(2)}\}$, respectively, are distinct if 
$\beta_{Ii}$ do not vanish for all $I$ and $i$.  For example, the
expectation value of the number operator
$\hat{N}^{(1)}_i = \hat{a}_i^{(1)\dagger}\hat{a}_i^{(1)}$ for the state
$|0_{(1)}\rangle$ is zero by definition but for the state
$|0_{(2)}\rangle$ it can be calculated by using Eq.~(\ref{sec2:bog1}) as
\begin{equation}
\langle 0_{(2)}|N^{(1)}_i|0_{(2)}\rangle  = \sum_I |\beta_{Ii}|^2.
\end{equation}
We similarly find for the number operator 
$N^{(2)}_I = \hat{a}_I^{(2)\dagger}\hat{a}_I^{(2)}$,
\begin{equation}
\langle 0_{(1)}|N^{(2)}_I|0_{(1)}\rangle = \sum_i |\beta_{Ii}|^2.
\end{equation}

Although the choice of the vacuum state is not 
unique in general, 
{\em there is a natural vacuum state if the spacetime is static}, 
i.e.~if the spacetime 
metric is of the form (\ref{sec2:met}) with the lapse function $N(x)$ 
and the metric $G_{ab}$ being independent of $t$.\footnote{In fact, if a
globally hyperbolic 
spacetime is stationary, i.e.~if the metric is $t$-independent with
$(\partial/\partial t)^\mu$ being everywhere timelike but with
the cross term $g_{ti}$, $i\neq t$, not necessarily vanishing, one has a
natural vacuum state in this spacetime under certain
conditions~\cite{Ashtekaretal75,Kay78}.}
In such a case the 
equation for determining the mode functions becomes
\begin{equation}
\partial_t^2 f_i = 
NG^{-1/2}\partial_a (N G^{1/2}G^{ab} \partial_b f_i) - N^2 m^2 f_i.
\label{sec2:t-indepeq}
\end{equation}
Then, it is natural to let the positive-frequency solutions 
$f_i$ have a $t$-dependence of the form $e^{-i\omega_i t}$,
where $\omega_i$ are positive constants interpreted as the energy of the
particle with respect to the (future-directed) 
Killing vector\footnote{A Killing vector 
$K^\mu$ is a vector satisfying
$\nabla_\mu K_\nu + \nabla_\nu K_\mu = K^\alpha\partial_\alpha
g_{\mu\nu} + g_{\alpha\nu}\partial_\mu K^\alpha +
g_{\mu\alpha}\partial_\nu K^\alpha = 0$.  In a coordinate system such
that $K^\mu = (\partial/\partial\theta)^\mu$, one has
$\partial g_{\mu\nu}/\partial\theta = 0$. See, e.g., 
\textcite{Waldbook84}.} 
$\partial/\partial t$.
If the spacetime is globally hyperbolic
and static, then this choice of
positive-frequency modes leads to a well-defined and natural 
vacuum state that preserves the time-translation symmetry.  We call this
state the {\em static vacuum}.

Minkowski spacetime has global timelike Killing vector fields, 
which generate
time translations in various inertial frames.  The sets of
positive-frequency modes corresponding to these Killing vectors are the
same and are the usual positive-frequency modes proportional to
$e^{-ik_0 t}$ with $k_0 > 0$, where $t$ is the time parameter with
respect to one of the inertial frames.  Thus, all these Killing vector
fields define the same vacuum state.\footnote{It has been shown by
\textcite{Chmielowski94} that two commuting 
global timelike Killing vector fields define the same vacuum state.}

Now, in the region defined by $|t|<z$ 
in Minkowski spacetime
(here, $z$ is one of the space coordinates), the
boost Killing vector $z(\partial/\partial t) + t(\partial/\partial z)$,
i.e.~the vector with
$t$- and $z$-components being $z$ and $t$, respectively,
is timelike and future-directed.  Hence, this region,
called the {\em right Rindler wedge}, is a static spacetime with this
Killing vector playing the role of time translation. 
Thus, one can define the corresponding static vacuum state.  
As was observed by
\textcite{Fulling73}, this vacuum state is not the same as the state
obtained by restricting the usual Minkowski vacuum to this region.  
This observation is
crucial in understanding the Unruh effect, as we shall explain in the
next subsections.

\subsection{Rindler wedges}
\label{subsection:Rindler}

\begin{figure}[t]
\epsfig{file=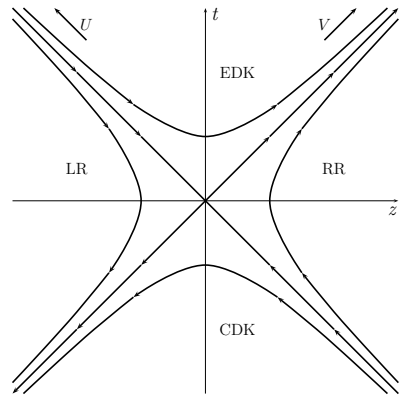,angle=0,width=0.60\linewidth,clip=}
\caption{\label{MilneRindler} The regions with $|t|<z$, $|t| < -z$,
$t>|z|$ and $t< -|z|$, denoted RR, LR, EDK and CDK, respectively, are
the left Rindler wedge, right Rindler wedge, expanding degenerate Kasner 
universe and
contracting degenerate Kasner universe, respectively.
The curves with arrows are integral
curves of the boost Killing vector $z(\partial/\partial t) +
t(\partial/\partial z)$.
The direction of increasing $U=t-z$ and that of
increasing $V=t+z$ are also indicated.}
\end{figure}

As we have seen in the previous subsection, one has a natural static 
vacuum state in a static globally hyperbolic spacetime.  
Minkowski spacetime with the metric
\begin{equation}
ds^2 = dt^2 - dx^2 - dy^2 - dz^2
\end{equation}
is of course a static globally hyperbolic 
spacetime.  As mentioned above, 
the part of this
spacetime defined by $|t| < z$, 
called the right Rindler wedge, is also
a static globally hyperbolic spacetime.  The
region with the condition $|t| < -z$ is called 
the {\em left Rindler wedge},
and is also a static globally hyperbolic spacetime.  The region with 
$t > |z|$, also a globally hyperbolic spacetime though not a static one, 
is called the {\em expanding
degenerate Kasner universe} and the globally hyperbolic spacetime with the condition 
$t < -|z|$ is called the {\em contracting degenerate Kasner universe}.  
These regions are shown
in Fig.~\ref{MilneRindler}. 

Minkowski spacetime is invariant under the boost
\begin{eqnarray}
t & \mapsto & t\cosh \beta + z\sinh\beta, \label{sec2:boost1}\\
z & \mapsto & t\sinh \beta + z \cosh \beta, \label{sec2:boost2}
\end{eqnarray}
where $\beta$ is the boost parameter. 
These transformations are generated by the
Killing vector
$z(\partial/\partial t) + t(\partial/\partial z)$.
The boost invariance of Minkowski spacetime
motivates the following coordinate transformation:
\begin{equation}
t  =  \rho\sinh\eta, \;\;\;
z  =  \rho\cosh\eta, 
\label{sec2:Rindcoord}
\end{equation}
where $\rho$ and $\eta$ takes any real value.  Then, the Killing vector
is $\partial/\partial\eta$, and the metric takes
the form
\begin{equation}
ds^2 = \rho^2d\eta^2 - d\rho^2 - dx^2 - dy^2. \label{sec2:RindlerRL}
\end{equation}
This metric is independent of $\eta$ as expected.  The world lines
with fixed values of $\rho$, $x$, $y$ are the trajectories of
the boost transformation given by Eqs.~(\ref{sec2:boost1}) and
(\ref{sec2:boost2}). Each world line has a constant proper
acceleration given by $\rho^{-1} = {\rm const}.$

The coordinates $(\eta,\rho,x,y)$ cover only the regions with $z^2 >
t^2$, i.e.~the left and right Rindler wedges,
as can readily be seen from Eq.~(\ref{sec2:Rindcoord}).  
To discuss QFT in the right
Rindler wedge, it is convenient to make a further coordinate
transformation $\rho = a^{-1}e^{a\xi}$, $\eta = a\tau$, i.e.
\begin{equation}
t = a^{-1}e^{a\xi}\sinh a\tau ,\;\;\;
z = a^{-1}e^{a\xi}\cosh a\tau,  
\label{sec2:rightcoords}
\end{equation}
where $a$ is a positive constant~\cite{Rindler66}.
Then, the metric takes the form
\begin{equation}
ds^2 = e^{2a\xi}(d\tau^2 - d\xi^2) - dx^2 - dy^2. 
\label{sec2:rightmetric}
\end{equation}
This coordinate system will be useful because the world line with $\xi=0$
has a constant acceleration of $a$.  
The coordinates $(\bar{\tau},\bar{\xi})$ for the left Rindler wedge
are given by
\begin{equation}
t = a^{-1}e^{a\bar{\xi}}\sinh a\bar{\tau},\;\;\;
z = - a^{-1}e^{a\bar{\xi}}\cosh a\bar{\tau}.
\label{sec2:leftcoords} 
\end{equation}

The Killing vector $z(\partial/\partial t) + t(\partial/\partial z)$ is
timelike in the two Rindler wedges and spacelike in the
degenerate Kasner universes.  It becomes null on 
the hypersurfaces 
$t=\pm z$ dividing Minkowski spacetime into the four regions. 
These hypersurfaces are examples of {\em Killing horizons}, 
which are defined as null hypersurfaces to which the Killing
field is normal \cite{WaldQFTCS}.

Since the right (or left) Rindler wedge
is a static spacetime in its own right, it has a
natural static vacuum state as we noted before.
{\em The Unruh effect is defined in this review as the fact that
the usual vacuum state for QFT in Minkowski spacetime
restricted to the right Rindler wedge is a thermal state 
with $\tau$ playing 
the role of time, and similarly for the left Rindler wedge.}  
The correlation between the right
and left Rindler wedges in the usual Minkowski vacuum state plays an
important role in the Unruh effect.

\subsection{Two dimensional example}
\label{twodimensions}

The two dimensional massless scalar field in Minkowski spacetime is
problematic because of infrared divergences~\cite{Coleman73}. 
Nevertheless, this theory
is a very good model for explaining the Unruh effect,
and it is not necessary to deal with the infrared divergences for this
purpose.  It also turns out that the Unruh effect in scalar
field theory in higher dimensions can be derived in essentially the same
manner as in this model.

The massless scalar field in two dimensions, 
$\hat{\Phi}(t,z)$, satisfies
\begin{equation}
     ( {\partial^2}/{\partial t^2}
        -
      {\partial^2}/{\partial z^2}) \hat{\Phi} 
= 0. 
\label{sec2:twodimMink}
\end{equation}
This field can be expanded as
\begin{eqnarray}
\hat{\Phi}(t,z) & = & \int_{0}^\infty \frac{dk}{\sqrt{4\pi k}} \left(
\hat{b}_{-k}e^{-ik(t-z)} + \hat{b}_{+k}e^{-ik(t+z)}\right.\nonumber \\
&& \ \ \ \ \ \ \ \ \ \ \left. + \hat{b}_{-k}^\dagger
e^{ik(t-z)} +
\hat{b}_{+k}^\dagger e^{ik(t+z)}\right).
\end{eqnarray}
The annihilation and creation operators satisfy
\begin{equation}
[ \hat{b}_{\pm k}, \hat{b}_{\pm k'}^\dagger ] = \delta(k-k')
\end{equation}
with all other commutators vanishing.
By using the definitions
\begin{equation}
U  =  t - z,\;\;\;
V  =  t + z,
\end{equation}
we can write
\begin{equation}
\hat{\Phi}(t,z) = \hat{\Phi}_{-}(U) + \hat{\Phi}_{+}(V),
\end{equation}
where
\begin{equation}
\hat{\Phi}_{+}(V) = \int_0^\infty dk \left[ \hat{b}_{+k}f_k(V) +
\hat{b}_{+k}^\dagger f_k^*(V)\right]
\end{equation}
with
\begin{equation}
f_k(V) = (4\pi k)^{-1/2}e^{-ikV},
\end{equation}
and similarly for $\hat{\Phi}_{-}(U)$. 
Since the left and right-moving sectors of the field, 
$\hat{\Phi}_{+}(V)$ and
$\hat{\Phi}_{-}(U)$, do not interact with one another,  
we discuss only the left-moving sector $\hat{\Phi}_{+}(V)$.  
(Thus, we will
discuss the Unruh effect for the theory consisting only of the
left-moving sector.) The Minkowski vacuum state $|0_{\rm M}\rangle$ is
defined by $\hat{b}_{+k}|0_{\rm M}\rangle = 0$ for all $k$.

Using the metric in the right Rindler wedge given by
Eq.~(\ref{sec2:rightmetric}), one finds a field equation of the same
form as Eq.~(\ref{sec2:twodimMink}):
\begin{equation}
\left( {\partial^2}/{\partial\tau^2} 
- {\partial^2}/{\partial\xi^2}\right)\hat{\Phi} = 0.
\end{equation}
(This is a result of the 
conformal invariance of the massless scalar field
theory in two dimensions.)  The solutions to this differential equation
can be classified again into the left- and right-moving modes which
depend only on $v=\tau+\xi$ and $u=\tau-\xi$, respectively.  These
variables are related to $U$ and $V$ as follows:
\begin{eqnarray}
U & = & t-z = -a^{-1}e^{-au},\\
V & = & t+z = a^{-1}e^{av}.
\end{eqnarray}
The Lagrangian density is invariant under the coordinate transformation
$(t,z)\mapsto (\tau,\xi)$.  As a result, 
by going through the quantization procedure laid out in
Sec.~\ref{section2:curvedspace} one finds exactly the same theory
as in the whole of Minkowski spacetime with $(t,z)$ replaced by
$(\tau,\xi)$.  Thus, we have, for $0 < V$,
\begin{eqnarray}
\hat{\Phi}_{+}(V) & = & \int_0^\infty d\omega\,\left[ 
\hat{a}_{+\omega}^Rg_{\omega}(v)
+ \hat{a}_{+\omega}^{R\dagger}g_{\omega}^*(v) \right],\label{sec2:phi+}
\end{eqnarray}
where
\begin{equation}
g_{\omega}(v) = (4\pi\omega)^{-1/2}\,e^{-i\omega v}, \label{sec2:gomegav}
\end{equation}
and where
\begin{equation}
[ \hat{a}_{+\omega}^R,\hat{a}_{+\omega'}^{R\dagger}] =
\delta(\omega-\omega')
\label{sec2:Acommutators}
\end{equation}
with all other commutators vanishing.
Notice that the functions $g_\omega(v)$ are eigenfunctions of the boost
generator $\partial/\partial\tau$.

The field $\hat{\Phi}_{+}(V)$ 
can be expressed in the left
Rindler wedge with the condition $V < 0 < U$, by using the
left Rindler coordinates $(\bar{\tau},\bar{\xi})$ defined by
Eq.~(\ref{sec2:leftcoords}).  Defining 
$\bar{v} = \bar{\tau} - \bar{\xi}$,  one obtains
Eqs.~(\ref{sec2:phi+})--(\ref{sec2:Acommutators}) with $v$
replaced by $\bar{v}$ and with the annihilation and
creation operators $\hat{a}_{+\omega}^R$ and 
$\hat{a}_{+\omega}^{R\dagger}$ replaced
by a new set of operators $\hat{a}_{+\omega}^L$ and
$\hat{a}_{+\omega}^{L\dagger}$.  The variable $\bar{v}$ 
is related to $V$ by
\begin{equation}
V  =  - a^{-1}e^{-a\bar{v}}.
\end{equation}
The static vacuum state in the left and right Rindler wedges,
the Rindler vacuum state $|0_{\rm R}\rangle$, is defined by 
$\hat{a}_{+\omega}^R|0_{\rm R}\rangle 
= \hat{a}_{+\omega}^L|0_{\rm R}\rangle = 0$ for
all $\omega$.

To understand the Unruh effect we need to find the Bogolubov
coefficients $\alpha^R_{\omega k}$, $\beta^R_{\omega k}$,
$\alpha^L_{\omega k}$ and $\beta^L_{\omega k}$, where
\begin{eqnarray}
\theta(V)g_\omega(v)
& = & 
\int_0^\infty \frac{dk}{\sqrt{4\pi
k}}\left(\alpha^R_{\omega k}e^{-ikV}
+\beta^R_{\omega k}e^{ikV}\right),\nonumber \\
 \label{sec2:gomegavRR}\\
\theta(-V)g_\omega(\bar{v})
& = &  
\int_0^\infty \frac{dk}{\sqrt{4\pi
k}}\left(\alpha^L_{\omega k}e^{-ikV}
+\beta^L_{\omega k}e^{ikV}\right). \nonumber \\
\label{sec2:gomegavLR}
\end{eqnarray}
Here, $\theta(x)=0$ if $x<0$ and $\theta(x)=1$ if $x>0$, i.e.~$\theta$
is the Heaviside function.  
To find $\alpha^R_{\omega k}$ we multiply Eq.~(\ref{sec2:gomegavRR}) by
$e^{ikV}/2\pi$, $k>0$, and integrate over $V$. Thus, we find
\begin{eqnarray}
\alpha^R_{\omega k}
& = & \sqrt{4\pi k}\int_0^\infty \frac{dV}{2\pi} g_{\omega}(V)e^{ikV}\nonumber \\
& = & \sqrt{\frac{k}{\omega}}\int_0^\infty
\frac{dV}{2\pi}(aV)^{-i\omega/a}e^{ikV}. \label{sec2:prealphaR}
\end{eqnarray}
We introduce a cut-off for this integral for
large $V$ by letting $V \to V+i\varepsilon$, 
$\varepsilon \to 0+$.\footnote{A cut-off of this kind
is always understood
in these calculations in field theory, as exemplified by the definition
of the delta function $\delta(k) =
\int (dx/2\pi)e^{ikx -\varepsilon |x|} = (2\pi
i)^{-1}[(k-i\varepsilon)^{-1} - (k+i\varepsilon)^{-1}]$.}
Then, changing the integration path to the
positive imaginary axis by letting $V=ix/k$, we find
\begin{eqnarray}
\alpha^R_{\omega k}
 & = &
\frac{ie^{\pi\omega/2a}}{\sqrt{\omega k}}\left(\frac{a}{k}\right)^{-i\omega/a}
\int_0^\infty 
\frac{dx}{2\pi}x^{-i\omega/a}e^{-x}\,dx\nonumber \\
& = & \frac{ie^{\pi\omega/2a}}{2\pi\sqrt{\omega k}}
\left(\frac{a}{k}\right)^{-i\omega/a}\Gamma(1-i\omega/a).
\label{sec2:alphaR}
\end{eqnarray}
To find the coefficients $\beta^R_{\omega k}$ we replace $e^{ikV}$ in
Eq.~(\ref{sec2:prealphaR}) by $e^{-ikV}$.  Then, the appropriate 
substitution is
$V=-ix/k$. As a result we obtain
\begin{equation}
\beta^R_{\omega k}
= - \frac{ie^{-\pi\omega/2a}}{2\pi\sqrt{\omega k}}
\left(\frac{a}{k}\right)^{-i\omega/a}\Gamma(1-i\omega/a).
\end{equation}
A similar calculation leads to
\begin{eqnarray}
\alpha^L_{\omega k} & = & -\frac{ie^{\pi\omega/2a}}
{2\pi\sqrt{\omega k}}
\left(\frac{a}{k}\right)^{i\omega/a}\Gamma(1+i\omega/a),\\
\beta^L_{\omega k} & = & \frac{ie^{-\pi\omega/2a}}{2\pi\sqrt{\omega
k}}
\left(\frac{a}{k}\right)^{i\omega/a}\Gamma(1+i\omega/a).
\end{eqnarray}
We find that these coefficients obey the following relations crucial to
the derivation of the Unruh effect:
\begin{equation}
\beta^L_{\omega k}  =  - e^{-\pi\omega/a}\alpha^{R*}_{\omega k}, \;\;\;
\beta^R_{\omega k}  =  - e^{-\pi\omega/a}\alpha^{L*}_{\omega k}.
\label{sec2:betaalpha}
\end{equation}
By substituting these relations in Eqs.~(\ref{sec2:gomegavRR}) and
(\ref{sec2:gomegavLR}) we find that the following functions are linear
combinations of positive-frequency modes $e^{-ikV}$ in Minkowski spacetime:
\begin{eqnarray}
G_\omega(V) & = & \theta(V)g_\omega(v) +
\theta(-V)e^{-\pi\omega/a}g_{\omega}^*(\bar{v}),
\label{sec2:cruc1a}
\\
\bar{G}_\omega(V) & = & 
\theta(-V)g_\omega(\bar{v}) + \theta(V)e^{-\pi\omega/a}g_{\omega}^*(v).
\label{sec2:cruc2a}
\end{eqnarray}
One can show that these functions 
are purely positive-frequency solutions in Minkowski
spacetime by analyticity argument as well: since a positive-frequency
solution is analytic in the lower-half plane on the complex $V$-plane,
the solution $g_\omega(v) = (4\pi\omega)^{-1/2}(V)^{-i\omega/a}$, 
$V>0$, should be continued to the negative real line avoiding the
singularity at $V=0$ around a small circle in the lower half-plane, thus
leading to $(4\pi\omega)^{-1/2}e^{-\pi\omega/a}(-V)^{-i\omega/a}$ for
$V<0$. This was the original argument by \textcite{Unruh76}.

Eqs.~(\ref{sec2:cruc1a}) and (\ref{sec2:cruc2a}) can be inverted as
\begin{eqnarray}
\theta(V)g_\omega(v) & \propto & G_\omega(V) -
e^{-\pi\omega/a}\bar{G}_\omega^*(V),\label{sec2:cruc1}\\
\theta(-V)g_\omega(\bar{v}) & \propto & \bar{G}_\omega(V) -
e^{-\pi\omega/a}G_\omega^*(V). \label{sec2:cruc2}
\end{eqnarray}
By substituting these equations in
\begin{eqnarray}
\hat{\Phi}_{+}(V) & = & \int_0^\infty d\omega\,\left\{\theta(V)
[\hat{a}_{+\omega}^R g_{\omega}(v) +
\hat{a}_{+\omega}^{R\dagger}g_{\omega}^*(v) ]\right.
\nonumber \\
&& 
\left. + \theta(-V) [\hat{a}_{+\omega}^L  g_{\omega}(\bar{v}) +
\hat{a}_{+\omega}^{L\dagger} g_{\omega}^*(\bar{v}) ]\right\},
\end{eqnarray}
we find that the integrand here is proportional to
\begin{eqnarray}
&& G_\omega(V) [ \hat{a}_{+\omega}^R - 
e^{-\pi\omega/a}\hat{a}_{+\omega}^{L\dagger} ]
\nonumber \\
&& + \bar{G}_{\omega}(V) [\hat{a}_{+\omega}^L 
- e^{-\pi\omega/a}\hat{a}_{+\omega}^{R\dagger} ]
+ {\rm H.c.}\nonumber
\end{eqnarray}
Since the functions $G_\omega(V)$ and $\bar{G}_\omega(V)$ are
positive-frequency solutions (with respect to the usual time translation)
in Minkowski spacetime, the operators $\hat{a}_{+\omega}^R -
e^{-\pi\omega/a} \hat{a}_{+\omega}^{L\dagger}$ and $\hat{a}_{+\omega}^L -
e^{-\pi\omega/a}\hat{a}_{+\omega}^{R\dagger}$ annihilate the
Minkowski vacuum state $|0_{\rm M}\rangle$.  Thus,
\begin{eqnarray}
(\hat{a}_{+\omega}^R-e^{-\pi\omega/a}
\hat{a}_{+\omega}^{L\dagger})|0_{\rm M}\rangle &
= & 0,\label{sec2:vacannihi1}\\
(\hat{a}_{+\omega}^L-e^{-\pi\omega/a}
\hat{a}_{+\omega}^{R\dagger})|0_{\rm M}\rangle &
= & 0. \label{sec2:vacannihi2}
\end{eqnarray}
These relations uniquely determine the Minkowski vacuum 
$|0_{\rm M}\rangle$ as we explain below.

To explain how the state $|0_{\rm M}\rangle$ 
is formally expressed in the Fock space on the Rindler
vacuum state $|0_{\rm R}\rangle$ and to show that the state $|0_{\rm
M}\rangle$ is a thermal state when it is probed only in the right (or
left) Rindler wedge,
we use the approximation where the Rindler
energy levels $\omega$ are discrete.\footnote{We comment on how one
can discuss thermal states in field theory without discretization in 
Sec.~\ref{section2:extension}.} 
Thus, we write $\omega_i$ in place of $\omega$ and let
\begin{equation}
[ \hat{a}_{+\omega_i}^R, \hat{a}_{+\omega_{j}}^{R\dagger} ]
= 
[ \hat{a}_{+\omega_i}^L, \hat{a}_{+\omega_j}^{L\dagger} ]
= \delta_{ij}
\label{sec2:disccommu}
\end{equation}
with all other commutators among $\hat{a}_{+\omega_{i}}^R$, 
$\hat{a}_{+\omega_{i}}^L$ and their Hermitian conjugates vanishing. 
Using the discrete version of Eqs.~(\ref{sec2:vacannihi1}) 
and the commutators (\ref{sec2:disccommu}), we find
\begin{eqnarray}
\langle 0_{\rm M}|\hat{a}^{R\dagger}_{+\omega_i}\hat{a}_{+\omega_i}^R
|0_{\rm M}\rangle & = & 
e^{-2\pi\omega_i /a} \langle 0_{\rm M}|\hat{a}^{L\dagger}_{+\omega_i}
\hat{a}_{+\omega_i}^L
|0_{\rm M}\rangle \nonumber \\
& & + e^{-2\pi\omega_i /a}.
\end{eqnarray}
The same relation with $\hat{a}^R_{+\omega_i}$ and
$\hat{a}^{R\dagger}_{+\omega_i}$ replaced by $\hat{a}^{L}_{+\omega_i}$ and
$\hat{a}^{L\dagger}_{+\omega_i}$, respectively and vice versa, 
can be found using
Eq.~(\ref{sec2:vacannihi2}).  By solving these two relations as
simultaneous equations, we find
\begin{eqnarray}
\langle 0_{\rm M}|\hat{a}^{R\dagger}_{+\omega_i} \hat{a}_{+\omega_i}^R
|0_{\rm M}\rangle  & = & 
\langle 0_{\rm M}|\hat{a}^{L\dagger}_{+\omega_i}\hat{a}_{+\omega_i}^L
|0_{\rm M}\rangle \nonumber \\
& = & (e^{2\pi \omega_i/a} - 1)^{-1}. 
\label{sec2:improp1}
\end{eqnarray}
Hence, the expectation value of the Rindler-particle number is that of a
Bose-Einstein particle in a thermal bath of temperature $T =
a/2\pi$.  
This indicates that the Minkowski vacuum can be expressed as a
thermal state in the Rindler wedge with the boost generator as the
Hamiltonian. 

Eq.~(\ref{sec2:improp1}) can be expressed 
without discretization.  Define 
\begin{equation}
\hat{a}_{+f}^R \equiv \int_0^\infty d\omega\, f(\omega)
\hat{a}_{+\omega}^R,
\end{equation}
where 
$
\int_0^\infty d\omega\,|f(\omega)|^2 = 1
$.
Then,
\begin{equation}
\langle 0_{\rm M}|\hat{a}^{R\dagger}_{+f} \hat{a}_{+f}^R|0_{\rm M}\rangle
= \int_0^\infty d\omega \frac{|f(\omega)|^2}{e^{2\pi\omega/a} -1}.
\end{equation}
Exactly the same formula applies to the left Rindler number operator.

It should be emphasized that showing the correct properties of 
the expectation value of the
number operators $\hat{a}_{+f}^{R\dagger} \hat{a}_{+f}^R$ and
$\hat{a}_{+f}^{L\dagger}\hat{a}_{+f}^L$ is not
enough to conclude that the Minkowski vacuum state restricted to the
right or 
left Rindler wedge is a thermal state.  It is necessary to show that the
probability of each right/left Rindler-energy eigenstate corresponds 
to the grand canonical
ensemble if the other Rindler wedge is disregarded.  One can
show this fact by using the discrete version of 
Eqs.~(\ref{sec2:vacannihi1}) and (\ref{sec2:vacannihi2}). 
First we note that these equations imply
\begin{equation}
(\hat{a}_{+\omega_i}^{R\dagger}\hat{a}_{+\omega_i}^R
 - \hat{a}_{+\omega_i}^{L\dagger}\hat{a}_{+\omega_i}^L)|0_{\rm M}\rangle 
= 0.
\end{equation}
Thus, the number of the left Rindler particles is the same as that of the
right Rindler particles for each $\omega_i$.  This implies that 
we can write
\begin{equation}
|0_{\rm M}\rangle \propto 
\prod_{i}\sum_{n_i=0}^\infty \frac{K_{n_i}}{n_i!} 
(\hat{a}_{+\omega_i}^{R\dagger}
\hat{a}_{+\omega_i}^{L\dagger})^{n_i}|0_{\rm R}\rangle. 
\label{sec2:nonrigorous}
\end{equation}
One can readily find the recursion formula satisfied by $K_{n_i}$ using
the discrete version of Eqs.~(\ref{sec2:vacannihi1}) and 
(\ref{sec2:vacannihi2}).  The result is
\begin{equation}
K_{n_i+1} - e^{-\pi\omega_i/a}K_{n_i} = 0.
\end{equation}
Hence, $K_{n_i} = e^{-\pi n_i\omega_i/a}K_0$ and
\begin{equation}
|0_{\rm M}\rangle =  
\prod_{i}\left(C_{i}\sum_{n_{i}=0}^\infty
e^{-\pi n_{i} \omega_i/a}|n_i,R\rangle\otimes |n_i,L\rangle\right), 
\label{sec2:heuristic1}
\end{equation}
where $C_i = \sqrt{1- \exp(-2\pi \omega_i/a)}$.
Here the state with $n_i$ left-moving particles with Rindler energy
$\omega_i$ in each of 
the left and right Rindler wedges is denoted
$|n_i,R\rangle \otimes |n_i,L\rangle$, i.e.
\begin{equation}
\prod_i |n_i,R\rangle\otimes |n_i,L\rangle \equiv
\left\{ \prod_i \frac{1}{n_i!}(\hat{a}_{+\omega_i}^{R\dagger}
\hat{a}_{+\omega_i}^{L\dagger})^{n_{i}}\right\}|0_{\rm R}\rangle.
\end{equation}
If one probes only the right Rindler wedge, then the Minkowski vacuum
is described by the density matrix obtained by tracing out the left
Rindler states, i.e.
\begin{equation}
\hat{\rho}_R =  \prod_i \left( C_i^2 \sum_{n_{i}=0}^\infty \exp(-2\pi
n_{i}\omega_i/a)|n_{i},R\rangle\langle n_{i},R|
\right). \label{sec2:discdensity}
\end{equation}
This is the density matrix for the system of free bosons with
temperature $T = a/2\pi$.  Thus, the Minkowski vacuum state $|0_{\rm
M}\rangle$ for the
left-moving particles restricted to the left (or right) Rindler wedge is
the thermal state with temperature $T=a/2\pi$ with the boost
generator normalized on $z^2-t^2=1/a^2$ as the Hamiltonian. This is the
Unruh effect for the left-moving sector.  It is clear
that the Unruh effect for the right-moving sector can be derived
in a similar manner.

\subsection{Massive scalar field in Rindler wedges}
\label{section2:massive}

The Unruh effect for scalar field theory in four-dimensional Minkowski
spacetime can be derived in the same way as for the
two-dimensional example.  Nevertheless, in view of the skepticism on
the Unruh effect expressed recently by some 
authors~\cite{Belinskiietal97,Fedotovetal99,Belinskiietal02,Belinskiietal04,Oriti00}
 we review the Unruh effect in this theory~\cite{Fulling73,Unruh76}, drawing attention to
some aspects that appear to have caused the skepticism. [See
\textcite{rebuttal} for an explanation as to why this skepticism is
unfounded.]

The free quantized 
massive scalar field $\hat{\Phi}(t,z,{\bf x}_\perp)$, 
${\bf x}_\bot \equiv (x,y)$, can be expanded as
\begin{equation}
\hat{\Phi}(t,z,{\bf x}_\bot) = 
\int d^3{\bf k}( \hat{a}_{k_z {\bf k}_\perp}^M f_{k_z{\bf k}_\perp} + 
\hat{a}_{k_z {\bf k}_\perp}^{M\dagger} f_{k_z{\bf k}_\perp}^*),
\end{equation}
where
the positive-frequency mode functions are 
\begin{equation}
f_{k_z{\bf k}_\perp}(t,z,{\bf x}_\bot) = 
[(2\pi)^3 2k_0]^{-1/2}
e^{-ik_0t + ik_z z + i{\bf k}_\perp\cdot {\bf x}_\perp}
\end{equation}
with ${\bf k}_\perp \equiv 
(k_x,k_y)$ and $k_0 \equiv \sqrt{k_z^2 + {\bf k}_\perp^2 + m^2}$.  
The Klein-Gordon inner product can readily be calculated as
\begin{eqnarray}
(f_{k_z{\bf k}_\perp},f_{k_z^\prime{\bf k}_\perp^\prime})_{\rm KG} 
& = & \delta(k_z - k_z^\prime)\delta^2({\bf k}_\perp - {\bf k}_\perp^\prime),\\
(f_{k_z{\bf k}_\perp}^*,f_{k_z^\prime{\bf k}_\perp^\prime})_{\rm KG} 
& = & 0.
\end{eqnarray}
Hence, quantizing the scalar field $\hat{\Phi}(t,z,{\bf x}_\bot)$ 
by imposing the equal-time commutation relations (\ref{sec2:canonical1})
and (\ref{sec2:canonical2}), we find 
\begin{equation}
[\hat{a}_{k_z{\bf k}_\perp}^M,
\hat{a}_{k_z^\prime{\bf k}_\perp^\prime}^{M\dagger}] = 
\delta(k_z-k_z^\prime)\delta^2({\bf k}_\perp - {\bf k}_\perp^\prime)
\end{equation}
with all other commutators among annihilation and creation operators
vanishing.

The field equation in the right Rindler wedge with the metric 
(\ref{sec2:rightmetric}) can readily be found from
Eq.~(\ref{sec2:t-indepeq}) by letting 
$N=e^{a\xi}$ and the metric of the hypersurfaces with constant $\tau$
be diagonal 
with $G_{\xi\xi} = e^{2a\xi}$ and $G_{xx}=G_{yy} = 1$.  Thus, 
\begin{equation}
\frac{\partial^2\hat{\Phi}}{\partial \tau^2}
= \left[ \frac{\partial^2\ }{\partial \xi^2}
+ e^{2a\xi}\left(\frac{\partial^2\ }{\partial x^2}
+ \frac{\partial^2\ }{\partial y^2}\right) - m^2 e^{2a\xi}
\right]\hat{\Phi}.
\end{equation}
The positive-frequency solutions
are chosen to be proportional to $e^{-i\omega\tau}$, 
where $\omega$ is a positive 
constant. This choice corresponds to the static vacuum state with
respect to the $\tau$-translation, i.e.~the Rindler vacuum state.
It is also clear that one may assume that
they are proportional to $e^{i{\bf k}_\perp\cdot{\bf x}_\perp}$.  
Thus, we write the positive-frequency modes as 
\begin{equation}
v^R_{\omega{\bf k}_\perp} 
= \frac{1}{2\pi\sqrt{2\omega}} 
g_{\omega  k_\perp}(\xi)e^{-i\omega\tau+ 
i{\bf k}_\perp\cdot{\bf x}_\perp}
\end{equation}
with the function $g_{\omega k_\perp}(\xi)$ satisfying
\begin{equation}
\left[ -\frac{d^2\ }{d\xi^2} 
 + e^{2a\xi}(k_\perp^2 + m^2)\right]
g_{\omega k_\perp}(\xi) = \omega^2 g_{\omega k_\perp}(\xi). 
\label{sec2:bessel}
\end{equation}
This equation is analogous to a time-independent Schr\"{o}dinger 
equation with an exponential potential.  Thus, the physically relevant 
solutions $g_{\omega{\bf k}_\perp}(\xi)$ tend to zero as 
$\xi\to +\infty$ and oscillate like $e^{\pm i\omega \xi}$ as 
$\xi \to -\infty$.  Note in particular that there is no distinction
between the left- and right-moving modes.
We choose $g_{\omega k_\perp}(\xi)$ to satisfy, for $\xi <0$ and 
$|\xi| \gg 1$,
\begin{equation}
g_{\omega k_\perp}(\xi) \approx 
\frac{1}{\sqrt{2\pi}}( e^{i[\omega\xi+\gamma(\omega)]} + 
e^{-i[\omega\xi + \gamma(\omega)]}), \label{sec2:normalization}
\end{equation}
where $\gamma(\omega)$ 
is a real constant. This choice of normalization implies
[see, e.g., \textcite{Fullingbook89}]
\begin{equation}
\int_{-\infty}^{\infty}d\xi\, g^*_{\omega k_\perp}(\xi)
g_{\omega' k_\perp}(\xi) = 
\delta(\omega-\omega').  \label{sec2:delta-normal}
\end{equation}
We present the derivation of this formula in 
Appendix~\ref{Appendix A} for completeness. 
As a result we have
\begin{eqnarray}
( v_{\omega{\bf k}_\perp}^R,v_{\omega^\prime{\bf
k}_\perp^\prime}^R )_{\rm KG}
& = & \delta(\omega-\omega^\prime)
\delta^2({\bf k}_\perp-{\bf k}_\perp^\prime),
\\
(v_{\omega{\bf k}_\perp}^{R*},
v_{\omega^\prime{\bf k}_\perp^\prime}^R )_{\rm KG} & = & 0.
\end{eqnarray}
The Klein-Gordon inner product here is defined taking the hypersurface
$\Sigma$ in Eq.~(\ref{sec2:KGinner}) to be a $\tau= {\rm const}$ 
Cauchy surface of the right Rindler wedge.  It
can also be defined taking $\Sigma$ to be the entire $t=0$ hypersurface
of the Minkowski spacetime by defining $v_{\omega{\bf k}_\perp}^R = 0$
in the left Rindler wedge (and on the plane $t=z=0$ for definiteness).
The functions $g_{\omega k_\perp}(\xi)$ satisfying the differential 
equation (\ref{sec2:bessel}) and normalization condition
(\ref{sec2:normalization}) are
\begin{equation}
g_{\omega k_\perp}(\xi)
= \left[\frac{2\omega\sinh(\pi\omega/a)}{\pi^2 a}\right]^{1/2}
K_{i\omega/a}\left(
\frac{\kappa}{a}e^{a\xi}\right)
\label{sec2:normalizedgomega}
\end{equation}
with $\kappa \equiv (k_\bot^2 + m^2)^{1/2}$,
where $K_\nu(x)$ is the modified Bessel function~\cite{Gradshteynbook}.
Hence, 
\begin{equation}
v_{\omega{\bf k}_\perp}^R
= \left[\frac{\sinh(\pi\omega/a)}{4\pi^4 a}\right]^{1/2}
K_{i\omega/a}\left(\frac{\kappa}{a}e^{a\xi}\right)
e^{i{\bf k}_\perp\cdot {\bf x}_\perp - i\omega\tau}.
\label{sec2:normalizedvR}
\end{equation}
We present the derivation of this result in Appendix~\ref{Appendix A}
as well.
Thus, we can expand the field $\hat{\Phi}$ in the right Rindler wedge
as 
\begin{equation}
\hat{\Phi}(\tau,\xi,{\bf x}_\perp) =
\int_{-\infty}^{\infty}d\omega \int d^2{\bf k}_\perp
( \hat{a}_{\omega{\bf k}_\perp}^R v^R_{\omega {\bf k}_\perp}
+ \hat{a}_{\omega{\bf k}_\perp}^{R\dagger}v^{R*}_{\omega {\bf k}_\perp} 
).
\end{equation}
Then,  according to the general results presented
in Sec.~\ref{section2:curvedspace}, we have
\begin{equation}
[ \hat{a}_{\omega {\bf k}_\perp}^R,
\hat{a}_{\omega^\prime {\bf k}_\perp^\prime}^{ R \dagger} 
]
= \delta(\omega-\omega^\prime)\delta^{2}({\bf k}_\perp 
- {\bf k}_\perp^\prime)
\end{equation}
with all other commutators among $\hat{a}_{\omega{\bf k}_\perp}^R$ and 
$\hat{a}_{\omega{\bf k}_\perp}^{R\dagger}$ vanishing.

Quantization of the field $\hat{\Phi}$ in the left Rindler wedge
proceeds in exactly the same way.  The positive-frequency modes
$v^L_{\omega{\bf k}_\perp}(\bar{\tau},\bar{\xi},{\bf x}_\perp)$ are
obtained from $v^R_{\omega {\bf k}_\perp}(\tau,\xi,{\bf x}_\perp)$
simply by replacing $\tau$ and $\xi$ by $\bar{\tau}$ and $\bar{\xi}$,
respectively.  The coefficient operators $\hat{a}^L_{\omega {\bf
k}_\perp}$ and $\hat{a}^{L\dagger}_{\omega {\bf k}_\perp}$ satisfy the
commutation relations
\begin{equation}
[ 
\hat{a}^{L}_{\omega {\bf k}_\perp},
\hat{a}^{L\dagger}_{\omega' {\bf k}'_\perp}
]
= \delta(\omega-\omega')\delta^2({\bf k}_\perp-{\bf k}'_\perp)
\end{equation}
with all other commutators vanishing.
Thus, one can 
expand the field $\hat{\Phi}$ in the left and right Rindler wedges as
\begin{widetext}
\begin{eqnarray}
\hat{\Phi} & = & \int_{0}^{+\infty}d\omega \int d^2{\bf k}_\perp
\left[ \hat{a}_{\omega{\bf k}_\perp}^R
v^R_{\omega {\bf k}_\perp}(\tau,\xi,{\bf x}_\perp)
+ \hat{a}_{\omega{\bf k}_\perp}^{R\dagger}
v_{\omega {\bf k}_\perp}^{R*}(\tau,\xi,{\bf x}_\perp) 
+ \hat{a}_{\omega{\bf k}_\perp}^L
v^L_{\omega{\bf k}_\perp}(\bar{\tau},\bar{\xi},{\bf x}_\perp)
+ \hat{a}_{\omega{\bf k}_\perp}^{L\dagger}v_{\omega{\bf k}_\perp}^{L*}
(\bar{\tau},\bar{\xi},{\bf x}_\perp)
\right].\nonumber \\
\label{sec2:rindler}
\end{eqnarray}
\end{widetext}
The Rindler vacuum state $|0_{\rm R}\rangle$ is defined by requiring that 
$\hat a^R_{\omega{\bf k}_\perp}|0_{\rm R}\rangle 
= \hat a^L_{\omega{\bf k}_\perp}|0_{\rm R}\rangle=0 $ 
for all $\omega$ and ${\bf
k}_\perp$.
As it stands, this expansion makes sense only in the Rindler wedges. 
However, it will be shown that 
the modes $v^R_{\omega{\bf k}_\bot}$ and 
$v^L_{\omega{\bf k}_\bot}$ can naturally be extended to the whole of 
Minkowski spacetime
[see Eqs.~(\ref{sec2:wexplicit}), (\ref{sec2:uw}) and (\ref{sec2:vw})]. 
After this extension we shall see that Eq.~(\ref{sec2:rindler}) gives
just another valid mode expansion of the field $\hat{\Phi}$ in Minkowski
spacetime.\footnote{This point is emphasized in
\textcite{Birrelletal82}.} 
In particular, in Sec.~\ref{section2:completeness} 
the two-point function calculated using this
expansion in the state $|0_{\rm M}\rangle$ 
will be shown to give the standard result in Minkowski spacetime.
 
\subsection{Bogolubov coefficients and the Unruh effect}

In this subsection we find the Bogolubov coefficients between the two 
expansions of the
massive scalar field $\hat{\Phi}$ 
in Minkowski spacetime and derive the Unruh effect, 
i.e.~the fact that the Minkowski vacuum state is a thermal state with 
temperature $T = a/2\pi$ on the right or left Rindler wedge.

It is clear that the Bogolubov 
coefficients between modes with different ${\bf k}_\perp$ are zero. 
Thus, we can write in general
\begin{eqnarray}
v^R_{\omega{\bf k}_\perp} & = & 
\int_{-\infty}^{\infty} \frac{dk_z}{\sqrt{4\pi k_0}} \left[ 
\alpha_{\omega k_z k_\perp}^R e^{-ik_0 t + ik_z z} \right.\nonumber \\
&& \left. \ \ \ \ \ \ \ \  
+ \beta_{\omega k_z k_\perp}^R e^{ik_0t - ik_z z}\right]
\frac{e^{i{\bf k}_\bot\cdot{\bf x}_\bot}}{2\pi},
\label{sec2:vRexpansion}\\
v^L_{\omega{\bf k}_\perp} & = & \int_{-\infty}^{\infty} 
\frac{dk_z}{\sqrt{4\pi k_0}} \left[ 
\alpha_{\omega k_z k_\perp}^L e^{-ik_0 t + ik_z z}  \right.\nonumber \\
&& \left. \ \ \ \ \ \ \ \ 
+ \beta_{\omega k_z k_\perp}^L e^{ik_0t - ik_z z}\right]
\frac{e^{i{\bf k}_\bot\cdot{\bf x}_\bot}}{2\pi}.
\end{eqnarray}
We are assuming here that the modes
$v^{\rm R}_{\omega{\bf k}_\bot}$ and $v^{\rm L}_{\omega {\bf k}_\bot}$
have been suitably extended to the whole of Minkowski spacetime.
The relation between $(\tau,\xi)$ and $(t,z)$ given by
Eq.~(\ref{sec2:rightcoords}) is the same as that
between $(\bar{\tau},\bar{\xi})$ and $(t,-z)$ given by 
Eq.~(\ref{sec2:leftcoords}).  Hence,
$v^L_{\omega{\bf k}_\perp}$ is obtained from 
$v^R_{\omega{\bf k}_\perp}$ by letting $z \mapsto -z$.  {}From this
observation we find the following relations:
\begin{equation}
\alpha_{\omega k_z k_\perp}^L  
=  
\alpha_{\omega\, -k_z k_\perp}^R, \;\;\;
\beta_{\omega k_z k_\perp}^L  
=  
\beta_{\omega\, -k_z k_\perp}^R.
\label{sec2:alpha_betaLR}
\end{equation}
These Bogolubov coefficients will be found explicitly later, 
but it is clear from
the discussion of the massless scalar field theory in 
two dimensions that
the Unruh effect will follow if
\begin{eqnarray}
(\hat{a}^R_{\omega {\bf k}_\perp}
- e^{-\pi\omega/a}\hat{a}^{L\dagger}_{\omega \, -{\bf k}_\perp})
|0_{\rm M}\rangle & = & 0,\label{sec2:annihilate1}\\
(\hat{a}^L_{\omega {\bf k}_\perp}
- e^{-\pi\omega/a}\hat{a}^{R\dagger}_{\omega \, -{\bf k}_\perp})
|0_{\rm M}\rangle & = & 0. \label{sec2:annihilate2}
\end{eqnarray}
[See the corresponding equations 
(\ref{sec2:vacannihi1}) and (\ref{sec2:vacannihi2}) in the 
two-dimensional model.]  These
relations in turn will result if the following modes are purely
positive-frequency in Minkowski spacetime:
\begin{eqnarray}
w_{-\omega {\bf k}_\perp} & \equiv &
\frac{v_{\omega {\bf k}_\perp}^R + 
e^{-\pi\omega/a}v_{\omega\,-{\bf k}_\perp}^{L*}}
{\sqrt{1-e^{-2\pi\omega/a}}},
\label{sec2:wpositive1}\\
w_{+\omega {\bf k}_\perp} & \equiv &
\frac{v_{\omega {\bf k}_\perp}^L + e^{-\pi\omega/a}v_{\omega\, 
-{\bf k}_\perp}^{R*}}{\sqrt{1-e^{-2\pi\omega/a}}}.
\label{sec2:wpositive2}
\end{eqnarray}
[See the corresponding equations (\ref{sec2:cruc1a}) and
(\ref{sec2:cruc2a}) in the two-dimensional model.]
This fact in turn will follow if
\begin{equation}
\beta^R_{\omega k_z k_\perp}  =  
- e^{-\pi\omega/a}\alpha^{L*}_{\omega k_z k_\perp}, \;\;\;
\beta^L_{\omega k_z k_\perp}  = 
- e^{-\pi\omega/a}\alpha^{R*}_{\omega k_z k_\perp}.
\label{sec2:Bogrelation}
\end{equation}
[See the corresponding equation (\ref{sec2:betaalpha}).]
We shall show Eq.~(\ref{sec2:Bogrelation}) by 
explicit evaluation of the Bogolubov coefficients, 
which were originally computed by \textcite{Fulling73}. 

To calculate the Bogolubov coefficients it is 
convenient to examine the behavior of the solutions on the future
Killing horizon, $t=z$, $t>0$.  There we have
\begin{eqnarray}
v_{\omega{\bf k}_\perp}^R & \to
& \int_{-\infty}^{\infty} \frac{dk_z}{\sqrt{4\pi k_0}}
\nonumber \\
&& \times \left[ 
\alpha_{\omega k_z k_\bot}^R e^{-i(k_0 -k_z)V/2} + \right.\nonumber \\
&& \left. + \beta_{\omega k_z k_\bot}^R e^{i(k_0-k_z)V/2}\right]
\frac{e^{i{\bf k}_\bot\cdot{\bf x}_\bot}}{2\pi}.
\end{eqnarray}
On the other hand, using the small-argument approximation 
(\ref{appendixA:small-argumentK}) for the
modified Bessel function, we have for $\xi \to -\infty$
\begin{eqnarray}
v_{\omega{\bf k}_\perp}^R & \to & 
\frac{i}{4\pi}\left[a\sinh(\pi\omega/a)\right]^{-1/2}
e^{i{\bf k}_\bot\cdot{\bf x}_\bot}
\nonumber \\
&& \times
\left( \frac{(\kappa/2a)^{i\omega/a}e^{-i\omega u}}{\Gamma(1+i\omega/a)} 
- \frac{(\kappa/2a)^{-i\omega/a}e^{-i\omega v}}{\Gamma(1-i\omega/a)}
\right),\nonumber \\
\label{sec2:smallargument}
\end{eqnarray}
where we recall $\kappa = (k_\perp^2 + m^2)^{1/2}$.  
The first term inside the parentheses in this equation 
oscillates infinitely many times as
$u\to \infty$, where the future Killing horizon is, and is bounded. 
Such a term should be regarded as zero.  Hence, the
Bogolubov coefficient $\alpha_{\omega k_z k_\bot}^R$ is obtained 
by multiplying Eq.~(\ref{sec2:smallargument}) by
$e^{i(k_0-k_z)V/2}$ and integrating over $V$ as 
\begin{eqnarray}
\alpha_{\omega k_z k_\perp}^R & = &
- \frac{i(\kappa/2a)^{-i\omega/a}(k_0-k_z)}{4
\sqrt{\pi ak_0\sinh(\pi\omega/a)}\Gamma(1-i\omega/a)}
\nonumber \\
&& \times \int_0^\infty dV\,
(aV)^{-i\omega/a}e^{i(k_0-k_z)V/2}\nonumber \\
& = & 
\frac{e^{\pi\omega/2a}}{\sqrt{4\pi k_0 a \sinh(\pi\omega/a)}}
\left(\frac{k_0+k_z}{k_0-k_z}\right)^{-i\omega/2a},\nonumber \\
\end{eqnarray}
where we have used $\kappa = \sqrt{(k_0-k_z)(k_0+k_z)}$.
Note that 
we have implicitly chosen 
a particular (and natural) extension of the modes
$v^{\rm R}_{\omega {\bf k}_\bot}$ to the whole of Minkowski spacetime.
[Otherwise it should not be possible to find the coefficients Bogolubov
coefficients $\alpha^{\rm R}_{\omega k_z k_\bot}$ and
$\beta^{\rm R}_{\omega k_z k_\bot}$ in Eq.~(\ref{sec2:vRexpansion})].
In particular, we have excluded any delta-function 
contribution at $V=0$.

By multiplying Eq.~(\ref{sec2:smallargument}) by $e^{-i(k_0-k_z)V/2}$
and integrating over $V$ we find
\begin{equation}
\beta_{\omega k_z k_\perp}^R =  - \frac{e^{-\pi\omega/2a}}
{\sqrt{4\pi k_0 a \sinh(\pi\omega/a)}}
\left(\frac{k_0+k_z}{k_0-k_z}\right)^{-i\omega/2a}.
\end{equation}
Introducing the {\em rapidity} $\vartheta(k_z)$ defined as
\begin{equation}
\vartheta(k_z) \equiv  \frac{1}{2}\log \left(
\frac{k_0+k_z}{k_0-k_z}\right), \label{sec2:rapidity}
\end{equation}
and using Eq.~(\ref{sec2:alpha_betaLR}), we
have
\begin{eqnarray}
\alpha^R_{\omega k_z k_\perp} & = &
\alpha^L_{\omega\, -k_z\, k_\perp}\nonumber \\
& = & \frac{e^{-i\vartheta(k_z)\omega/a}}{\sqrt{2\pi k_0 a
(1-e^{-2\pi\omega/a})}},\\
\beta^R_{\omega k_z k_\perp} & = &
\beta^L_{\omega\,-k_z k_\perp}\nonumber \\
& = & 
- \frac{e^{-\pi\omega/a}e^{-i\vartheta(k_z)\omega/a}}{\sqrt{2\pi k_0 a
(1-e^{-2\pi\omega/a})}}.
\end{eqnarray}
Hence, Eq.~(\ref{sec2:Bogrelation}) is
satisfied and as a result the vacuum state $|0_{\rm M}\rangle$
restricted to the left (or right) Rindler wedge is a thermal state with
temperature $T= a/2\pi$ with the boost generator normalized on
$t^2-z^2=1/a^2$ as the Hamiltonian.

Although we have now established
the Unruh effect, it is useful to examine
the modes natural to the Rindler wedges further for later discussion.
The purely positive-frequency modes in Minkowski spacetime defined by
Eqs.~(\ref{sec2:wpositive1}) and (\ref{sec2:wpositive2}) are
\begin{eqnarray}
w_{\pm \omega {\bf k}_\perp} & = & 
\int_{-\infty}^{\infty} \frac{dk_z}{\sqrt{8a}\,\pi k_0}
e^{\pm i\vartheta(k_z)\omega/a}\, e^{-ik_0t+ik_z z}\nonumber \\
&& \ \ \ \ \ \ \ \ \ \ \times
\frac{e^{i{\bf k}_\perp\cdot {\bf x}_\perp}}{2\pi}.
\label{sec2:wexplicit}
\end{eqnarray}
The modes $v^R_{\omega{\bf k}_\perp}$ and $v^L_{\omega{\bf k}_\perp}$,
which vanish in the left and right Rindler wedges,
respectively, are expressed in terms of these modes as
\begin{eqnarray}
v^R_{\omega {\bf k}_\perp} & = & \frac{w_{-\omega {\bf k}_\perp}
- e^{-\pi\omega/a}w^{*}_{+\omega\,-{\bf k}_\perp}}
{\sqrt{1-e^{-2\pi\omega/a}}}, \label{sec2:uw}\\
v^L_{\omega {\bf k}_\perp} & = & \frac{w_{+\omega {\bf k}_\perp}
- e^{-\pi\omega/a}w^{*}_{-\omega\,-{\bf k}_\perp}}
{\sqrt{1-e^{-2\pi\omega/a}}}. \label{sec2:vw}
\end{eqnarray}
These formulas and Eq.~(\ref{sec2:wexplicit}) give the modes
$v^R_{\omega{\bf k}_\bot}$ and $v^L_{\omega{\bf k}_\bot}$ as
distributions in the whole of Minkowski spacetime.
One can verify that the modes $w_{\pm \omega {\bf k}_\perp}$ satisfy 
\begin{eqnarray}
(w_{\pm \omega{\bf k}_\perp},w_{\pm \omega^\prime{\bf k}_\perp^\prime})_{\rm KG}
& = & \delta(\omega-\omega^\prime)\delta^2({\bf k}_\perp 
- {\bf k}_\perp^\prime),\nonumber \\
\label{sec2:posmodes}\\
(w_{\pm\omega{\bf k}_\perp}^*,w_{\pm\omega^\prime{\bf k}_\perp^\prime}^*)_{\rm KG}
& = & 
 - \delta(\omega-\omega^\prime)\delta^2({\bf k}_\perp 
- {\bf k}_\perp^\prime)
\nonumber \\
\label{sec2:negmodes}
\end{eqnarray}
with all other Klein-Gordon inner products 
among $w_{\pm \omega{\bf k}_\perp}$ and their complex conjugates 
vanishing. Here the Klein-Gordon inner product is defined as an
integral over the $t=0$ hypersurface in Minkowski spacetime.
The following formula, which can be shown by using
$d\vartheta(k_z) = dk_z/k_0$,
is useful in calculating these Klein-Gordon inner 
products:
\begin{equation}
\int_{-\infty}^\infty 
\frac{dk_z}{2\pi a k_0}e^{i\vartheta(k_z)(\omega-\omega')/a}
= \delta(\omega-\omega^\prime).
\end{equation}

It is worth emphasizing that these modes form a complete
set of solutions to the Klein-Gordon equation, 
not only in the left and right Rindler
wedges but also in the whole of Minkowski spacetime. 
This fact can readily be seen by inverting the relation 
(\ref{sec2:wexplicit}):
\begin{eqnarray}
&& \frac{1}{\sqrt{2k_0(2\pi)^3}} 
e^{-ik_0t +ik_z z + i{\bf k}_\perp\cdot{\bf x}_\perp}\nonumber \\
&& =  \frac{1}{\sqrt{2\pi a k_0}}
\int_0^\infty d\omega [ 
e^{i\vartheta(k_z)\omega/a}
w_{-\omega {\bf k}_\perp} 
\nonumber \\
&& \ \ \ \ \ \ \ \ \ \ \ \ \ \ \ \ \ \ 
+ e^{-i\vartheta(k_z)\omega/a}
w_{+\omega{\bf k}_\perp} ].  \label{sec2:unique!}
\end{eqnarray}
One may object to this conclusion as do \textcite{Belinskiietal97}
because the
modes $w_{\pm \omega{\bf k}_\bot}$ 
were originally defined only on the left and right Rindler wedges,
which are open regions; in particular, these modes are not defined 
on the plane $t=z=0$.  However, the formula (\ref{sec2:wexplicit}) 
gives the positive-frequency modes $w_{\pm\omega {\bf k}_\bot}$ 
in terms of the momentum eigenfunctions unambiguously
as a distribution over the whole of Minkowski spacetime.  
In other words, if $f(t,{\bf x})$ 
is a compactly-supported smooth function on
Minkowski spacetime, whose support may intersect the plane $t=z=0$, the
mode functions $w_{\pm \omega{\bf k}_\bot}$ smeared with $f$ is well
defined and unique.  That is,
\begin{eqnarray}
\hat{f}^{R}(\pm \omega,{\bf k}_\bot)
& \equiv & \int d^4 x\,w^*_{\pm \omega {\bf k}_\bot}(t,z,{\bf x}_\bot)
f(t,z,{\bf x}_\bot)\nonumber \\
& = & \int_{-\infty}^\infty \frac{dk_z}{\sqrt{2\pi ak_0}}
e^{\mp i\vartheta(k_z)\omega/a}\hat{f}^{M}(k_z,{\bf k}_\bot),
\nonumber \\ \label{sec2:fR}
\end{eqnarray}
where
\begin{equation}
\hat{f}^{M}(k_z,{\bf k}_\bot)
\equiv \frac{1}{\sqrt{(2\pi)^32k_0}}\int d^4 x 
e^{ik_0 t - i{\bf k}\cdot{\bf x}}
f(t,{\bf x}).  \label{sec2:fM}
\end{equation}
We have used $w^*_{\pm \omega {\bf k}_\bot}$ rather than $w_{\pm \omega
{\bf k}_\bot}$ here for later convenience. Note that
the function $\hat{f}^{M}(k_z,{\bf k}_\bot)$ tends to 
$0$ as $k_z \to \pm\infty$ faster than any powers of
$|k_z|^{-1}$ due to the smoothness of $f(t,{\bf x})$.  This implies that
the integral in Eq.~(\ref{sec2:fR}) is absolutely
convergent. [In fact Eq.~(\ref{sec2:fR}) should be taken as the
definition of the modes $w^*_{\pm \omega{\bf k}_\bot}$ as distributions
over the full Minkowski spacetime.]

Since the modes $w_{\pm \omega {\bf k}_\bot}$ and 
$w^*_{\pm \omega {\bf k}_\bot}$ form a complete set of solutions in
Minkowski spacetime, the Rindler modes
$v^R_{\omega{\bf k}_\perp}$ and $v^L_{\omega{\bf k}_\perp}$ and their
complex conjugates form a complete set as is clear from 
Eqs.~(\ref{sec2:wpositive1}) and (\ref{sec2:wpositive2}).  Related
comments will be made in the next two subsections.

\subsection{Completeness of the Rindler modes in Minkowski spacetime}
\label{section2:completeness}

In the previous subsection 
we commented that the Rindler modes form a
complete set of solutions to the Klein-Gordon equation in Minkowski
spacetime.  To emphasize this point again 
we show in this subsection that the
Wightman 
two-point function is correctly reproduced everywhere in Minkowski
spacetime even if we use the Rindler modes.
It is our hope that the calculation here will 
dispel any suspicion that the Rindler modes may be incomplete due 
to the singularity on the hypersurfaces $t=\pm z$. 

The two-point function in the Minkowski vacuum state is well known to be
\begin{eqnarray}
\Delta(x;x') & = &
\langle 0_{\rm M}|\hat{\Phi}(x)
\hat{\Phi}(x')|0_{\rm M}\rangle\nonumber \\
& = & \int \frac{dk_z d^2{\bf k}_\bot}{2k_0(2\pi)^3}
e^{-ik\cdot(x-x')}, \label{sec2:two-pt}
\end{eqnarray}
where $x = (t,z,{\bf x}_\bot)$ and similarly for $x'$.
To calculate the two-point functions with the Rindler modes we use the
expansion (\ref{sec2:rindler}) with the Rindler modes
$v^{L}_{\omega{\bf k}_\bot}$ and $v^{R}_{\omega{\bf k}_\bot}$ given by
Eqs.~(\ref{sec2:wexplicit}), (\ref{sec2:uw}) and (\ref{sec2:vw}).
By Eqs.~(\ref{sec2:annihilate1}) and 
(\ref{sec2:annihilate2}) we see that the Rindler annihilation operators
can be written as
\begin{eqnarray}
\hat{a}_{\omega{\bf k}_\perp}^R & = & 
\frac{\hat{b}_{-\omega{\bf k}_\perp} +
e^{-\pi\omega/a}\hat{b}^{\dagger}_{+\omega\,-{\bf k}_\perp}}
{\sqrt{1-e^{-2\pi\omega/a}}},\label{sec2:barelation1}\\
\hat{a}_{\omega{\bf k}_\perp}^L & = & 
\frac{\hat{b}_{+\omega{\bf k}_\perp} +
e^{-\pi\omega/a}\hat{b}^{\dagger}_{-\omega\,-{\bf k}_\perp}}
{\sqrt{1-e^{-2\pi\omega/a}}},\label{sec2:barelation2}
\end{eqnarray}
where the operators $b_{\pm \omega{\bf k}_\perp}$ annihilate the
Minkowski vacuum $|0_{\rm M}\rangle$ and have the following standard
commutation relations:
\begin{equation}
[b_{\pm \omega {\bf k}_\perp},b^\dagger_{\pm \omega' {\bf
k}_{\perp}'}] 
= \delta(\omega-\omega')\delta^2({\bf k}_\perp -
{\bf k}^\prime_\perp)
\end{equation}
with all other commutators vanishing.  
Eqs.~(\ref{sec2:barelation1}) and (\ref{sec2:barelation2}) can be used
to find the following expectation values:
\begin{widetext}
\begin{eqnarray}
 \langle 0_{\rm M}|a_{\omega{\bf k}_\bot}^{R\dagger}
a_{\omega'{\bf k}'_\bot}^R|0_{\rm M}\rangle
& = &
\langle 0_{\rm M}|a_{\omega{\bf k}_\bot}^{L\dagger}
a_{\omega'{\bf k}'_\bot}^L|0_{\rm M}\rangle
= (e^{2\pi\omega/a} -1)^{-1}
\delta(\omega-\omega')\delta^2({\bf k}_\bot - {\bf k}'_\bot),\\
\langle 0_{\rm M}|a_{\omega{\bf k}_\bot}^{R}
a_{\omega'{\bf k}'_\bot}^{R\dagger}|0_{\rm M}\rangle
& = &
\langle 0_{\rm M}|a_{\omega{\bf k}_\bot}^{L}
a_{\omega'{\bf k}'_\bot}^{L\dagger}|0_{\rm M}\rangle
= (1-e^{-2\pi\omega/a})^{-1} 
\delta(\omega-\omega')
 \delta^2({\bf k}_\bot - {\bf k}'_\bot),\\
\langle 0_{\rm M}|a_{\omega{\bf k}_\bot}^L
a_{\omega'{\bf k}'_\bot}^R|0_{\rm M}\rangle 
& = & \langle 0_{\rm M}|a_{\omega{\bf k}_\bot}^{L\dagger}
a_{\omega'{\bf k}'_\bot}^{R\dagger}|0_{\rm M}\rangle
= (e^{\pi\omega/a} - e^{-\pi\omega/a})^{-1}
\delta(\omega-\omega')
 \delta^2({\bf k}_\bot + {\bf k}'_\bot).
\end{eqnarray}
The vacuum expectation values of the other products of two
creation/annihilation operators vanish.  Then, the two-point function
of the field $\hat{\Phi}(x)$ given in Eq.~(\ref{sec2:rindler}) is
\begin{eqnarray}
\Delta(x;x')
& = & \int_0^\infty d\omega \int d^2{\bf k}_\bot 
\left\{ \left[ v^R_{\omega{\bf k}_\bot}(x)v^{R*}_{\omega{\bf
k}_\bot}(x') + v^L_{\omega{\bf k}_\bot}(x)v^{L*}_{\omega{\bf
k}_\bot}(x')\right]
(1-e^{-2\pi\omega/a})^{-1}\right.\nonumber \\
&& \ \ \ \ \ \ \ \ \ \ +  \left[ v^{R*}_{\omega{\bf k}_\bot}(x)
v^R_{\omega{\bf
k}_\bot}(x') + v^{L*}_{\omega{\bf k}_\bot}(x)v^L_{\omega{\bf
k}_\bot}(x')\right]
(e^{2\pi\omega/a}-1)^{-1}\nonumber \\
&& \ \ \ \ \ \ \ \ \ \ 
+ 2\left[ v^R_{\omega{\bf k}_\bot}(x)v_{\omega\,-{\bf k}_\bot}^L(x^{\prime})
+ v^{R*}_{\omega{\bf k}_\bot}(x)v^{L*}_{\omega\,-{\bf k}_\bot}(x^{\prime})\right] 
(e^{\pi\omega/a}-e^{-\pi\omega/a})\nonumber \\
&& \ \ \ \ \ \ \ \ \ \ \left.
+ 2\left[ v^L_{\omega{\bf k}_\bot}(x)
v_{\omega\,-{\bf k}_\bot}^R(x^{\prime})
+ v^{L*}_{\omega{\bf k}_\bot}(x)
v^{R*}_{\omega\,-{\bf k}_\bot}(x^{\prime})\right] 
(e^{\pi\omega/a}-e^{-\pi\omega/a})\right\}.
\end{eqnarray}
This expression can be simplified using Eqs~(\ref{sec2:uw})
and (\ref{sec2:vw}) as
\begin{equation}
\Delta(x;x^{\prime})
 = \int_0^\infty d\omega \int d^2{\bf k}_\bot
\left[ w_{+\omega{\bf k}_\bot}(x)w^*_{+\omega{\bf k}_\bot}(x^{\prime})
+ w_{-\omega{\bf k}_\bot}(x)w^*_{-\omega{\bf
k}_\bot}(x^{\prime})\right],
\label{sec2:undefined?}
\end{equation}
where $w_{\pm\omega{\bf k}_\bot}$ are given by
Eq.~(\ref{sec2:wexplicit}).  Thus,
\begin{equation}
\Delta(x;x^{\prime}) = 
\frac{1}{32\pi^4 a}
\int_{-\infty}^\infty d\omega
\int_{-\infty}^{\infty}\frac{dk_z}{k_0}
\int_{-\infty}^{\infty}d\vartheta(k_z')
\int d^2{\bf k}_\bot e^{i\left[\vartheta(k_z)-\vartheta(k_z')
\right]\omega/a}
e^{-ik_0 t + ik'_0t' + ik_z z - ik'_z z'+
i{\bf k}_\bot\cdot({\bf x}_\bot -{\bf x}'_\bot)}.\label{sec2:intermed2}
\end{equation}
\end{widetext}
The $\omega$- and $\vartheta(k_z')$-integration can
readily be performed, and we find that the two-point function indeed
takes the form given by Eq.~(\ref{sec2:two-pt}).

The expression (\ref{sec2:undefined?}) is undefined if either of the two
points is on the hyperplane $t=\pm z$.  
However, since the two-point function
$\Delta(x;x')$ is defined as a distribution, it is well defined on the
whole of Minkowski spacetime if the following integral exists for all
compactly-supported functions $f(x)$ and $g(x)$:
\begin{equation}
F(f,g) \equiv \int d^4x d^4 x'f^*(x)g(x')\Delta(x;x').
\end{equation}
We find using Eq.~(\ref{sec2:undefined?})
\begin{eqnarray}
F(f,g) & = & \int_0^\infty d\omega \int d^2{\bf k}_\bot\nonumber \\
&&\times 
\left[
\hat{f}^{R*}(-\omega,{\bf k}_\bot)
\hat{g}^{R}(-\omega,{\bf k}_\bot)\right. \nonumber \\
&& \left. + \hat{f}^{R*}(+\omega,{\bf k}_\bot)
\hat{g}^{R}(+\omega,{\bf k}_\bot)\right],
\end{eqnarray}
where $\hat{f}^{R}$ is defined by Eq.~(\ref{sec2:fR}), and
$\hat{g}^{R}$ is defined similarly.  It can readily be shown that this
agrees with the standard expression for the smeared two-point function,
\begin{equation}
F(f,g) = \int d^3{\bf k}\, \hat{f}^{M*}({\bf k})\hat{g}^{M}({\bf k}),
\end{equation}
where $\hat{f}^{M}$ 
is defined by Eq.~(\ref{sec2:fM}) and the Fourier transform
$\hat{g}^{M}$ is defined similarly.

\subsection{Unruh effect and quantum field theory in the expanding 
degenerate Kasner universe}
\label{section2:Milne} 

In this subsection 
we review the relation between the modes in the Rindler
wedges and those in 
the expanding degenerate Kasner universe.  It is
well known that there is a choice of the positive-frequency modes in the
degenerate Kasner universes that
gives a state identical to the Minkowski
vacuum~\cite{Fullingetal74}.  We first show that
these positive-frequency modes are in fact the modes $w_{\pm
\omega{\bf k}_\bot}$ in the Rindler wedges~\cite{Gerlach88}.  Then, we
show that the Rindler vacuum state $|0_{\rm R}\rangle$ is identical to
one of the states in the expanding
degenerate Kasner universe found in the 
literature~\cite{Birrelletal82,Fullingetal74}. 

We introduce the following coordinate transformation:
\begin{equation}
t  =  T\cosh a\zeta, \;\;\;
z  =  T\sinh a\zeta.
\end{equation}
With $T>0$ the coordinate system $(T,\zeta,{\bf x}_\bot)$ covers the
region with the condition $t > |z|$, i.e.~the expanding degenerate Kasner
universe. Then, the Minkowski metric becomes
\begin{equation}
ds^2 = dT^2 - a^2 T^2d\zeta^2 - dx^2 - dy^2.
\end{equation}
The hyperplanes 
of constant $T$ are spacelike, and the variable $T$ plays the
role of time.
Hence, the $T= {\rm const}$ space expands in the $\zeta$-direction
linearly. We note that
\begin{equation}
\frac{t+z}{t-z} = e^{2a\zeta}, \label{sec2:2azeta}
\end{equation}
and that $(t^2-z^2)^{1/2}=T$.

Let us change the integration variable in the expression
(\ref{sec2:wexplicit}) for modes $w_{\pm \omega {\bf k}_\perp}$
from $k_z$ to the rapidity $\vartheta = \vartheta(k_z)$ 
[see Eq.~(\ref{sec2:rapidity})].
Then, we have
\begin{equation}
k_0  =  \kappa\cosh \vartheta,\;\;\;
k_z  =  \kappa\sinh \vartheta,
\end{equation}
where $\kappa = (k_\perp^2 + m^2)^{1/2}$ as before.
Thus, we obtain, using Eq.~(\ref{sec2:2azeta}) 
after shifting of the integration variable as $\vartheta \to \vartheta +
a\zeta$,
\begin{eqnarray}
w_{\pm\omega{\bf k}_\bot}
& = & 
\frac{e^{i{\bf k}_\bot\cdot{\bf x}_\bot \pm i\omega\zeta}}{2\pi}
\int_{-\infty}^\infty \frac{d\vartheta}{\sqrt{8a}\,\pi}
e^{\pm i \omega \vartheta/a} \nonumber \\
&& \ \ \ \ \ \ \ \ \ \ 
\times \exp\left( - i\kappa T\cosh\vartheta \right).
\end{eqnarray}
The $\vartheta$-integral is the same for both signs of 
$e^{\pm i\omega\vartheta/a}$ 
because the imaginary part of the integrand is odd in
$\vartheta$.  Adopting the minus sign and using the
formula~\cite{Gradshteynbook}
\begin{equation}
H_\nu^{(2)}(x) = - \frac{e^{i\nu\pi/2}}{\pi i}
\int_{-\infty}^\infty e^{-ix\cosh t - \nu t}\, dt
\end{equation}
with $\nu = i\omega/a$, we find
\begin{equation}
w_{\pm\omega{\bf k}_\bot} =
-i \frac{e^{i{\bf k}_\bot\cdot{\bf x}_\bot 
\pm i\omega\zeta}}{2\pi\sqrt{8a}}e^{\pi\omega/2a}
H_{i\omega/a}^{(2)}(\kappa T).
\end{equation}
These modes are well known to form a complete set of positive-frequency
modes which correspond to the Minkowski vacuum
state~\cite{Fullingetal74}.

Now, from Eq.~(\ref{sec2:uw}) we find that the positive-frequency modes
with respect to the boost generator 
in the right Rindler wedge corresponding to the Rindler vacuum state
take the following form in the expanding degenerate Kasner universe:
\begin{eqnarray}
v^R_{\omega{\bf k}_\bot} & = & -i
\frac{e^{-i\omega\zeta}
e^{i{\bf k}_\bot\cdot{\bf x}_\bot}}
{2\pi\sqrt{8a(e^{\pi\omega/a}-e^{-\pi\omega/a})}}
\nonumber \\
&& \times \left\{ e^{\pi\omega/a} H_{i\omega/a}^{(2)}
\left(\kappa T\right)
+ \left[H_{
i\omega/a}^{(2)}\left(\kappa T\right)\right]^*\right\}.\nonumber\\
\label{sec2:milnrindler}
\end{eqnarray}
Then, recalling the fact that
$[H^{(2)}_\nu(x)]^* = H_{-\nu}^{(1)}(x)$ if $\nu$ is purely
imaginary and if $x$ is real and using 
the formulas~\cite{Gradshteynbook}
\begin{eqnarray}
&& e^{\nu\pi i}H^{(2)}_{-\nu} (z) = H^{(2)}_{\nu}(z),\\
&& H^{(1)}_{\nu}(z) + H^{(2)}_{\nu}(z)  =  2J_{\nu}(z)
\end{eqnarray}
with $\nu=-i\omega/a$ in Eq.~(\ref{sec2:milnrindler}), we find
\begin{equation}
v_{\omega{\bf k}_\bot}^R  = -i 
\frac{e^{-i\omega\zeta}
e^{i{\bf k}_\bot\cdot{\bf x}_\bot}}
{2\pi\sqrt{4a\sinh(\pi\omega/a)}}J_{-i\omega/a}(\kappa T).
\label{sec2:thisexp}
\end{equation}
In exactly the same manner we find that the left Rindler modes
$v^L_{\omega {\bf k}_\bot}$ are given by Eq.~(\ref{sec2:thisexp}) with
$e^{-i\omega\zeta}$ replaced by $e^{i\omega\zeta}$
in the expanding degenerate Kasner universe.
These modes 
have been identified as the positive-frequency modes corresponding
to a state which is inequivalent to the Minkowski
vacuum~\cite{Birrelletal82,Fullingetal74}.  
Thus, the Rindler vacuum state $|0_{\rm R}\rangle$ is in
fact one of the states in the expanding 
degenerate Kasner universes given in the literature.

\subsection{Unruh effect and classical field theory}
\label{subsection:classicalfieldtheory}

Although the Unruh effect, like the Hawking effect, is a quantum effect,
its derivation does not involve any loop calculations.  
It is also the result of 
properties of classical solutions to the field equation.  These
observations naturally lead to the following question:
``Are there any aspects of the Unruh effect that can be described
entirely in the framework of classical field theory?" 
In this context, it is useful to note that, although the Unruh 
temperature $T = \hbar a/(2\pi c)$ (at $\xi=0$) is proportional 
to $\hbar$, since the energy of a particle can
be written as $E= \hbar \omega$, where $\omega$ is the angular frequency,
the Boltzmann factor $\exp(-E/ T) = \exp(-2\pi\omega c /a )$ is
independent of
$\hbar$.  This is consistent with the fact that the Bogolubov
transformation encoding the Unruh effect 
is derived using only classical solutions.
It is indeed
possible to define some quantities in classical field theory which 
exhibit what one may call the classical Unruh
effect~\cite{Higuchietal93b} as we briefly describe here.
\footnote{However, we
find the claim by~\textcite{Barutetal90} that the Unruh effect can be
explained without invoking a thermal bath rather misleading.  If one
were to describe physics in the Rindler wedge with the boost generator as the Hamiltonian, then the thermal bath with the temperature $a/2\pi$ 
would definitely be a necessary ingredient.}

We consider the classical scalar field $\phi$ in Minkowski spacetime
satisfying $(\Box +m^2)\phi = 0$.  The energy-momentum tensor is
\begin{equation}
T_{\mu\nu} = \nabla_\mu\phi\nabla_\nu\phi - g_{\mu\nu}
(\nabla_\alpha\phi\nabla^\alpha \phi - m^2\phi^2)/2.
\end{equation}
Now, if 
$X^\mu$ is a Killing vector, then the current $J^\mu_{(X)}$ defined by
\begin{equation}
J^\mu_{(X)} = X_\nu T^{\mu\nu}
\end{equation}
is conserved because of the Killing equation and the equation
$\nabla_\mu T^{\mu\nu} = 0$.  Hence, the energy associated with the
Killing vector $X^\mu$ defined by
\begin{equation}
E_X = \int d\Sigma\, n_\mu J^\mu_{(X)},
\end{equation}
where $\Sigma$ is a Cauchy hypersurface and $n_\mu$ is the future-directed
unit vector normal to
$\Sigma$, is conserved.  
If $T^\mu$ is the time-translation vector, then the energy
$E_T$ with $X^\mu=T^\mu$ is the ordinary energy.  If 
$R^\mu = a [z(\partial/\partial t)^\mu + t(\partial/\partial z)^\mu]$, 
i.e.~the boost Killing vector (normalized at $\xi=0$), 
then $E_R$ with $X^\mu=R^\mu$ is the Rindler energy.

It is convenient for our purposes to rewrite the energy $E_X$ as
\begin{equation}
E_X = (i/2) (\phi,X^\mu\nabla_\mu \phi)_{KG}.  \label{sec2:KGenergy}
\end{equation}
This can readily be established, by using the equality
\begin{eqnarray}
&& \phi\nabla_\mu(X^\alpha \nabla_\alpha\phi) -
X^\alpha\nabla_\alpha\phi\nabla_\mu\phi + 2X^\alpha
T_{\alpha\mu}\nonumber \\
&& = \nabla^\alpha 
\left[\phi(X_\alpha\nabla_\mu\phi - X_\mu\nabla_\alpha\phi)\right].
\end{eqnarray}

Now, one can divide the scalar field into the positive- and
negative-frequency parts with respect to the time-translation Killing
vector as
\begin{equation}
\phi(x) = \phi^{(T+)}(x) + \phi^{(T-)}(x),
\end{equation}
where the negative-frequency part is the complex conjugate of the
positive-frequency part, $\phi^{(-T)}(x) = [\phi^{(+T)}(x)]^*$, and
where the positive-frequency part is given as
\begin{equation}
\phi^{(T+)}(x) = \int
\frac{d^3{\bf k}}{\sqrt{2k_0}(2\pi)^{3/2}} c_T({\bf k})e^{-ik_0t + i{\bf
k}_\perp\cdot {\bf x}_\perp},
\end{equation}
for some function $c_T({\bf k})$.
Then, since $T^\mu\partial_\mu =\partial_t$, we find the energy by using
Eq.~(\ref{sec2:KGenergy}) as
\begin{equation}
E_T  = \int d^3{\bf k}\,k_0|c_T({\bf k})|^2.
\end{equation}
It is natural to define the quantity $N_T$ by dividing the integrand 
$k_0|c_T({\bf k})|^2$ by
$k_0$ as
\begin{equation}
N_T = \int d^3{\bf k}\,|c_T({\bf k})|^2,
\end{equation}
because the expected quantum-mechanical particle number is $N_T/\hbar$. We
call $N_T$ the {\em classical Minkowski particle number}.  It
is clear that
\begin{equation}
N_T = (\phi^{(T+)},\phi^{(T+)})_{\rm KG}. \label{sec2:Minkowskinumber}
\end{equation}

Now, if the field $\phi$ vanishes in the left Rindler wedge, then it can
be expanded in terms of the right Rindler modes $v^{R}_{\omega {\bf
k}_\bot}$. 
Thus, we have
\begin{equation}
\phi(x) = \phi^{(R+)}(x) + \phi^{(R-)}(x),
\end{equation}
where the positive-frequency part with respect to the boost Killing
vector $R^\mu$ is defined by
\begin{equation}
\phi^{(R+)}(x) = \int_0^\infty d\omega \int d^2{\bf k}_\perp
c_R(\omega,{\bf k}_\perp)v^R_{\omega{\bf k}_\bot}
\end{equation}
for some function $c_R(\omega,{\bf k}_\bot)$,
and the negative-frequency part is $\phi^{(R-)}(x) =
[\phi^{(R+)}(x)]^*$.  The Rindler energy is found by letting
$X^\mu=R^\mu$ in Eq.~(\ref{sec2:KGenergy}) as
\begin{equation}
E_R = \int_0^\infty d\omega \int d^2{\bf k}_\perp \,\omega
|c_R(\omega,{\bf k}_\perp)|^2.
\end{equation}
We can define the {\em classical Rindler particle number} as
\begin{equation}
N_R \equiv \int_0^\infty d\omega \int d^2{\bf k}_\perp\,
|c_R(\omega,{\bf k}_\perp)|^2.  \label{sec2:classicalRindler}
\end{equation}
Then we have
\begin{equation}
N_R = (\phi^{(R+)},\phi^{(R+)})_{\rm KG}. \label{sec2:Rindlernumber}
\end{equation}

It is possible to express the Minkowski particle number $N_T$ in terms
of $c_R(\omega, {\bf k}_\perp)$ as follows.  {}From Eq.~(\ref{sec2:uw})
we find
\begin{eqnarray}
\phi & = & \int_0^\infty d\omega \int d^2{\bf k}_\bot
\left[ c_R(\omega,{\bf k}_\bot)v^R_{\omega{\bf k}_\bot}
+ c_R^*(\omega,{\bf k}_\bot)v^{R*}_{\omega{\bf k}_\bot}\right]\nonumber
\\
& = &  \phi^{(T+)} + \phi^{(T-)},
\end{eqnarray}
where
\begin{eqnarray}
\phi^{(T+)} & = & 
\int_0^\infty d\omega \int d^2{\bf k}_\bot
\left[ \frac{c_R(\omega,{\bf k}_\bot)}{\sqrt{1-e^{-2\pi\omega/a}}}
\,w_{-\omega{\bf k}_\bot}\right. \nonumber \\
&& \ \ \ \ \ \ \ \ 
\left. - \frac{e^{-\pi\omega/a}c_R^*(\omega,{\bf k}_\bot)}
{\sqrt{1-e^{-2\pi\omega/a}}}\,w_{+\omega{\bf k}_\bot}\right].
\end{eqnarray}
Then, using 
Eq.~(\ref{sec2:posmodes}), we obtain the classical Minkowski particle
number as
\begin{eqnarray}
N_T & = & (\phi^{(T+)},\phi^{(T+)})_{\rm KG}\nonumber \\
& = & \int_0^\infty d\omega\int d^2{\bf k}_\bot
|c_R(\omega,{\bf k}_\bot)|^2\coth\frac{\pi\omega}{a}.\nonumber \\
\end{eqnarray}
Comparing this expression with that for the classical Rindler
particle number (\ref{sec2:classicalRindler}), we find that 
the Fourier components with respect to the Rindler time $\tau$ 
of the classical Minkowski particle number
is enhanced by a factor of $\coth(\pi\omega/a)$ in comparison to those
of the classical Rindler particle number.  We refer the reader 
to \textcite{Higuchietal93b} for the interpretation of this formula in
the context of the Unruh effect.

\subsection{Unruh effect for interacting theories and 
in other spacetimes}
\label{section2:extension}

In this subsection we briefly mention some works which extend the Unruh
effect to interacting field theory and other spacetimes.

Let us first discuss the work of 
\textcite{Bisognanoetal75,Bisognanoetal76}, who derived
the Unruh effect for (interacting) 
quantum field theory satisfying Wightman
axioms~\cite{Wightman56,Streateretal64,Jost65}.  The Unruh effect was
not presented as the main result in their work, and it was only 
several years after its publication that its connection to the Unruh
effect was discovered by Sewell, who also extended their derivation of
the Unruh effect to a class of spacetimes
including Schwarzschild and de~Sitter spacetimes~\cite{Sewell82}.

In order to discuss the work of Bisognano and Wichmann, 
it is necessary to review a mathematically more satisfactory way to
define a thermal state in quantum field theory, which is called the KMS
condition~\cite{Haagetal67}.  [The initials KMS stand 
for~\textcite{Kubo57} and \textcite{Martinetal59}.]  
We first explain (one version of) the KMS condition 
for a quantum system with a finite number of energy levels
with a Hamiltonian $\hat{H}$ and a complete set of eigenstates
$|n\rangle$ with energy $E_n$. 
The expectation value of an operator $\hat{A}$ 
in a thermal state with inverse temperature $\beta = 1/T$ is
\begin{equation}
\langle \hat{A}\rangle_\beta = \frac{\sum_{n}e^{-\beta E_n}\langle
n|\hat{A}|n\rangle}{\sum_{m} e^{-\beta E_m}}
= \frac{{\rm Tr}(e^{-\beta \hat{H}}\hat{A})}{{\rm Tr}
( e^{-\beta \hat{H}})}.
\end{equation}
Let ${\cal H}$ be the Hilbert space spanned by $|n\rangle$.  This
thermal state is realized as a pure state in the Hilbert space
${\cal H}\otimes {\cal H}$ as 
\begin{equation}
|\beta\rangle = \frac{\sum_n e^{-\beta E_n/2}
|n\rangle \otimes |n\rangle}{\sqrt{\sum_{m}e^{-\beta E_m}}},
\label{sec2:state-beta}
\end{equation}
if the operators $\hat{A}$ on ${\cal H}$ are
identified with 
$
\hat{A}^{(e)} = \hat{I} \otimes \hat{A}
$,
where $\hat{I}$ is the identity operator.  That is,
\begin{equation}
\langle \beta|\hat{A}^{(e)}|\beta\rangle = 
\langle \hat{A}\rangle_\beta.
\end{equation}  
The time-evolution operator is taken to be
\begin{equation}
\exp(-i\hat{H}^{(e)}\tau) = 
\exp(i\hat{H}\tau)\otimes \exp(-i\hat{H}\tau). \label{sec2:extendedH}
\end{equation}

Now, let us define an anti-unitary involution $\hat{J}^{(e)}$ by
\begin{equation}
\hat{J}^{(e)}\alpha |n\rangle\otimes |m\rangle = \alpha^*|m\rangle \otimes
|n\rangle,
\end{equation}
where $\alpha$ is any c-number.  Then, the operator $\hat{J}^{(e)}$ commutes
with the time-evolution operator:
\begin{equation}
\hat{J}^{(e)}\exp(-i\hat{H}^{(e)}\tau) 
= \exp(-i\hat{H}^{(e)}\tau)\hat{J}^{(e)},\ \ \forall 
\tau\in \mathbb{R}.  \label{sec2:KMS1}
\end{equation}
One can also show by an explicit calculation that,
for any operator $\hat{A}$ given by
a matrix as $\hat{A}|n\rangle = \sum_m |m\rangle A_{mn}$,
\begin{equation}
\exp(-\hat{H}^{(e)}\beta/2)\hat{A}^{(e)}|\beta\rangle = \hat{J}^{(e)} 
\hat{A}^{(e)\dagger}|\beta\rangle.
\label{sec2:KMS2}
\end{equation}
It can readily be seen that, in our model
with a finite number of energy levels, Eq.~(\ref{sec2:KMS2})
implies that the state $|\beta\rangle$ must be given by
Eq.~(\ref{sec2:state-beta}) up to an overall phase factor.

In algebraic field theory a state\footnote{In algebraic field
theory ``a state" means, roughly speaking, ``a density matrix" 
in general.} that allows a Hilbert space
representation satisfying the conditions
(\ref{sec2:KMS1}) and (\ref{sec2:KMS2}) is called a KMS state at 
inverse temperature $\beta$.  Thus, 
the Unruh effect in algebraic field theory is the statement that the
Minkowski vacuum restricted to the right Rindler wedge is a KMS state
at inverse temperature $\beta = 2\pi/a$ if the time-evolution is
identified with a boost, which is the $\tau$-translation in the Rindler
coordinates~(\ref{sec2:rightcoords}).
Remarkably, the doubling of the Hilbert space 
and the involution $\hat{J}^{(e)}$ in the above construction, 
which might look somewhat
artificial in the context of statistical mechanics, naturally arise
here.  Thus, given the QFT in the right Rindler wedge
with a boost generator as the Hamiltonian
we `extend' it by including the left Rindler wedge
and operators acting there.  The extended boost generator automatically 
takes the form (\ref{sec2:extendedH}) 
since the corresponding Killing vector
field is past-directed in the left Rindler wedge.

In the two-dimensional model (with only the left movers)
the involution $\hat{J}^{(e)}$ is defined by requiring 
\begin{eqnarray}
\hat{J}^{(e)}|0_{\rm M}\rangle & = & |0_{\rm M}\rangle,\\
\hat{J}^{(e)}\hat{a}_{+\omega}^R\hat{J}^{(e)} & = & 
\hat{a}_{+\omega}^L,\\
J^{(e)}\hat{a}_{+\omega}^{R\dagger} J^{(e)} 
& = & \hat{a}_{+\omega}^{L\dagger}.
\end{eqnarray}
Note that $[\hat{J}^{(e)}]^2 = \hat{I}\otimes \hat{I}$. The involution 
$\hat{J}^{(e)}$ is in fact the
$PCT$ transformation, i.e.~the anti-unitary transformation 
$\hat{\Phi}(t,z)\mapsto \hat{\Phi}(-t,-z)$ 
in this two-dimensional model. For the
four-dimensional scalar field it is the $\pi$-rotation about the
$z$-axis times the $PCT$ transformation 
[see \textcite{Bisognanoetal75}]. With these definitions one can
readily verify that Eqs.~(\ref{sec2:vacannihi1}) and
(\ref{sec2:vacannihi2}) imply Eq.~(\ref{sec2:KMS2}).  The commutation
relation (\ref{sec2:KMS1}) follows from the fact that the Lorentz boost
commutes the $PCT$ transformation.

The derivation of the Unruh effect by~\textcite{Bisognanoetal75} 
using the algebraic approach
was for any interacting scalar field satisfying the Wightman 
axioms. They also generalized this result to quantum fields of arbitrary
spins~\cite{Bisognanoetal76}.  They showed that the
Minkowski vacuum restricted to the right or left Rindler wedge is a KMS
state as explained above.  For the four-dimensional scalar field theory,
for example, 
if $\exp(-i\hat{K}\alpha)$ 
is the boost operator corresponding to
\begin{eqnarray}
t & \mapsto & t(\alpha) \equiv t\cosh\alpha a + z\sinh\alpha a, 
\label{sec2:talpha} \\
z & \mapsto & z(\alpha) \equiv t\sinh\alpha a + z\cosh\alpha a,
\label{sec2:zalpha}
\end{eqnarray}
then, for $\alpha = i\pi/a$, one has $(t,z)\mapsto (-t,-z)$.
Bisognano and Wichmann proved, roughly speaking, that this fact
translates to
\begin{eqnarray}
&& \exp(-\hat{K}\pi/a)
\hat{\Phi}(t^{(1)},z^{(1)},{\bf x}_\perp^{(1)})
\cdots \hat{\Phi}(t^{(n)},z^{(n)},{\bf x}_\perp^{(n)})
|0_{\rm M}\rangle \nonumber \\
&& = \hat{\Phi}(-t^{(1)},-z^{(1)},{\bf x}_\perp^{(1)})
\cdots \hat{\Phi}(-t^{(n)},-z^{(n)},{\bf x}_\perp^{(n)})|0_{\rm
M}\rangle,\nonumber \\
\label{sec2:Bison}
\end{eqnarray}
where $|0_{\rm M}\rangle$ is a unique Poincar\'{e} invariant vacuum,
which is assumed to exist, if
$(t^{(i)},z^{(i)},{\bf x}_\perp^{(i)})$, $i=1,2,\ldots,n$, are spatially
separated points in the right Rindler
wedge.\footnote{Bisognano and Wichmann showed that the rigorous
version of Eq.~(\ref{sec2:Bison}) makes sense,
i.e.~that states 
obtained by multiplying $|0_{\rm M}\rangle$ by a finite number of
operators of the form $\int d^4 x f(x)\hat{\Phi}(x)$, where $f(x)$ 
has
support in the right Rindler wedge, is in the domain of the operator
$\exp(-\beta\hat{K})$ for $0\leq \beta \leq \pi/a$.}
Then, they converted the
relation (\ref{sec2:Bison}) to the KMS condition (\ref{sec2:KMS2}) with
$\hat{H}^{(e)} = \hat{K}$, $\beta = 2\pi/a$ and with 
$\hat{J}^{(e)}$ being the
$PCT$ operator times the $\pi$-rotation about the $z$-axis for operators
$\hat{A}^{(e)}$ acting in the left Rindler wedge by a result similar to
the Reeh-Schlieder theorem~\cite{Reehetal61}.
\footnote{This theorem states,
roughly speaking,
that any state in the Hilbert space of 
the scalar field theory can be
approximated by applying polynomials of operators of the form
$\int d^4 x\,f(x)\hat{\Phi}(x)$ on the vacuum state $|0_{\rm M}\rangle$,
where $f(x)$ have
support in a finite spacetime region.} [See \textcite{Kay85} for
a discussion of the Bisognano-Wichmann theorem in the context of free
field theory.]

Let us describe how Eq.~(\ref{sec2:Bison}) can be derived in the
simplest case with $n=1$ and with free (four-dimensional) scalar field.  
Using 
$\hat{K}|0_{\rm M}\rangle = 0$ and $\hat{a}^M_{k_z{\bf k}_\bot}|0_{\rm
M}\rangle = 0$, we have for a real parameter $\alpha$
\begin{eqnarray}
&& \exp(i\alpha\hat{K})\hat{\Phi}(t,z,{\bf x}_\bot)|0_{\rm
M}\rangle\nonumber \\
& & = 
\hat{\Phi}(t(\alpha),z(\alpha),{\bf x}_\bot)|0_{\rm M}\rangle
\nonumber \\
& & = \int \frac{d^3{\bf k}}{\sqrt{(2\pi)^32k_0}}\,
e^{i(k_0z-k_z t)\sinh\alpha a - i{\bf k}_\bot\cdot{\bf
x}_\bot}\nonumber \\
&& \times e^{i(k_0t-k_z z)\cosh\alpha a}
\hat{a}^{M\dagger}_{k_z{\bf k}_\bot}|0_{\rm M}\rangle, 
\label{sec2:modulusless}
\end{eqnarray}
where $t(\alpha)$ and
$z(\alpha)$ are defined by Eqs.~(\ref{sec2:talpha}) and 
(\ref{sec2:zalpha}), respectively.  It can be shown that the variable
$\alpha$ can be analytically continued from $0$ to $i\pi/a$ if $z >
|t|$, i.e.~if the point $(t,z,{\bf x}_\bot)$ is in the right Rindler
wedge.\footnote{To be precise, one needs to consider the inner product
of the state in Eq.~(\ref{sec2:modulusless}) with a normalized
one-particle state. 
Note that the modulus of $e^{i(k_0z-k_z z)\sinh\alpha a}$ in 
this equation is
always less than or equal to $1$ if $\alpha$ is between $0$ and $i\pi/a$.
This fact is essential in showing that this
analytic continuation is possible.}
Thus,
\begin{equation}
\exp(-\hat{K}\pi/a)\hat{\Phi}(t,z,{\bf x}_\bot)|0_{\rm M}\rangle
= \hat{\Phi}(-t,-z,{\bf x}_\bot)|0_{\rm M}\rangle, \label{sec2:Wich}
\end{equation}
if the point $(t,z,{\bf x}_\bot)$ is in the right Rindler wedge.  This
is indeed Eq.~(\ref{sec2:Bison}) with $n=1$ for a free field.
Noting that the point $(-t,-z,{\bf x}_\bot)$ is in the left Rindler
wedge and using the expansion (\ref{sec2:rindler}) of the field
$\hat{\Phi}$ in terms of the Rindler modes, one can readily deduce from 
Eq.~(\ref{sec2:Wich})
the relations
(\ref{sec2:annihilate1}) and (\ref{sec2:annihilate2}), which were
crucial in showing the Unruh effect.

\textcite{Unruhetal84} derived the Unruh effect for the scalar field
theory with arbitrary potential term $V(\hat{\Phi})$ in the path
integral approach. [They also discussed the Unruh effect for spinors.
See also~\textcite{Gibbonsandso78}.]  Here we present
their argument, for the
two-dimensional scalar field for simplicity of notation,
in a slightly modified manner.  What needs to be shown is that
\begin{equation}
\langle 0_{\rm M}|T\left[\hat{\Phi}(x)\hat{\Phi}(x^{\prime})\right]
|0_{\rm M}\rangle \\
= \frac{{\rm Tr}\left\{ e^{-\beta\hat{K}}
T\left[\hat{\Phi}(x)\hat{\Phi}(x^{\prime})\right]\right\}}
{{\rm Tr}(e^{-\beta \hat{K}})}, \label{sec2:UnruhWeiss}
\end{equation}
where the trace is over all states, $\hat{K}$ is the boost operator
defined above and $\beta = 2\pi/a$.  
The argument for a similar equality involving an $n$-point function with
arbitrary $n$ is almost identical.

The Lagrangian density for the scalar field with potential $V(\phi)$ is
\begin{equation}
{\cal L} = 
[     ( {\partial\phi}/{\partial t} )^2 
 -    ( {\partial\phi}/{\partial z} )^2]/2 - V(\phi).
\end{equation}
In the Rindler coordinates given by Eq.~(\ref{sec2:Rindcoord})  
with $\eta = a\tau$,
i.e.
\begin{equation}
t  =  \rho \sinh a\tau, \;\;\;
z  =  \rho \cosh a\tau, 
\label{sec2:Lorentz}
\end{equation}
this Lagrangian density
is given by
\begin{equation}
{\cal L} = a\rho\left[ \frac{1}{2a^2 \rho^2}
\left(\frac{\partial\phi}{\partial \tau}\right)^2 
-\frac{1}{2}\left(\frac{\partial\phi}{\partial \rho}\right)^2 
- V(\phi)\right].
\end{equation}
Define the Euclidean action by letting $\tau = -i\tau_e$ as
\begin{eqnarray}
S_E^R(\beta) & \equiv & -\int_0^\beta d\tau_e\int_0^\infty d\rho
\,{\cal L}_{(\tau=-i\tau_e)}
\nonumber \\
& = & \int_0^\beta
a\,d\tau_e \int_0^\infty d\rho\,\rho \nonumber \\
&& \times 
\left[ \frac{1}{2}\left(\frac{\partial\phi}{\partial\rho}\right)^2
+ \frac{1}{2\rho^2}
\left(\frac{1}{a}\frac{\partial\phi}{\partial\tau_e}\right)^2
+ V(\phi)\right],
\nonumber\\
\end{eqnarray}
where $\phi(\tau_e+\beta,\rho) = \phi(\tau_e,\rho)$.
It is well known [see, e.g., \textcite{Bernard74}] that the right-hand
side of Eq.~(\ref{sec2:UnruhWeiss}) for an arbitrary value of $\beta$ 
is obtained by the analytic continuation
$\tau_e = i\tau$ of the following expression:
\begin{eqnarray}
&& D_\beta(x_e,x_e') \nonumber \\
&& \equiv
\frac{\int_{\phi(\tau_e=0)=\phi(\tau_e=\beta)} 
\left[D\phi\right]
\phi(x_e)\phi(x_e')\exp\left[-S_E^R(\beta)\right]}
{\int_{\phi(\tau_e=0)=\phi(\tau_e=\beta)} 
\left[D\phi\right]
\exp\left[-S_E^R(\beta)\right]},\nonumber \\
\end{eqnarray}
where $x_e = (t_e,z_e)$ is obtained from Eq.~(\ref{sec2:Lorentz}) as
\begin{equation}
t_e  =  \rho\sin a\tau_e,\;\;\;
z_e  = \rho\cos a\tau_e.
\end{equation}
These equations show that 
the Euclideanized right Rindler wedge is the two-dimensional
Euclidean space expressed in polar coordinates if $0\leq a\tau_e \leq
2\pi$, i.e.~if $\beta = 2\pi/a$.  Thus, one has
\begin{eqnarray}
S_E^R(2\pi/a) & = & S_E\nonumber \\
& \equiv & \int_{-\infty}^\infty dt_e \int_{-\infty}^\infty
dz_e\nonumber \\
&& \times \left[ \left(\frac{\partial\phi}{\partial t_e}\right)^2
+ \left(\frac{\partial\phi}{\partial z_e}\right)^2 + V(\phi)\right].
\nonumber \\
\end{eqnarray}
Hence,
\begin{equation}
D_{2\pi/a}(x_e;x_e')
= \frac{\int \left[D\phi\right]\phi(x_e)\phi(x_e')\exp(-S_E)}{
\int\left[D\phi\right] \exp(-S_E)}. \label{sec2:expSE}
\end{equation}
It is well known that the time-ordered two-point function in
(two-dimensional) Minkowski spacetime, i.e.~the left-hand side of
Eq.~(\ref{sec2:UnruhWeiss}), is obtained from the right-hand
side of Eq.~(\ref{sec2:expSE}) 
by the analytic continuation $t_e = it$.  Since both sides of
Eq.~(\ref{sec2:UnruhWeiss}) are obtained by the analytic continuation of
the same function $D_{2\pi/a}(x_e;x_e')$ 
with $x_e = (t_e,z_e) = (it,z)$, Eq.~(\ref{sec2:UnruhWeiss}) holds.

The analog of the Unruh effect in Schwarzschild spacetime
was first derived by
\textcite{Hartleetal76} using analytic properties of the time-ordered 
two-point function for scalar and other free fields.
They showed that the
physically-acceptable\footnote{The condition for ``physical
acceptability" here is essentially 
the so-called {\em Hadamard condition}.  See
\textcite{Wald78} and \textcite{Fullingetal78} 
for an early use of this condition.}
vacuum state invariant under the time-translation in the Kruskal
extension~\cite{Kruskal60} of Schwarzschild spacetime with mass $M$ is
a thermal state of temperature $1/8\pi M$.  
This result obviously has very close connection
to the Hawking effect~\cite{Hawking74}. 
A similar method was used by \textcite{Gibbonsetal77} to show that the
physically-acceptable de~Sitter-invariant vacuum state
of the free scalar field in de~Sitter
spacetime with Hubble constant $H$ is a thermal state of temperature
$H/2\pi$ of the theory inside the cosmological horizon with the
de~Sitter boost generator fixing the horizon as the Hamiltonian.  
\textcite{Narnhoferetal96} found that an accelerated detector with
acceleration $a$ in de~Sitter spacetime responds as if it was in a
thermal bath of temperature $(H^2 + a^2)^{1/2}/2\pi$, and
\textcite{Deseretal97} obtained a similar result in anti-de~Sitter
spacetime.  Interestingly, they found that the temperature is equal to 
the Unruh
temperature corresponding to the 
acceleration of the detector in $5$-dimensional
Minkowski spacetime in which (anti-)de~Sitter spacetime is embedded.
\textcite{Jacobson98} gives a simple explanation of these results, and
\textcite{Buchholzetal07} discuss them in the context of their definition
of a local temperature.  For
some work related to the response rate of the Unruh-DeWitt detector in
de~Sitter spacetime, see, e.g., \textcite{Higuchi87} and
\textcite{Garbrechtetal04a,Garbrechtetal04b}.
The Bisognano-Wichmann result was also extended to Schwarzschild and
de~Sitter spacetimes by Sewell as mentioned before.

\textcite{Kayetal91} proved the analog of the Unruh effect in a class of
spacetimes with bifurcate Killing horizons~\cite{Boyer69} adopting the
viewpoint that 
Hadamard states are the only physical 
states for the free
scalar field theory.  They showed that the Wightman two-point function
$\Delta(x;x')$ on the horizon satisfies
\begin{equation}
\partial_{U}\partial_{U'}\Delta(U,s;U',s') = -
\frac{1}{4\pi}\frac{1}{(U-U'-i\varepsilon)^2}\delta^2(s,s')
\end{equation}
with $x=(U,s)$, where $s$ parametrizes 
the null geodesics on the Killing
horizon and where $U$ is an affine parameter on each geodesic, for a
Hadamard state invariant under the Killing symmetry.  This
formula allowed them to show that, if such a state exists, it must be
unique. Then they applied essentially the same argument as for the
massless scalar field theory in the two-dimensional Rindler wedges to
derive the Unruh-like effect. 
[See also~\textcite{Kay93,Kay01} for further
developments and a brief account of this result.]

\section{Applications}
\label{section:Applications}

In this section we review some works using
the Unruh effect to examine some 
selected phenomena. We begin by discussing
each phenomenon using plain quantum field theory adapted to inertial
observers, and then we show how the same observables can be recalculated
from the point of view of Rindler observers with the help of the 
Unruh effect. The first example is connected with the excitation of 
accelerated detectors and atoms, the second one with the weak decay of 
non-inertial protons and the third one with the interpretation of 
the radiation emitted by charges from the point of view of uniformly 
accelerated observers. In particular we clarify the traditional 
question whether or not uniformly accelerated charges emit radiation 
from the point of view coaccelerated observers. 

\subsection{Unruh-DeWitt detectors}
\label{subsection:detector}

Models of photon detectors have been discussed 
for some time in
quantum optics \cite{Glauber63}. \textcite{Unruh76} has introduced
a detector model consisting of a small box containing
a non-relativistic particle satisfying the Schr\"odinger equation. 
The system is said to have detected a quantum
if the particle 
in the box jumps from the ground state to some excited state.
In the same paper, Unruh also discusses a relativistic detector model
[see also \textcite{Sanchez81} for a similar model]. Here, we 
consider in more detail the detector model introduced by 
\textcite{DeWitt79}, which consists of a {\em two-level point 
monopole}.  
We call generically
two-level point monopoles as {\em Unruh-DeWitt detectors}
following the literature.  
A discussion 
on particle detectors with finite spatial extent can be found in 
\textcite{Groveetal83}.

Particle detectors have 
often been used to probe the Unruh 
thermal bath. Sometimes, however, distinct detector designs may 
lead to contrasting conclusions about the same given feature of the bath. 
For instance, \textcite{Israeletal83}, \textcite{Sanchez85},
\textcite{Hintonetal83} and \textcite{Hinton83} 
have argued that the Unruh thermal bath is anisotropic while 
\textcite{Kolbenstvedt87}, \textcite{Gerlach83} and 
\textcite{Groveetal85} have argued the opposite. 
It is not surprising that, in general, directionally sensitive detectors 
will respond differently 
if they are given distinct orientations. Nevertheless,
the Unruh thermal bath is as isotropic as a thermal bath in equilibrium 
in a general static spacetime can be in the sense 
that Killing observers will see 
no net energy flux, etc.~in 
any space direction, as is well known. In general, the temperature
$\beta^{-1}|_i$ measured by a Killing observer following a curve $i$
generated by a Killing vector 
will be position dependent. Two Killing observers 
following curves $i=1,2$ will have their temperatures related as 
$$
{\beta^{-1}|_1}/{\beta^{-1}|_2} =  
[(\zeta_{\mu} \zeta^{\mu})|_2/(\zeta_{\mu} \zeta^{\mu})|_1]^{1/2},
$$
where
$\zeta^{\mu}$  is the Killing vector tangent to the world line of
the corresponding observer~\cite{Tolman}.

Let us consider a two-level Unruh-DeWitt detector in 
Minkowski spacetime.
The detector will be represented by a Hermitian operator $\hat m$ acting
on a 
two-dimensional Hilbert space.  The excited state, $|E \rangle$, and the
unexcited state, $|E_0 \rangle$, are assumed to be eigenstates of 
the detector's Hamiltonian $\hat H$:
\begin{equation}
\hat H | E   \rangle = E   | E    \rangle,\;\;
\hat H | E_0 \rangle = E_0 | E_0  \rangle
\end{equation}
with eigenvalues $E$ and $E_0$, respectively ($E>E_0$). The
monopole is time evolved as usual:
\begin{equation}
\hat m(\tau )\equiv e^{i\hat H \tau} \hat m_0 e^{-i\hat H \tau}, 
\label{m}
\end{equation}
where $\tau$ is the detector's proper time. The matrix element
$q \equiv \langle E | \hat m_0 | E_0 \rangle $
depends on the detector 
design.\footnote{Two-level point monopoles have been also 
successfully used to model the excitation and deexcitation of 
atoms  \cite{Audretschetal94, Zhuetal07}.}

Now, let us couple our Unruh-DeWitt detector to a real massive 
scalar field $\hat{\Phi}(x)$
satisfying the Klein-Gordon equation 
$\Box \hat{\Phi} +m^2 \hat{\Phi} = 0$ through the interaction action
\begin{equation}
     \hat S_I =
       \int_{-\infty}^{\infty}
       d\tau \; \hat m (\tau )\; 
\hat{\Phi}[x(\tau)],
  \label{SI1}
\end{equation}
where $x^\mu (\tau)$ is the detector's world line. 
Next, we analyze the response of the detector from the point of view 
of inertial and Rindler observers separately. Related investigations 
for detectors coupled with electromagnetic and Dirac fields can 
be found in \textcite{Boyer80,Boyer84} and \textcite{Iyeretal80}, 
respectively.

\subsubsection{Uniformly accelerated detectors in Minkowski vacuum:
               Inertial observer perspective}
\label{subsubsection:uadinertial}

\begin{figure}[t]
\epsfig{file=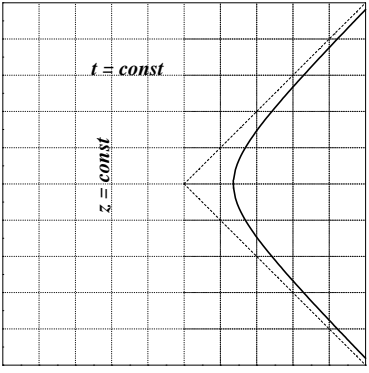,angle=0,width=0.6\linewidth,clip=}
\caption{\label{Cartesiancoordinates} 
The world line of a uniformly accelerated detector moving 
along the $z$-axis in the Minkowski spacetime covered with 
Cartesian  coordinates is shown.}
\end{figure}

In Cartesian coordinates, $x^\mu= (t,x,y,z)$, 
of Minkowski spacetime 
the world line $x^\mu =x^\mu (\tau )$ of a 
uniformly accelerated detector along the $z$-axis with proper 
acceleration $a$ is given by 
\begin{equation}
t(\tau ) = a^{-1} \sinh a \tau, \;\;\; z(\tau) = a^{-1} \cosh a \tau\;
\label{Rindlercoordinates}
\end{equation}
and $x(\tau ), y(\tau ) = {\rm const}$~(see Fig.~\ref{Cartesiancoordinates}).

Let us expand  $\hat{\Phi}(x)$ 
in terms of positive-
and negative-energy eigenstates of the Hamiltonian 
$\hat H = i \partial/\partial t$, associated with inertial
observers, as [see Sec.~(\ref{section2:massive})]
\begin{equation}\label{Phimassive}
\hat{\Phi}(x) 
=
        \int d^3 {\bf k}
        \left(
        u_{{\bf k}} \hat{a}_{\bf k}^M + {\rm H.c.}
        \right),
\end{equation}
where 
\begin{equation}\label{positivefrequencymodePhimassive}
u_{\bf k} = [2\omega (2\pi)^{3}]^{-1/2} e^{- i k_\mu x^\mu}
\end{equation}
with
$k^\mu = (\omega, {\bf k}) $, $\omega = \sqrt{{\bf k}^2 + m^2}$ 
and
$$
[ \hat{a}_{\bf k}^M, \hat{a}_{\bf k'}^{M\dagger} ] = 
\delta^3( {\bf k} - {\bf k'} ).
$$

The proper excitation rate, i.e.~the excitation probability divided by
the total detector {\em proper} time $T$, associated with the 
uniformly accelerated 
detector in the inertial vacuum is given 
by\footnote{ Often, the excitation 
rate is alternatively expressed in terms of the golden 
rule~(\ref{RMgeneral}).}
\begin{equation}
^{\rm exc}\! R =
          T^{-1}
          \int d^{3} {\bf k}
          |^{\rm exc}\!\! {\cal A}^{\rm em}_{\bf k}|^2,
\label{RMinertial}
\end{equation}
where the excitation amplitude is (up to an arbitrary phase)
\begin{eqnarray}
^{\rm exc}\!\! {\cal A}^{\rm em}_{\bf k} 
    & = & i \langle E |\otimes \!\langle{\bf k}{}_{\rm M}
      | \hat S_I | 0_{\rm M} \rangle \otimes|E_0 \rangle
\nonumber \\
    & = &  \frac{q}{(16 \pi^3 \omega)^{1/2}} 
    \int_{-\infty}^{\infty} d\tau 
    \exp (i \Delta E \tau)
\nonumber \\
     & \times &  \exp[(i\omega/a) \sinh a\tau - (i k_z/a) 
\cosh a\tau ]
\nonumber \\
\label{UDinertialA}
\end{eqnarray}
with $\Delta E\equiv E-E_0$. We have adopted here the subscript $M$ to label 
states defined by inertial observers in Minkowski spacetime. 
(Note here 
that we are using the convention that space components of the momentum
$k^\mu$ are given with lower indices.  That is, $k_x$, $k_y$ and 
$k_z$ are the $x$- $y$- and
$z$-components, respectively, of the contravariant vector $k^\mu$.)
We note 
that because Eq.~(\ref{SI1}) is
linear in $\hat{\Phi}[x(\tau)]$,
the detector excitation is accompanied 
by the emission of a particle\footnote{ 
This combination, i.e.~excitation with particle emission, can be also 
observed in the anomalous Doppler effect where atoms move in media with 
refractive index $n$ with velocity $v>1/n$  \cite{Frolovetal86}.}
with momentum ${\bf k}$. 
By  using
Eq.~(\ref{UDinertialA}) in Eq.~(\ref{RMinertial}), we obtain
\begin{equation}
^{\rm exc}\! R \equiv \int d^2 {\bf k}_\perp R^\perp,
\label{RMinertialintermediary}
\end{equation}
where 
${\bf k}_\perp = (k_x, k_y)$ denotes the transverse momentum with
respect to the direction of the acceleration, the quantity $R^\perp$ is
given by  
\begin{eqnarray}
R^\perp & = & \frac{|q|^2}{16\pi^3 T} 
             \int^{\infty}_{-\infty} \frac{ d k_z }{ \omega }
             \int^{\infty}_{-\infty} d\tau' 
             \int^{\infty}_{-\infty} d\tau''  
             e^{i\Delta E (\tau' - \tau'')}     
\nonumber \\   
& \times & e^{ i \omega [\sinh a\tau' - \sinh a\tau'']/a }
           e^{-i k_z    [\cosh a\tau' - \cosh a\tau'']/a }
\nonumber \\
&=&  \frac{|q|^2}{16\pi^3 T}
     \int^{\infty}_{-\infty} \frac{dk_z}{\omega}
     \int^{\infty}_{-\infty} d \tau
     \int^{\infty}_{-\infty} d \sigma 
     e^{i\Delta E \sigma}
\nonumber \\
& \times & e^{ (2 i/a) \sinh a\sigma 
              (\omega \cosh a\tau - k_z \sinh a\tau ) },
\nonumber
\end{eqnarray}  
and we have defined
$
\tau \equiv (\tau' + \tau'')/2
$ 
and 
$
\sigma \equiv \tau' - \tau'' 
$. 
Because the interaction is kept turned on for an arbitrarily long time 
interval, 
the total time $T$ diverges. To obtain 
explicitly the excitation rate per unit time,
the total time $T$ must be 
canceled by factoring out the divergent part 
$\int_{-\infty}^{\infty}d\tau$ from the integrals above. 
To this end, we first
note that the momentum of the emitted 
particle is boosted due to the nonzero
velocity of the detector, which is $\tau$-dependent.  
Hence, it is expected
that the integrand can be made $\tau$-independent by
boosting back the momentum variables.  
Motivated by this physical picture, we introduce a new momentum
variable as 
\begin{equation}
k_z \mapsto {k'}_z \equiv 
k_z \cosh a \tau - \omega \sinh a\tau,
\label{k->k'}
\end{equation}
which can be inverted as
\begin{equation}
{k'}_z \mapsto k_z = 
{k'}_z \cosh a \tau + {\omega'} \sinh a\tau.
\label{k'->k}
\end{equation}
Here
$
{\omega'} \equiv [({k'}_z)^{2} +k_\perp^2 +m^2]^{1/2} 
$
can be expressed as
$$
{\omega'} = \omega \cosh a \tau - k_z \sinh a \tau,
$$ 
where 
$
k_\perp \equiv \sqrt{(k_x)^2 + (k_y)^2}
$.
It can be shown that 
$ 
d {k'}_z/{\omega'} = d k_z/ \omega
$. 
This transformation indeed allows us to factor out
$T=\int_{-\infty}^\infty d\tau$, and we obtain
\begin{equation}
R^\perp =\frac{|q|^2}{16\pi^3 }
     \int^{\infty}_{-\infty} \frac{d {k'}_z}{{\omega'}}
     \int^{\infty}_{-\infty} d \sigma 
     e^{i\Delta E \sigma}
     e^{ (2 i {\omega'}/a)  \sinh (a\sigma/2)}.
\end{equation}
By making now a further change of variables as
\begin{eqnarray}
{k'}_z &\mapsto& \zeta \equiv ({\omega'} + 
                      {k'}_z)/\sqrt{k_\perp^2 + m^2}
                      \nonumber \\
\sigma      &\mapsto& \lambda \equiv \exp (a \sigma/2),
\end{eqnarray}
we obtain
\begin{eqnarray}
R^\perp &=&\frac{|q|^2}{8 \pi^3 a  }
          \int^{\infty}_{1} \frac{d\zeta}{\zeta}
          \int^{\infty}_{0} d\lambda\, \lambda^{2i \Delta E /a -1}
\nonumber \\          
          &\times& 
          \exp [i (k_\perp^2+m^2)^{1/2}
          (\lambda -\lambda^{-1}) (\zeta + \zeta^{-1})/(2a)  ].
\nonumber          
\end{eqnarray}
Now, we make the change of variables
$
\{\zeta, \lambda \} \mapsto \{ \kappa, \varpi \}
$ 
with
$
\kappa = \zeta \lambda 
$ 
and  
$
\varpi = \zeta/\lambda 
$,
and write
\begin{eqnarray}
R^\perp &= &\frac{|q|^2}{16 \pi^3 a}
          \left| 
          \int^{\infty}_{0} \frac{d\kappa}{\kappa^{1-i\Delta E/a}}
          e^{ i (k_\perp^2+m^2)^{1/2} ( \kappa- \kappa^{-1})/(2a)}
          \right|^2
\nonumber \\
        &= & \frac{|q|^2 e^{-\pi \Delta E/a } }{4 \pi^3 a  } 
\left\{K_{i \Delta E/a} [(k_\perp^2+m^2)^{1/2}/ a]\right\}^2,
\end{eqnarray}
where 
$K_{\nu}(x)$ 
is the modified Bessel function \cite{Gradshteynbook}.
(We recall here that the function $K_{i\lambda}(x)$ is real if $x$ and
$\lambda$ are real.) Finally, we obtain for the proper excitation 
rate~(\ref{RMinertialintermediary}) 
\begin{figure}[t]
\epsfig{file=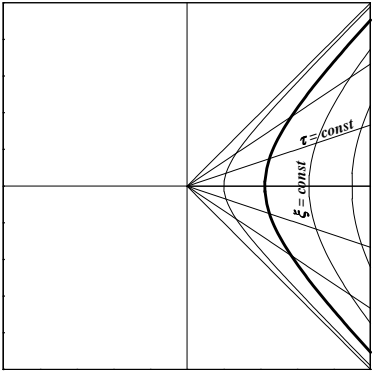,angle=0,width=0.60\linewidth,clip=}
\caption{\label{RindlercoordinatesFig} The world line of a uniformly 
accelerated detector moving along the $z$-axis in Minkowski 
spacetime covered with Rindler coordinates is shown.}
\end{figure}
\begin{eqnarray}
^{\rm exc}\! R     
& & 
= 
\frac{|q|^2 a e^{-\pi \Delta E/a}}{2 \pi^2 }
          \int^{\infty}_{0} d\mu \; \mu\;  
\nonumber \\
& & 
\times \left( K_{i\Delta E/a} [\sqrt {\mu^2+(m/a)^2 \,} ]
       \right)^2.
\label{RMfinal}
\end{eqnarray}
The angular distribution of the corresponding emitted particles 
in the massless case can be found in \textcite{Kolbenstvedt88}. 
Next, we shall reproduce this detector response from the point 
of view of
Rindler observers and discuss it.

\subsubsection{Uniformly accelerated detectors in Minkowski vacuum:
               Rindler observer perspective}
\label{subsubsection:uadRindler}
The spacetime appropriate for investigating
the excitation rate of our detector
with proper acceleration $a$ from the point of view of 
uniformly accelerated observers is the Rindler wedge. We choose the right 
Rindler wedge ($ z > |t| $) to work with, where we recall that it has 
a global timelike isometry associated with the Killing field 
$z \partial/\partial t + t \partial/\partial z$. 
By covering it with Rindler coordinates $(\tau,\xi, y, z)$
($-\infty<\tau, \xi, x, y <+\infty$), 
which are related with $(t,x,y,z)$ by Eq.~(\ref{sec2:rightcoords}), 
we obtain the line element of the Rindler wedge as written in 
Eq.~(\ref{sec2:rightmetric}).

The world lines of the Rindler observers are given 
by $\xi,x,y = {\rm const}$ and are hyperbolas 
in the two-dimensional diagram of Minkowski spacetime 
with $x$ and $y$ suppressed~(see Fig.~\ref{RindlercoordinatesFig}).
The corresponding four-velocity and four-acceleration 
are 
$u^\mu = e^{- a \xi}(1,0,0,0) $ 
and 
$a^\mu = e^{-2 a \xi }(0, a ,0,0)$,
respectively,
where $a^{\mu} = u^{\nu}\nabla_{\nu} u^{\mu}$ 
[see, e.g., \textcite{Waldbook84}]. Thus, the proper acceleration
of the Rindler observers is  
$\sqrt{-a^\mu a_\mu} = a e^{- a \xi}  = {\rm const}$.
Our uniformly accelerated detector 
with proper acceleration $a$ will lie 
at $\xi = 0$ (for some $x, y = {\rm const}$).

Next, we expand $\hat{\Phi}(x)$
in terms of positive- and negative-energy eigenstates of the Hamiltonian 
$\hat H = i \partial/\partial \tau$, associated with the Rindler 
observers, as [see Sec.~(\ref{section2:massive})]
\begin{equation}\label{Phimassiveprime2}
\hat{\Phi}(x) =
        \int d\omega d^2 {\bf k}_\perp
        [
        v_{\omega {\bf k}_\perp}^R \hat{a}^R_{\omega {\bf k}_\perp} + {\rm H.c.}
        ],
\end{equation}
where 
\begin{equation}\label{positivefrequencymodePhimassiveprime2}
v_{\omega {\bf k}_\perp}^R = 
\left[
\frac{\sinh (\pi \omega /a)}{4\pi^4 a}
\right]^{1/2}
\!\!
K_{i\omega/a} \left[ \frac{ \sqrt{k_\perp^2+m^2}}{ a e^{-a\xi} } \right]
e^{i {\bf k}_\perp\cdot {\bf x}_\perp - i \omega \tau}
\end{equation}
are  Klein-Gordon orthonormalized, and we recall that the
creation and annihilation operators of Rindler particles
satisfy the commutation relations
\begin{equation}\label{creationanihilationcomutatorprime2}
[\hat{a}_{\omega {\bf k}_\perp}^R, 
 \hat{a}_{\omega' {\bf k'}_\perp}^{R \dagger} ] 
= \delta (\omega-\omega') \delta^2( {\bf k}_\perp - {\bf k'}_\perp).
\end{equation}
The Rindler vacuum $|0_{\rm R}\rangle$
is defined by $\hat{a}_{\omega {\bf k}_\perp}^R |0_{\rm R}\rangle = 0$.
A detector lying at rest within a uniformly accelerated 
cavity prepared in the Rindler vacuum 
is not excited~\cite{Levinetal92}.\footnote{An account on vacuum states 
                             in static spacetimes
                             with horizons can be found in 
                             \textcite{Fulling77}.}
We emphasize that the quantum numbers 
$\{\omega, {\bf k}_\perp\}$ 
associated with the timelike and spacelike global Killing 
fields 
$\partial/\partial \tau$ 
and 
$\partial/\partial x$, $\partial/\partial y$, respectively,
are independent of each other (see Sec.~\ref{subsubsection:E<mc2}).

Before we analyze the behavior of the detector in the {\em Minkowski}
vacuum, we formally consider the detector's excitation probability with 
simultaneous emission of a Rindler particle in the {\em Rindler} vacuum. 
The amplitude associated with this process in first order of perturbation 
is
\begin{equation}
{^{\rm exc} \! {\cal A}_{\omega {\bf k}_\perp}^{\rm em}}
\equiv 
i \langle E | \otimes  \! \langle \omega {\bf k}_\perp\, {}_{\rm R}
| \hat S_I | 
0_{\rm R} \rangle \otimes | E_0 \rangle,
\label{AREM}
\end{equation}
where we recall that we use Eq.~(\ref{Phimassiveprime2}) 
in $\hat S_I$ as given in Eq.~(\ref{SI1}). The differential 
probability associated with this amplitude is
\begin{equation}
dW^{\rm em} = \vert ^{\rm exc} \! {\cal A}_{\omega {\bf k}_\perp}^{\rm em} \vert^2 
              d^2{\bf k}_\perp d\omega.
\label{DWREM}
\end {equation}
Now, we should take into account the fact
that due
to the Unruh effect
the Minkowski vacuum corresponds to a thermal bath of Rindler particles. 
We emphasize that {\em the Minkowski vacuum is indistinguishable from 
the thermal bath built on the Rindler vacuum as long as the detector 
stays in the Rindler wedge} since the Minkowski vacuum is a linear 
combination of products of the left and right Rindler states. For this 
reason, the detector's excitation rate with simultaneous emission of 
a Rindler particle into the {\em Minkowski} vacuum is given by 
Eq.~(\ref {DWREM}) combined with the proper thermal factor [see 
Eq.~(4.9) in \textcite{Higuchietal92b} for more details]:
\begin{equation}
^{\rm exc}\! R^{\rm em} = T^{-1} 
                      \int  dW^{\rm em} 
                      [1 + n(\omega) ],
\label{PMEM}
\end{equation}
where 
\begin{equation}
n(\omega) = 1/ \left[\exp(\beta \omega) -1\right]
\label{Planckfactor}
\end{equation}
is the 
Rindler scalar particle number density in the momentum space.
Here $\beta ^{-1} = a/2\pi$ is the Unruh temperature 
as measured by  Rindler observers at $\xi = 0$.
The first and second terms in the square brackets in Eq.~(\ref{PMEM})
are associated with spontaneous and induced emission, respectively.

Similarly, one can calculate the detector's excitation rate
with simultaneous absorption of a Rindler particle
from the Unruh thermal bath as
\begin{equation}
^{\rm exc}\! R^{\rm abs} = T^{-1} 
                       \int  dW^{\rm abs} 
                       n(\omega ),
\label{PMABS}
\end{equation}
where
\begin{equation}
dW^{\rm abs} \equiv \vert ^{\rm exc}\! {\cal A}_{\omega {\bf k}_\perp}^{\rm abs} \vert^2 
              d^2{\bf k}_\perp d\omega
\label{DWRABS}
\end{equation}
and
\begin{equation}
{^{\rm exc}\! {\cal A}_{\omega {\bf k}_\perp}^{ \rm abs}}
\equiv 
i \langle E | \otimes \langle 0_{\rm R} 
| \hat S_I | 
\omega {\bf k}_\perp\,{}_{\rm R} \rangle \otimes | E_0 \rangle
\label{ARABS}
\end{equation}
is the excitation amplitude with absorption  of a
Rindler particle $| \omega {\bf k}_\perp\,{}_{\rm R} \rangle$.
The excitation amplitudes~(\ref{AREM}) and~(\ref{ARABS}) can 
be shown to be 
\begin{eqnarray}
& & 
{^{\rm exc} \! {\cal A}_{\omega {\bf k}_\perp}^{ \rm em (abs) }} = 
q \int^{\infty}_{-\infty} 
d\tau 
\exp [i{\bf k}_\perp\cdot {\bf x}_\perp + i (\Delta E + (-) \omega) \tau ]
\nonumber \\
& &
\times 
     \left[ 
     \frac{\sinh (\pi \omega/a)}{4 \pi^4 a}
     \right]^{1/2}
K_{i\omega/a} \left(\! \frac{(k_\perp^2 + m^2)^{1/2} e^{a\xi}}{a} \! \right)
\nonumber \\ 
\label{amplitudedetectorrindler}
\end{eqnarray}
up to some multiplicative phase. It is easy to verify in this case
that  ${^{\rm exc}\! {\cal A}_{\omega {\bf k}_\perp}^{ \rm em }} =0 $, as expected,
since uniformly accelerated detectors are static according
to Rindler observers. Hence, according to these  observers
the only contribution to the detector response comes from the
absorption of Rindler particles from the Unruh thermal bath.

Now, since in first order of perturbation there is no interference,
the total detector excitation rate in the {\em Minkowski} 
vacuum is
\begin{equation}
^{\rm exc}\! R = ^{\rm exc}\! R^{\rm em} + ^{\rm exc}\! R^{\rm abs}.
\label{totalresponsedetector}
\end{equation}
{\em By using Eq.~(\ref{amplitudedetectorrindler}) to calculate
Eq.~(\ref{totalresponsedetector}), we get Eq.~(\ref{RMfinal}),
as expected.}
Of course, inertial and Rindler observers 
must agree on the value of scalar observables, 
such
as the {\em proper} 
excitation rate of a given detector, although they can differ in
how they describe 
the 
phenomenon.
Because inertial and Rindler observers 
would
expand the
quantum fields with different sets of normal modes, they 
would
end
up extracting different particle contents from the same field theory.
As a result, it is natural for {\em inertial} and {\em Rindler} 
observers to describe the detector excitation as being accompanied by the 
{\em emission of a Minkowski particle} and by the {\em absorption of 
a Rindler particle from the Unruh thermal bath} \cite{UnruhWald84}, 
respectively. This conclusion can be generalized for detectors 
confined in the Rindler wedge following {\em general world lines} by 
saying that the detector excitation which is associated with the
emission of a Minkowski particle as described by {\em inertial} 
observers corresponds in this case to the absorption {\em or} 
emission of a Rindler particle from {\em or} to the Unruh thermal 
bath according to Rindler observers \cite{Matsas96}.

Let us comment on one possible source of confusion concerning the
Unruh-DeWitt detector.  A naive (and wrong) application of
the equivalence principle might lead to the conclusion that an inertial
detector which has the same velocity as an accelerated one at a certain
time would detect Unruh radiation.  This is of course not the case: no
detector in an inertial motion detects any Unruh radiation.

Before proceeding further, we note for 
later purposes that in the particular case 
with $m=0$, Eq.~(\ref{RMfinal}) takes the form
\begin{equation}
^{\rm exc}\! R^{m=0} = \frac{|q|^2 }{2 \pi} 
            \frac{\Delta E}{e^{ \beta \Delta E }-1}.
\label{RPlanck}
\end{equation}

\subsubsection{Rindler particles with frequency $\omega <m $ }
\label{subsubsection:E<mc2}

Here we discuss in more detail the existence of Rindler particles with
frequencies $\omega <m $, which was crucial in the whole 
discussion above
(notice, e.g., that the range of the $\omega$ integrations in 
Eqs.~(\ref{PMEM}) and~(\ref{PMABS}) is $0<\omega< + \infty $).
The standard theory of quantum fields uses the fact that Minkowski
spacetime is 
invariant under time and space translations.
The linear three-momentum 
$(k_x,k_y,k_z)$ associated with the translational isometries on the 
spacelike hypersurfaces $t= {\rm const}$ constitutes a suitable set of quantum
numbers to label free particles.  In this simple case, the dispersion
relation $E \equiv \hbar \omega = \sqrt{| {\bf p} \, c |^2 + m^2 \, c^4 \;}$
imposes a simple constraint between the particle mass $m$, momentum ${\bf p}$
and energy $E$, and, thus, free particles with well-defined linear momenta must
have total energy $E \ge m\, c^2$. Moreover, 
in the classical context of General 
Relativity, the detection {\em in loco} of {\em point} particles satisfying 
$ E < m c^2 $ by direct capture is ruled out by the fact that an 
observer with 
four-velocity $u^\mu$ intercepting a particle with four-momentum 
$p^\mu = m v^\mu$ assigns to the particle an energy
$E = m v^\mu u_\mu \ge mc^2$. 

On the other hand, it is well known that the field quantization 
carried over arbitrary spacetime does not lead in general to 
any dispersion  relation connecting the frequency with other 
quantum numbers, avoiding thus the flat spacetime constraint 
$E \ge mc^2$. This can be understood by recalling that, strictly
speaking, the concept of 
{\em point particle} has no place in Quantum Field Theory. 
This raises the following question: What is the probability 
density associated with the detection of particles with $E < m c^2$, 
i.e.~$\omega < m$, at different space points of the Rindler wedge?
By answering this question, we can also extract some information
about the particle distribution of the Hawking radiation near
the event horizon of black holes. Indeed, much insight into the 
Hawking effect can be obtained in the simplified context provided 
by the Rindler wedge as we shall see next. [We refer the reader to 
\textcite{Castineirasetal02} for more details.]

Let us start by considering the line element of a two-dimensional 
Schwarzschild spacetime:
\begin{equation}
ds^2= \left( 1 - 2M/r \right) dt^2 - \left( 1 - 2M/r \right)^{-1} dr^2.
\label{SS}
\end{equation}
This can be seen as describing a two-dimensional 
black hole\footnote{The vacuum expectation value of the energy-momentum 
                    tensor for a massless scalar field in this spacetime 
                    was analyzed by \textcite{Daviesetal76} [see also
                    \textcite{Davies76} and \textcite{Daviesetal77b}]. 
                    See 
                    \textcite{Christensenetal77} for a comprehensive discussion
                    on this issue and \textcite{Candelasetal79} for further 
                    considerations.} 
with mass $M$. Close to the horizon, $r \approx 2M$, it can be written as
\begin{equation}
ds^2=  (\rho/4M)^2 dt^2 -  d\rho^2,
\label{RW}
\end{equation}
where $\rho(r) \equiv \sqrt{8M(r-2M)}$.  (Note that in these coordinates the
horizon is at $\rho=0$.) One can identify Eq.~(\ref{RW}) with the line 
element of the Rindler wedge~(\ref{sec2:rightmetric}) with $x,y = {\rm const}$ 
by letting $t=4Ma \tau$ and $\rho=e^{a \xi}/a$ provided that 
$0<\rho<+\infty$ and $-\infty < t < +\infty$. 
\begin{figure}
\begin{center}
\mbox{\epsfig{file=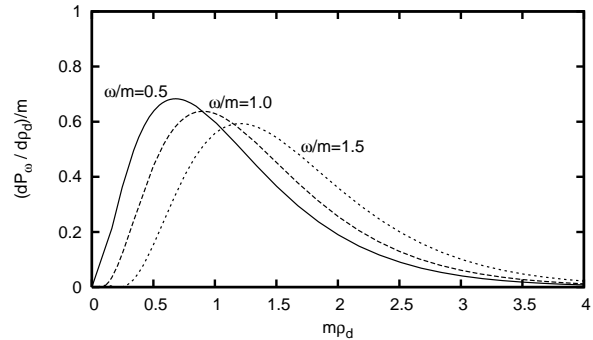,width=0.45\textwidth,angle=0}}
\end{center}
\caption{We plot the probability density $d{\cal P}_\omega/d\rho_d$ 
for different $\omega/m$ ratios, where we have let $Mm=1/4$. We 
note that the smaller the $\omega/m$ ratio,  the closer to the
horizon (on average) the particle lies, where the 
``gravitational potential'' is lower.}
\label{fig2ccmv}
\end{figure}

From here to the end of this subsection we shall be considering the 
spacetime of the Rindler wedge with line element~(\ref{RW}), 
where $0< \rho < +\infty$ and $ -\infty < t < + \infty$.
Now, let us choose a {\em fiducial observer} at $\rho = \rho_0 = 4 M$, 
whose proper time is $t$ [see Eq.~(\ref{RW})], with respect to whom the 
particle's energy is 
to be measured. He/she defines the total probability 
${P}_\omega(\rho_d) $ of detecting a particle at some point 
$\rho= \rho_d$ with 
energy $\omega$ per (detector) proper time $s^{\rm tot}_d$ as
$\Gamma_\omega (\rho_d) \equiv  {P}_\omega(\rho_d) / s^{\rm tot}_d$.
Then, the normalized probability density is
\begin{equation}
\frac{d{\cal P}_\omega}{d\rho_d} 
\equiv 
{\Gamma}_\omega (\rho_d)\left[
\int_0^{+\infty}{\Gamma}_\omega (\rho'_d) d\rho'_d\right]^{-1},
\label{density}
\end{equation}
where $ (d{\cal P}_\omega/d\rho_d) d\rho_d$ 
is the probability that a particle with energy 
$\omega$ is found between $\rho_d$ and $\rho_d + d\rho_d$.
Observers far away from the horizon will  be able to interact  
only with the ``tail'' of the ``wave functions'' associated with 
particles with small $\omega/m$. The smaller the 
$\omega/m$, the more difficult it is to detect these particles.

\begin{figure}
\begin{center}
\mbox{\epsfig{file=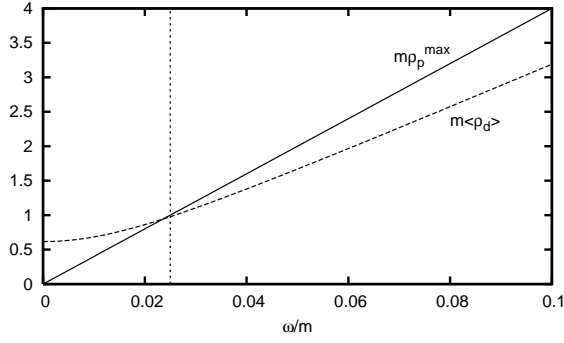,width=0.45\textwidth,angle=0}}
\end{center}
\caption{$\langle \rho_d \rangle$ is shown to be smaller than $\rho_p^{\rm max}
\equiv 4 M \omega / m$ {\em in the ``high-frequency'' regime 
$\omega > (4 M)^{-1} $} (i.e.~at the right of the vertical broken line) 
just as expected. (We have let $m M = 10$ here but this nice agreement is
verified for any $m M$.)}
\label{fig3ccmv}
\end{figure}

Now, in order to interpret Eq.~(\ref{density}) in the frame work of
General Relativity, 
let us first consider a row of detectors, each of them
lying at different $\rho_d$, and define the {\em average detection
position}
\begin{equation}
\langle \rho_d \rangle 
\equiv 
\int_0^{+\infty} \; d\rho_d \;\rho_d \;d {\cal P}_\omega / d \rho_d.
\label{averagedef}
\end{equation} 
By using Eq.~(\ref{density}), this can be shown to be  
(see Fig.~\ref{fig2ccmv})
\begin{eqnarray}
\langle \rho_d \rangle & = & 
\frac{\pi\tanh (4\pi M \omega) (64 M^2 \omega^2 +1 ) }{64 m M \omega}  
\nonumber \\
& \approx &
{\pi \;M  \omega}/{ m } 
\;\;\;\;\;\;\;\;\;\;\;\;\;\;\;\;\;\;\; (\omega \gg a),
\label{average}
\end{eqnarray} 
where $a \equiv 1/4M$ is the proper acceleration of the fiducial observer.
On the other hand, from General Relativity, a classical particle with 
mass $m$ {\em lying at rest} 
at some point $\rho_p$  has, according to our fiducial observer at  
$\rho_0 = 4M$, energy $\omega = m \rho_p/4M$. By considering that 
the particle  may have some kinetic energy in addition, 
the total energy would be $\omega \ge m \rho_p/4M$. 
{}From this equation, we obtain  
\begin{equation}
\rho_p  \le  {4 M \omega}/{ m } \equiv \rho_p^{\rm max}.
\label{particle}
\end{equation} 
This is expected to agree with $\langle \rho_d \rangle$, 
i.e.~$\langle \rho_d \rangle \le  \rho_p^{\rm max}$, at least
in the ``high-frequency'' regime $\omega \gg a$ (where the 
quantum and classical behaviors may be compared). 
This conclusion is indeed in agreement with Eqs.~(\ref{average}) 
and (\ref{particle}) as seen in Fig.~\ref{fig3ccmv}. 
In summary, the smaller the 
$\omega/m$ ratio, the more likely 
the observer is to detect the particle closer to the horizon.

\subsubsection{Static detectors in a thermal bath of Minkowski particles }
\label{subsubsection:staticdetectors}

Now, we shall show explicitly that the response rate (\ref{RMfinal}) 
does not
correspond to the one obtained when the detector lies at rest in a plain thermal bath 
of Minkowski particles heated up to the Unruh temperature $\beta^{-1} = a/(2\pi)$. 
In the latter case, the excitation rate is obtained by replacing 
Eq.~(\ref{RMinertial}) by
\begin{equation}
     ^{\rm exc}\! R^\beta \equiv 
     T^{-1} 
       \int d^3 {\bf k}       
       [ |^{\rm exc}\!\! {\cal A}_{\bf k}^{\rm em }|^2 [1+ n (\omega)] + 
         |^{\rm exc}\!\! {\cal A}_{\bf k}^{\rm abs}|^2 n (\omega) ], 
   \label{totalresponsedetectorinertial}
\end{equation}
where $n(\omega)= 1/ (\exp(\beta \omega ) -1)$ and
\begin{align}
{^{\rm exc}\!\! {\cal A}_{{\bf k}}^{\rm em}}
\equiv 
i \langle E | \otimes \!\langle {\bf k}\,{}_{\rm M} 
| \hat S_I | 
0_{\rm M} \rangle \otimes | E_0 \rangle,
\nonumber \\
{^{\rm exc}\!\! {\cal A}_{{\bf k}}^{\rm abs}}
\equiv 
i \langle E | \otimes \!\langle 0_{\rm M} 
| \hat S_I | 
{\bf k}\,{}_{\rm M} \rangle \otimes | E_0 \rangle	
\nonumber
\end{align}
are the excitation amplitudes with emission and absorption  of
Minkowski particles $| {\bf k}\,{}_{\rm M} \rangle$, respectively. 
In this case, the excitation amplitudes can be shown to be
(up to an arbitrary multiplicative phase) 
\begin{equation}
{^{\rm exc}\!\! {\cal A}_{{\bf k}}^{ \rm em (abs) }} = 
q  \delta(\omega + (-) \Delta E)/\sqrt{4 \pi \omega\,},
\label{amplitudedetectorminkowskithermalbath}
\end{equation}
where we have assumed with no loss of generality that the detector is at 
the origin ${\bf x}=0$. Clearly, 
${^{\rm exc}\!\! {\cal A}_{{\bf k}}^{ \rm em }} =0 $. 
Hence, the only contribution to the detector response is associated with 
the absorption of a Minkowski particle. 
By substituting
Eq.~(\ref{amplitudedetectorminkowskithermalbath}) in 
Eq.~(\ref{totalresponsedetectorinertial}), we obtain 
\begin{equation}
^{\rm exc}\! R^\beta = \frac{|q|^2 \; \Delta E }{2 \pi} 
            \frac{\theta (\Delta E - m) }{e^{ \beta \Delta E }-1}.
\label{RT}
\end{equation}
The presence of the step function $\theta (\Delta E -m)$ expresses the fact
that the detector can only be excited if its energy gap is large enough
to absorb massive scalar particles from the thermal bath. Clearly 
$^{\rm exc}\! R^\beta$ in Eq.~(\ref{RT}) with $\beta^{-1} = a/(2 \pi)$ and 
$^{\rm exc}\! R$ in  Eq.~(\ref{RMfinal}) {\em are} distinct. Indeed, 
there is no {\em a priori} reason 
why they should be the same.  Incidentally,
in the case $m=0$, 
Eqs.~(\ref{RT})  and~(\ref{RPlanck}) equal each other. 
However,
this is a coincidence, which has to do with the particular design 
of the detector and not with the Unruh effect.  As we have seen, 
what the Unruh effect does say is something else. In particular,
we have shown in Sec.~\ref{subsubsection:uadRindler}
how to recover Eq.~(\ref{RMfinal}) from the point of view of Rindler 
observers. 

In spite of 
the model dependence of the response rate,
the Unruh effect allows one to establish a 
relationship  between the excitation and deexcitation 
rates (assuming that they are 
well defined), which is independent of the detector model. Because of 
unitarity, the absolute value of the {\em excitation} amplitude 
(associated with the {\em absorption} of a Rindler particle),
\begin{equation}
|{^{\rm exc}\!\! {\cal A}^{ \rm abs }}|
\equiv 
|
\langle E | \otimes \langle 0_{\rm R} 
| \hat S_I | 
\omega {\bf k}_\perp\,{}_{\rm R} \rangle\otimes | E_0 \rangle
|
\propto \delta (\omega - \Delta E),
\label{excitationamplitude}
\end{equation}
must equal the absolute value of the {\em deexcitation} 
amplitude (associated with the {\em emission} of a Rindler
particle),
\begin{equation}
|{^{\rm deexc}\!\! {\cal A}^{ \rm em }}|
\equiv 
|
\langle E_0 | \otimes \langle  \omega {\bf k}_\perp\,{}_{\rm R}
| \hat S_I | 
0_{\rm R} \rangle \otimes | E \rangle
|
\propto \delta (\omega - \Delta E).
\label{deexcitationamplitude}
\end{equation}
Hence, we obtain from the discussion in 
Sec.~\ref{subsubsection:uadRindler} 
that the ratio of the excitation and deexcitation rates must be given by 
\begin{equation}
\frac{^{\rm exc}\! R}{^{\rm deexc}\! R} 
= 
\frac{n(\Delta E )}{1+n(\Delta E )} 
=
e^{-\beta \Delta E}.
\label{ratio}
\end{equation}
This quite universal relation is also called 
{\em principle of detailed balance}, and clearly holds 
for detectors lying at rest in a thermal bath of Minkowski particles at the
same temperature\footnote{
                       \textcite{Boyer80,Boyer84}  
                       has concluded in the context of {\em stochastic electrodynamics}
                       that a classical electric dipole oscillator accelerating
                       through {\em classical electromagnetic zero-point radiation}
                       responds as would a dipole oscillator in an inertial frame
                       in classical thermal radiation with temperature $a/(2\pi)$
                       in agreement with Eq.~(\ref{ratio})
                       [see also \textcite{Cole85}].}
provided that $\Delta E \geq m$. For
$\Delta E < m$, the ratio $^{\rm exc}\! R/ ^{\rm deexc}\! R$
is not well defined for the thermal bath in Minkowski spacetime 
since the excitation and deexcitation 
rates vanish. It is worthwhile to 
emphasize, however, that deviations from the hyperbolic motion 
because of backreaction or other effects will in general 
disturb Eq.~(\ref{ratio}) with no contradiction to the Unruh effect,
contrary to some recent claims 
\cite{Belinskiietal97,Fedotovetal99}.

\subsubsection{About the discussion whether or not uniformly 
               accelerated sources radiate}
\label{subsubsection:controversy1}

Some controversy has appeared in the literature about whether or not 
uniformly accelerated detectors and sources emit radiation according to
inertial observers.\footnote{See Sec.~\ref{subsubsection:iopbremsstrahlung}
for some related remarks concerning electric charges.}
This is sometimes called {\em Unruh radiation} (although, 
this terminology has also been used to mean something else 
[see Sec.~\ref{subsection:lasers}]). Indeed, the conclusion that 
the excitation of a uniformly accelerated detector is accompanied by 
the emission of a Minkowski particle according to inertial observers and 
absorption of a Rindler particle from the Unruh thermal bath according 
to Rindler observers \cite{UnruhWald84} was not unanimously accepted 
in the beginning \cite{Padmanabhan85}. \textcite{Grove86} claimed that a 
constantly accelerated object would emit 
negative- rather than positive-energy radiation as seen by 
inertial observers. Similar conclusions were reached by 
\textcite{Massaretal93}. Later, the excitation of uniformly 
accelerated detectors in the Minkowski vacuum was said to give 
rise to no energy flux\footnote{See, e.g.,
                                \textcite{Hinterleitner93,Huetal04,Huetal04b,Fordetal06}.
                                A reply to \textcite{Huetal04} can be found in
                                \textcite{Scullyetal04}. See also
                                Sec.~\ref{subsection:cavities} for some 
                                further remarks.}
according to inertial observers, supporting previous
conclusion due to \textcite{Raineetal91}. More recently 
\textcite{Parentani95} and \textcite{Massaretal96} stated 
that, although the mean flux vanishes,
once the thermal equilibrium is reached, each 
detector transition is indeed accompanied by the emission of a 
Minkowski particle.

The controversy above was revisited by \textcite{Unruh92}, who concluded 
that uniformly accelerated oscillators do
alter the measurable properties of the field. Ascribing these changes to 
{\em ``radiation''} or {\em ``vacuum fluctuation''} would be a terminology 
issue. This viewpoint 
is in agreement with \textcite{Audretschetal94b} who argued 
that the key to the seemingly contradictory results lies in the distinction 
between the different questions stated implicitly in the various 
approaches.\footnote{More discussions on this issue can
be found in \textcite{Linetal06}.}

\subsubsection{Other results concerning the Unruh-DeWitt detector}
\label{subsubsection:otherresults}

The emission rates obtained in the previous subsections,
which use detectors accelerated uniformly for all times, 
should be seen as approximations to those
obtained in real physical situations where the detectors
are accelerated and interacting for long enough. 
\textcite{Svaiteretal1992} 
considered the Unruh-DeWitt detector that was turned on only for a 
finite time. This model might appear physical but 
they obtained a transition probability which was logarithmically
divergent although the transition rate, the time-derivative of the
transition probability, was found to be finite.
[See also \textcite{Sriramkumaretal96}.]

In order to find the cause of
this divergence and confirm 
that the transition rate obtained with uniform
acceleration is a good approximation to that for a detector model turned
on only for a finite time, \textcite{Higuchietal93} 
modified the interaction
action~(\ref{SI1}) as follows:  
\begin{equation}
{\hat S}_I = \int_{-\infty}^{\infty} d\tau c(\tau)
\hat m(\tau ) 
\hat{\Phi}[x(\tau)]
\label{SIfinitetime}
\end{equation}
where the function
\begin{equation}
 c(\tau) \equiv  \left\{ \begin{array}{cc}
                 e^{\alpha (\tau + T)} &  \mbox{$\tau < -T$}  \\
                 1                     &  \mbox{$-T\leq
                                          \tau \leq  T$}, \\
                 e^{-\alpha (\tau - T)} &  \mbox{$\tau > T$}
                 \end{array}
                 \right.
\label{c(tau)}
\end{equation}
with $\alpha= {\rm const}$, has been inserted to switch on and off the
detector continuously as $\tau \to - \infty$ and 
$\tau \to \infty$, respectively. The field
$\hat{\Phi} (x)$ is assumed to be a {\em massless} scalar field for
simplicity. The excitation rate is calculated in the Rindler 
frame. 
Eq.~(\ref{totalresponsedetector}) for the interaction
(\ref{SIfinitetime}) takes the form 
\begin{eqnarray}
^{\rm exc}\! R_T 
         & = & ^{\rm exc}\! R^{\rm em}_T + ^{\rm exc}\! R^{\rm abs}_T
\nonumber \\
         & = & |q|^2 (I^{\rm sp} + I^{\rm in} + I^{\rm abs} )/(4\pi^2 T^{\rm tot}),
\label{TOTPM'}
\end{eqnarray}
where
$T^{\rm tot} \equiv 2 T$. Here
\begin{eqnarray}
I^{\rm sp} & = & \int_0^{\infty} d\omega \omega B(\omega),
\label{ISP}
\\
I^{\rm in} & = &\int_0^{\infty} d\omega g(\omega) B(\omega),
\label{IIN}
\\
I^{\rm abs} & = & \int_0^{\infty} d\omega g(\omega) B(-\omega), 
\label{IAB}
\end{eqnarray}
are associated with the spontaneous emission, induced emission and
absorption probabilities, respectively, with
\begin{eqnarray}
&  B(\omega) \equiv &
\frac{4\sin^2(\omega+\Delta E)T}{(\omega+\Delta E)^2}
- \frac{4\sin^2(\omega+\Delta E)T}{\alpha^2 + (\omega+\Delta E)^2}\nonumber \\
& & +\frac{4\alpha^2\cos 2(\omega +\Delta E) T}
                   {[\alpha^2+(\omega+\Delta E)^2]^2} \nonumber \\
& & +\frac{4\alpha^3\sin 2(\omega+\Delta E)T}
                   {(\omega+\Delta E)[\alpha^2+(\omega+\Delta E)^2]^2},
\label{BW}
\end{eqnarray}
and 
$g(\omega) \equiv \omega n(\omega)
 = \omega[e^{2\pi\omega/a} - 1]^{-1}$.
It is easy to see
that the integrands for $I^{\rm sp}$, $I^{\rm in}$ and $I^{\rm abs}$
in Eqs.~(\ref{ISP})-(\ref{IAB}) do not diverge at any value of $\omega$. 
Also, the integrands for $I^{\rm in}$ and $I^{\rm abs}$ tend to zero
exponentially as $\omega \to \infty$, and 
the leading term of the asymptotic expansion 
(for $\omega \gg \Delta E,\alpha ,a$) 
of the integrand for $I^{\rm sp}$ is 
$4 \alpha ^2\omega^{-3}  \cos ^2 \omega T$.
Thus, ${}^{\rm exc}\!R_T$ in Eq.~(\ref {TOTPM'}) 
is clearly finite. Now, suppose we switch 
on and off the detector instantaneously, in such a way that it 
only interacts with the field during the interval 
$-T < \tau < T$. 
This setup corresponds to the limit 
$\alpha \to +\infty$ 
[see Eq.~(\ref{c(tau)})]. 
In this regime the integrand for $I^{\rm sp}$ 
behaves asymptotically like
$4 \omega^{-1}\sin ^2 \omega T$, giving 
rise to a logarithmic ultraviolet divergence in $^{\rm exc}\!R_T$ 
found by Svaiter and Svaiter.
In a physical situation where we have a finite $\alpha$ and 
large $T$ (i.e.~$T \gg a^{-1}, \alpha ^{-1}, \Delta E ^{-1}$), 
one finds
\begin{equation}
^{\rm exc}\! R_T \approx 
\frac { |q|^2}{2 \pi }
\frac{\Delta E}{e^{\beta \Delta E} - 1},
\label{PM3}
\end{equation}
recovering the Planckian excitation rate~(\ref{RPlanck}).
Thus, the logarithmic divergence appears when we take
the $\alpha \to +\infty$ limit with finite $T$. This divergence would not
appear if we took $T\to +\infty$ from the beginning. In this case 
the absence of the logarithmic divergence could be attributed 
to the fact that the switching on and off would be 
moved away to infinite past and future, respectively.

The good ultraviolet behavior of the detector's total excitation
probability $^{\rm exc}\! R_T $ does  not depend sensitively on the particular
choice of the function $c(\tau )$ provided that 
$c(\tau)$ is at least continuous.
It would be interesting to see if the 
results obtained for finite-time {\em detectors} and
for finite-lifetime {\em observers} \cite{Martinettietal03}
are related.

Recently \textcite{Loukoetal06} 
have found a formula for the excitation
rate of the Unruh-DeWitt detector with any trajectory in Minkowski
spacetime for the massless scalar field, 
building on works by \textcite{Schlicht04} and
\textcite{Langlois05,Langlois06}.
If the trajectory is $x^\mu(\tau)$, where $\tau$ is the
proper time, and if the detector is
turned on at $\tau=\tau_0$, then the Louko-Satz formula for the 
excitation rate at proper time $\tau$ is
\begin{eqnarray}
R_{\rm LS}  & = & |q|^2\left\{ - \frac{\Delta E}{4\pi} +
 \frac{1}{2\pi^2}\int_0^{\tau-\tau_0}ds
\left[ \frac{1}{s^2}-\frac{\cos (s\Delta E)}{(\Delta x)^2}\right]\nonumber
\right. \\
&&\left.
+ \frac{1}{2\pi^2(\tau-\tau_0)}\right\}, \label{LoukoSatz}
\end{eqnarray}
where $\Delta x^\mu\equiv x^\mu(\tau) - x^\mu(\tau-s)$.  The transition
probability obtained by integrating $R_{\rm LS}$ from $\tau_0$ to a
given time is indeed logarithmically divergent because of the last term. 
In the limit $\tau_0\to -\infty$ they find
\begin{eqnarray}
\lim_{\tau_0\to -\infty}
R_{\rm LS} & & =  |q|^2\left\{
-\frac{\Delta E}{2\pi}\theta(-\Delta E)\right. 
\nonumber \\
   & & \left. +  \frac{1}{2\pi^2}\int_0^{\infty}ds\,\cos(s\Delta E)
\left[ \frac{1}{s^2}-\frac{1}{(\Delta x)^2}\right]\right\},
\nonumber \\
\end{eqnarray}
where $\theta(x)$ is the Heaviside function.  Louko and Satz have used this
formula to compute the excitation rate for a trajectory which is
inertial for $\tau\to-\infty$ and uniformly accelerated with
acceleration $a$ for $\tau \to +\infty$,
verifying that it vanishes as $\tau\to -\infty$ and
converges to the rate (\ref{RPlanck}) as $\tau \to
+\infty$.\footnote{Note that $R_{\rm LS}$ and its $\tau_0\to -\infty$
limit can become negative at some $\tau$.  
This may be due to the fact that quantum interference prevents one from
determining the exact time when the detector clicks.} Very 
recent discussions on the excitation of Unruh-DeWitt detectors 
with arbitrary trajectories can be found in \textcite{Obadiaetal07} and
\textcite{Satz07}.  We also note that \textcite{DeBievreetal06} have
shown that a uniformly accelerated 
Unruh-DeWitt detector will asymptotically 
have the Gibbs state with the Unruh temperature irrespective of its
initial state.

\subsubsection{Circularly moving detectors with constant velocity
 in the Minkowski vacuum}
\label{subsubsection:circular}

As we have seen, the Unruh effect is concerned with {\em uniformly 
accelerated observers}. In spite of this, some interesting 
questions can be addressed for the case of observers in uniform 
circular motion. 

Let us start by considering a circularly moving Unruh-DeWitt 
detector\footnote{For other stationary world lines, see 
\textcite{Letaw81} in conjunction with \textcite{Letawetal82}, 
and a recent work by \textcite{Korsbakkenetal04}.} 
\cite{Letawetal80}
in Minkowski spacetime\footnote{The symbols $r$ and $\theta$ correspond
to the usual polar coordinates.} at the radius $r=r_0$ with angular velocity 
$\Omega \equiv d\theta/dt > 0$. Using the interaction action~(\ref{SI1}), we may 
write the detector excitation amplitude as
\begin{eqnarray}
^{\rm exc}\!\! {\cal A}_{\bf k} 
 & = &  i \langle E |\otimes \!\langle{\bf k}\,{}_{\rm M}
      | \hat S_I | 0_{\rm M} \rangle \otimes|E_0 \rangle
\nonumber \\
& = &  i q 
    \int_{-\infty}^{\infty} d\tau 
    \exp (i \Delta E \,\tau)
 \!\langle{\bf k}\,{}_{\rm M}
      |  \hat \Phi [x^\mu(\tau) ] | 0_{\rm M} \rangle,
\nonumber \\
\label{UDgeneralA}
\end{eqnarray}
where $\tau$ is the detector proper time. 
Thus, the proper excitation rate~(\ref{RMinertial}) can 
be given as \cite{Broutetal95}
\begin{equation}
^{\rm exc}\! R^{m=0}_{\rm circ} =
    |q|^2 \int_{-\infty}^{\infty} d\sigma 
    \exp (-i \Delta E \,\sigma) \, 
    G^+ \left[ x(\tau), x(\tau') \right],
\label{RMgeneral}
\end{equation}
where $\sigma \equiv \tau - \tau'$. Here 
\begin{equation}
    G^+ \left[ x^\mu (\tau), x^\mu (\tau') \right] = 
    \!\langle 0_{\rm M} 
      |  \hat \Phi [x^\mu(\tau) ]
      \hat \Phi [x^\mu(\tau') ] | 0_{\rm M} \rangle 
\label{G+general}
\end{equation}
is the (positive-frequency) Wightman
function [see,  e.g., \textcite{Fullingbook89}] for the massless scalar
field in Minkowski spacetime.
In Cartesian coordinates this is written as 
\begin{eqnarray}
G^+ \left[ x^\mu (\tau), x^\mu (\tau') \right] 
&=& 
-1/(    4\pi^2[(t-t'-i\epsilon)^2-(x-x')^2
\nonumber \\
& & 
- (y-y')^2-(z-z')^2]),
\label{G+Cart}
\end{eqnarray}
which can be derived using the field
expansion~(\ref{Phimassive})-(\ref{positivefrequencymodePhimassive}) 
in terms of positive- and negative-energy modes with 
respect to inertial observers, as is well known.
We may write the world line of this detector as
\begin{equation}
    t = \gamma \tau  ,\,
    x = r_0 \cos (\Omega \gamma \tau)  ,\,
    y = r_0 \sin (\Omega \gamma \tau)  ,\,
    z = {\rm const},
\label{DWLCaC}
\end{equation}
where we impose the condition $r_0 \Omega < 1$ so that the world line is
timelike and 
$\gamma= (1-r_0^2\Omega^2)^{-1/2}$ is the Lorentz factor.
The proper acceleration of such a detector is 
$a \equiv {\sqrt {-a_{\mu}a^{\mu}}} = 
\Omega^2 \gamma^2 r_0$.
Then, the proper excitation rate can be written as 
\begin{eqnarray}
 && ^{\rm exc}\! R^{m=0}_{\rm circ} =
    \frac{|q|^2}{4\pi^2} \int_{-\infty}^{\infty} d\sigma \,
    e^{-i \Delta E \,\sigma}
\nonumber \\
 && \times  [ -\gamma^2(\sigma-i\epsilon)^2
              +4 r_0^2 \sin^2(\Omega \gamma \sigma/2) ]^{-1}.
\label{RMcirc}
\end{eqnarray}
This non-vanishing excitation rate has been 
evaluated numerically by \textcite{Letawetal80} and \textcite{Letaw81}
[see also \textcite{Kimetal87} for some further discussion].
For ultra-relativistic detectors ($\gamma\gg1$), 
one obtains 
\begin{equation}
^{\rm exc}\! R^{m=0}_{\rm circ} \approx
    \frac{}{} 
    \frac{|q|^2 a e^{-\sqrt{12} \Delta E/a}}
    {4 \pi \sqrt{12}}.
\label{RMcircultrarel}
\end{equation}
An attempt to give a physical interpretation of this formula
in terms of the depolarization of electrons in particle accelerators
will be discussed in Sec.~\ref{subsection:polarization}. 

Now, one could think of recalculating the excitation rate above
from the point of view of observers corotating with the detector.
A major difficulty appears, however.
In order to extract a natural particle content from the field 
theory, a global timelike Killing vector field $K$ 
associated with the rotating observers would be necessary 
[see, e.g., \textcite{WaldQFTCS}]. If such a Killing vector existed, then
the eigenvalue equation $
i K \phi^{\pm} = \pm \omega \phi^{\pm}
$
would separate positive-frequency modes ($\phi^+$) from
negative-frequency ones ($\phi^-$) 
(where $K$ is assumed to be future directed). 
The four-velocity of a circularly moving observer 
at $r=r_0$ with $\Omega = d\theta/dt =  {\rm const}$  can be written
as 
$$
u = d/d\tau = \gamma K,
$$
where
\begin{equation}
    K = ({\partial}/{\partial t}) + 
    \Omega \, ({\partial}/{\partial \theta})
\label{rot vector}
\end{equation}
is the associated Killing field. 
We notice now that for $r\Omega> 1\,$, $K$ is spacelike.
Thus, $K$ fails to be a {\em global} timelike Killing field.
If one ignored this fact and used $K$ to extract naively 
the particle content of the field theory, circularly moving 
observers would end up with identifying their vacuum state 
with the Minkowski vacuum itself \cite{Letawetal80}. 
This would lead to a puzzling situation since we know from
Eq.~(\ref{RMcirc}) that detectors carried by circularly moving 
observers do have a nonzero excitation rate.
Thus, either we have a suitable way to extract
the particle content from the theory \cite{Ashtekaretal75,Kay78} or 
it may be better not to introduce such a concept at 
all.\footnote{We recall that a detector acts as a ``vacuum fluctuometer" 
and that its response must not depend on the definition of the
particle \cite{Groveetal83}.}

In contrast to the case considered above, we now turn to a 
related but quite distinct physical situation, where the detector 
response {\em can} be naturally interpreted in terms of the 
particle content defined by the rotating observers. We consider the 
rotating detector with angular velocity $\Omega = {\rm const}$ 
confined inside a limiting surface \cite{Levinetal93, Daviesetal96}.
We assume a cylindrical surface at $r=\rho$ with $\rho < 1/\Omega $
and Dirichlet boundary conditions imposed on the scalar field 
$
\phi (t, r=\rho, \theta, z)= 0.
$

The positive-frequency orthonormal modes 
with respect to inertial observers are 
\begin{equation}
    u_{m n k_z} = C_{mn} 
    J_m\left( {\alpha_{mn} r}/{\rho} \right) 
    e^{im \theta} e^{ik_z z} 
    e^{-i\omega_{mn} t}.
\label{Cyl Dir modes}
\end{equation}
Here $m \in \mathbb{Z}$, $n \in \mathbb{N}_+$, 
$\alpha_{mn}$ is the $n$-th (non-vanishing) zero 
of the Bessel function
$J_m (x)$ ($J_m(\alpha_{mn})=0$), 
and the following dispersion relation is satisfied:
\begin{equation}
    \omega_{mn} = \sqrt{\alpha_{mn}^2/\rho^2+k_z^2}
    \,\,> 0.
\label{DR}
\end{equation}
The normalization constant
\begin{equation}
    C_{mn} = ( 2\pi \rho \sqrt{\omega_{mn}}\,
    |J_{m+1}(\alpha_{mn})|)^{-1}
\label{Cmn}
\end{equation}
has been chosen so that the modes $u_{mnk_z}$ satisfy the orthonormality
condition with respect to the Klein-Gordon inner product:
\begin{equation}
(u_{mnk_z},u_{m'n'k_z'})_{\rm KG}
= \delta_{mm'}\delta_{nn'}\delta(k_z-k_z').
\end{equation}
[See Eq.~(\ref{sec2:KGprod}) for the definition of the Klein-Gordon
inner product.]
The corresponding Wightman function (\ref{G+general}) is
\begin{eqnarray}
    G^+ \left( x^\mu, {x'}^\mu \right)\, & = & 
    \sum_{m=-\infty}^{+\infty} \sum_{n=1}^{+\infty} 
    \int_{-\infty}^{+\infty} dk_z C_{mn}^2 
    J_m ( {\alpha_{mn}r}/{\rho} ) 
    \nonumber \\
& &
    \times 
    J_m ( {\alpha_{mn}r'}/{\rho} ) \nonumber \\
&& \times
    e^{im(\theta-\theta')} e^{ik_z(z-z')} 
    e^{-i\omega_{mn}(t-t')}.
\label{G+Cyl}
\end{eqnarray}
In order to calculate the response rate~(\ref{RMgeneral})
we substitute the world line of the rotating detector
\begin{equation}
    t = \gamma \tau    ,\,
    r = r_0 = {\rm const}    ,\,
    \theta = \Omega t  ,\,
    z = {\rm const} 
\label{DWLCyC}
\end{equation}
in Eq.~(\ref{G+Cyl}), obtaining
\begin{eqnarray}
    G^+ \left( x^\mu, {x'}^\mu \right)\, &=& 
    \sum_{m=-\infty}^{+\infty} \sum_{n=1}^{+\infty} 
    \int_{-\infty}^{+\infty} dk_z C_{mn}^2 
    J_m^2 ( {\alpha_{mn} r_0}/{\rho} )
\nonumber \\
    &  &\times 
    e^{-i(\omega_{mn}-m\Omega)\gamma\sigma}.
\label{G+CylRD}
\end{eqnarray}
By substituting Eq. (\ref{G+CylRD}) into Eq. (\ref{RMgeneral}), 
we get
\begin{eqnarray}
    ^{\rm exc}\! R^{m=0}_{\rm circ}\, &=& |q|^2 
    \sum_{m=-\infty}^{+\infty} \sum_{n=1}^{+\infty} 
    \int_{-\infty}^{+\infty} dk_z C_{mn}^2 
    J_m^2 ( {\alpha_{mn}r_0}/{\rho} ) 
\nonumber \\
    &  &\times 
    \int_{-\infty}^{+\infty} d\sigma
    e^{-i[\Delta E +(\omega_{mn}-m\Omega)\gamma]\sigma}.
\label{RMCylDBC}
\end{eqnarray}
The result of the integration over $\sigma$ is proportional to
$ 
   \delta(\Delta E - (m\Omega -\omega_{m n})\gamma).
$
Assuming $\Omega >0$, we find that no contribution comes
from $m \leq 0$ in the sum of Eq.~(\ref{RMCylDBC})  (where we recall that 
$\Delta E >0$). Now, for $m>0$ there will be a lowest value of
$\omega_{mn}$  for each $m$, namely $\alpha_{m1}/\rho$ 
[corresponding to the first (non-vanishing) zero
of the Bessel function $J_m (x)$ and $k_z=0$]. Then, a necessary condition 
for a mode with a given $m$ to contribute in the
sum is that  
$
    \Omega \rho > \alpha_{m1}/m
$.
However, since $\alpha_{mn}>m$ [see, e.g., \textcite{Abramowitzbook}],
there is no integer value for $m$ that satisfies 
this condition because of our original constraint
that $\rho < 1/\Omega$. We conclude, thus, that the detector has 
a {\em vanishing response} when it is confined inside the limiting 
surface. [The same conclusion would hold if we had chosen Neumann 
rather than Dirichlet boundary conditions \cite{Daviesetal96}].

Now, let us show that in this case, namely for $\rho < 1/\Omega $,
it is possible to interpret the vanishing response in terms of 
the particle content defined by the rotating observers confined 
inside the boundary with angular velocity $\Omega$. This is so 
because in this case the Killing vector field $K$
associated with these observers is globally timelike. Let us 
rewrite Eq.~(\ref{Cyl Dir modes}) as 
\begin{equation}
    \tilde{u}_{mnk_z} = C_{mn}
    J_m\left( {\alpha_{mn}r}/{\rho} \right) 
    e^{im \theta'} 
    e^{ik_zz}
    e^{-i\tilde{\omega}_{mn} t}
\label{Cyl Dir rot modes}
\end{equation}
with $\tilde{\omega}_{mn} = \omega_{mn}-m\Omega >0$,
which are also positive-frequency modes with respect to
the corotating observers. We have defined 
$\theta' \equiv \theta-\Omega t$,
and $t$ can be interpreted here as the proper time 
of a rotating observer with angular velocity $\Omega$ lying at
$r=0$, i.e.~$K = \partial/\partial t|_{\theta'={\rm const}}$.
Clearly, by determining the Bogolubov transformation 
among the ``inertial" modes (\ref{Cyl Dir modes}) and 
``rotating" modes (\ref{Cyl Dir rot modes}) [see Eq.~(\ref{sec2:exp2})],
we obtain 
\begin{equation}
    \beta_{(i)(i')}= (\tilde{u}^*_{(i)}, u_{(i')})_{\rm KG} = 0,
\label{Bogolubov}
\end{equation}
where $(i)$ stands for the set $(m, n, k_z)\,$.
We conclude, thus, that there is no mixing between positive- and 
negative-energy modes between the two sets
[see, e.g., \textcite{Birrelletal82}]. 
As a result, the Minkowski vacuum coincides with the vacuum state 
defined by the rotating observers, who would correctly 
conclude that the response rate of the corotating detectors 
confined inside the limiting surface vanishes.
A similar analysis can be performed for the case of a compact space, 
like the one with topology $S^2\times\mathbb{R}^2$, 
or the Einstein static universe, wherein the field is automatically confined 
\cite{Daviesetal96}.\footnote{Other investigations on the response of particle 
detectors in spacetimes with non-trivial topology and endowed with 
boundaries can be found in \textcite{Copelandetal84}, \textcite{Abe90},
\textcite{Langlois06,Langlois05} and \textcite{Daviesetal89}.}

\subsection{Weak decay of non-inertial protons}
\label{subsection:decay}

As a second example, we discuss the weak decay of non-inertial protons. 
Although inertial protons are stable according to the standard 
particle model \cite{PDG}, non-inertial protons are not. This is so because 
the accelerating agent provides the required extra energy for 
the proton to decay. To the best of our knowledge, the first 
to consider the weak decay of accelerated protons were 
\textcite{Ginzburgetal64}, who described the baryons by classical 
currents while treating the other particles as quantized fields. 
At about the same 
time, \textcite{Zharkov65} investigated the weak and strong proton decays 
(and other processes) in the presence of a background electromagnetic field 
$A_\mu$ by using the formalism of 
\textcite{Nikishovetal64,Nikishovetal64b} 
[see also \textcite{Ritus69}], treating all particles as quantum fields.
More recently the weak decay of non-inertial protons under 
the influence of 
a gravitational field was studied by \textcite{Mueller97}, 
\textcite{Vanzellaetal00} and \textcite{Fregolenteetal06}.

Here we review a model of 
the weak decay of uniformly accelerated protons 
from the point of view of inertial and Rindler observers. For the sake 
of simplicity we present a model with 
a two-dimensional spacetime and massless 
neutrinos (using four-component spinors for the 
leptons)~\cite{Vanzellaetal01, Matsasetal99}, but a four-dimensional 
comprehensive calculation with massive neutrinos can be found in 
\textcite{Suzukietal03}. We evaluate the proton proper decay rate with 
respect to inertial and Rindler observers  and show that the results 
obtained are in agreement when the Unruh effect is taken into account 
in spite of the fact that uniformly accelerated protons are static 
according to Rindler observers. It will be interesting to see that 
what inertial observers interpret as being
$$
(i)\; p^+ \to n^0 e^+ \nu,
$$
are interpreted by Rindler observers as being the combination
of the following channels
$$
(ii)\;  p^+ e^- \to  n^0 \nu,\;
(iii)\; p^+ \bar\nu \to  n^0 e^+,\;
(iv)\;  p^+ e^- \bar\nu \to  n^0,
$$
where the $e^-$'s and $\bar \nu$'s on
the left-hand side are Rindler electrons 
and antineutrinos, respectively, absorbed from the Unruh thermal bath.
In our procedure, we take into account the proton-neutron mass difference
by introducing a {\em semiclassical} rather than {\em classical} current.
The current is ``classical" in the sense that the proton-neutron is associated 
with a well defined world line and ``quantum" in the sense that it behaves 
as a two-level quantum system.

The trajectory of a proton with proper acceleration 
$a = {\rm const}$ along the $z$-axis in Minkowski spacetime
is given in Cartesian coordinates by $ z = \sqrt{t^2 + a^{-2}} $. 
This can be written more simply as $\rho = a^{-1} = {\rm const}$, where
we are introducing a new set of Rindler coordinates 
$(\rho,\eta)$ which are related with $(t,z)$ by 
\begin{equation}
t = \rho \sinh \eta \;, \;\; z = \rho \cosh \eta,
\label{RCp}
\end{equation} 
and $0<\rho<+\infty$, $-\infty<\eta<+\infty$ 
[see Eq.~(\ref{sec2:Rindcoord})]. Thus, we describe our uniformly 
accelerated proton through the 
vector current 
\begin{equation}
j^\mu = q u^\mu \delta (\rho-a^{-1}),
\label{C}
\end{equation}
where $q$ will be associated with a small coupling constant and
$u^\mu$ is the nucleon's ``two-velocity":
$u^\mu = (a,0)$  and $u^\mu = (\sqrt{a^2 t^2 + 1}, at)$ 
in Rindler and  Minkowski coordinates, respectively.
 
The current~(\ref{C}) is suitable for describing stable accelerated 
protons but must be improved to allow proton-decay processes. For this 
purpose, we  consider the nucleon  as a two-level system. 
In this model, neutrons $|n \rangle$ and protons $|p  \rangle$ 
are going to be seen as excited and unexcited states of the nucleon, 
respectively, and are assumed to be eigenstates of the nucleon 
Hamiltonian $\hat H$:  
\begin{equation}
\hat H |n \rangle = m_n |n \rangle,\;\;
\hat H |p  \rangle = m_{p } |p  \rangle,
\end{equation}
where $m_n$ and $m_{p }$ are  the neutron and proton masses, 
respectively. Accordingly, to consider nucleon decay processes, 
we replace $q$ in Eq.~(\ref{C}) by the Hermitian monopole 
\begin{equation}
\hat q(\tau )\equiv e^{i\hat H \tau} \hat q_0 e^{-i\hat H \tau},
\label{Q}
\end{equation}
where $|\langle m_p | \hat q_0 | m_n \rangle |\equiv G_F  $, which is
dimensionless,
plays the role of an effective Fermi constant.
As a result, the current (\ref{C}) will be replaced by 
\begin{equation}
\hat j^\mu = \hat q(\tau) u^\mu \delta (\rho-a^{-1}) .
\label{CI}
\end{equation}

\subsubsection{Inertial observer perspective}
\label{subsubsection:DnipIop}

Let us first analyze the weak-decay process~$(i)$ 
of uniformly accelerated protons in the inertial frame. 
We shall describe electrons and neutrinos as fermionic fields:  
\begin{equation}
\hat \Psi(t,z) \!=\! \sum_{\sigma = \pm } \int_{-\infty}^{\infty} dk
\left( \hat b_{k \sigma} \psi^{(+\omega)}_{k \sigma} (t,z)
     + \hat d^\dagger_{k \sigma} \psi^{(-\omega)}_{-k -\sigma} (t,z) 
\right),
\label{FF}
\end{equation}
where $ \hat b_{k \sigma} $ and $ \hat d^\dagger_{k \sigma} $ 
are annihilation and creation operators of fermions
and antifermions, respectively, with momentum $k$ and
polarization $\sigma$. In the inertial frame, the frequency,
momentum and mass $m$ are related in the usual way as
$\omega=\sqrt{k^2+m^2}>0$, and
$ \psi^{(+\omega)}_{k \sigma} $ and $ \psi^{(-\omega)}_{k \sigma} $
are positive- and negative-frequency solutions of the Dirac equation
$i\gamma^\mu \partial_\mu \psi^{(\pm \omega)}_{k \sigma} 
 - m \psi^{(\pm \omega)}_{k \sigma} =0$.
By using the $\gamma^\mu$ matrices in the 
Dirac representation [see, e.g., \textcite{IZbook}], we find
\begin{equation}
\psi^{(\pm \omega)}_{k +} (t,z) =
 \frac{e^{i(\mp \omega t + kz)}}{\sqrt{2\pi}}
\left(
\begin{array}{c}
\pm \sqrt{(\omega \pm m)/2\omega} \\
0\\
k/\sqrt{2\omega(\omega \pm m)}\\
0
\end{array}
\right)  
\label{NM1}
\end{equation}
and
\begin{equation}
\psi^{(\pm \omega)}_{k -} (t,z) = 
\frac{e^{i(\mp \omega t + kz)}}{\sqrt{2\pi}}
\left(
\begin{array}{c}
0\\
\pm \sqrt{(\omega \pm m)/2\omega} \\
0\\
-k/\sqrt{2\omega(\omega \pm m)}
\end{array}
\right).
\label{NM2}
\end{equation}
In order to keep a unified procedure for inertial and
accelerated frame calculations, 
we have orthonormalized the modes (\ref{NM1})-(\ref{NM2})
according to the same inner-product 
definition that will be used in Sec.~\ref{subsubsection:DnipRop}: 
\begin{eqnarray}
( 
\psi^{(\pm \omega)}_{k \sigma} , \psi^{(\pm \omega')}_{k' \sigma'} 
)  
&\equiv &
\int_\Sigma d\Sigma_\mu 
\bar \psi^{(\pm \omega)}_{k \sigma} \gamma^\mu 
\psi^{(\pm \omega')}_{k' \sigma'}
\nonumber \\
&=&
\delta(k-k') \delta_{\sigma \sigma'} 
\delta_{\pm \omega \; \pm \omega'},
\label{IP}
\end{eqnarray}
where $\bar \psi \equiv \psi^\dagger \gamma^0$.
(In this section, we have chosen $t= {\rm const}$ for 
the hypersurface $\Sigma$.) Then, the  canonical 
anticommutation relations for fields and conjugate momenta 
lead to the following simple anticommutation relations for 
creation and annihilation operators:
\begin{equation}
\{\hat b_{k \sigma},\hat b^\dagger_{k' \sigma'}\}=
\{\hat d_{k \sigma},\hat d^\dagger_{k' \sigma'}\}=
\delta(k-k') \; \delta_{\sigma \sigma'},
\label{ACR}
\end{equation}
\begin{equation}
\{\hat b_{k \sigma},\hat b_{k' \sigma'}\}\!=\!
\{\hat d_{k \sigma},\hat d_{k' \sigma'}\}\!=\!
\{\hat b_{k \sigma},\hat d_{k' \sigma'}\}\!=\!
\{\hat b_{k \sigma},\hat d^\dagger_{k' \sigma'}\}\!=\!
0.
\end{equation}

Next, we model the relevant weak interaction by 
coupling the electron and neutrino fields, 
$\hat \Psi_e$ and $\hat \Psi_\nu$,
minimally to the nucleon current (\ref{CI}) 
using the parity-conserving Fermi action
\begin{equation}
\hat S_I = \int d^2x \sqrt{-g} \hat j_\mu 
           (\hat{\bar \Psi}_\nu \gamma^\mu \hat \Psi_e +
            \hat{\bar \Psi}_e \gamma^\mu \hat \Psi_\nu ),
\label{S}
\end{equation}
where $g$ is the determinant of the spacetime metric
components $g_{\mu \nu}$. Note that the second term inside 
the parentheses on the right-hand side of Eq.~(\ref{S}) 
does not contribute to 
the process~$(i)$. The vacuum transition amplitude is given by
\begin{equation}
{\cal A}^{p  \to n}_{(i)} =
\; \langle  n \vert \otimes \!\langle e^+_{k_e \sigma_e} , 
\nu_{k_\nu \sigma_\nu}\,{}_{\rm M} \vert \;
\hat S_I \;
\vert 0_{\rm M} \rangle \otimes \vert p  \rangle.
\label{AMP}
\end{equation}
By using the current (\ref{CI}) in Eq.~(\ref{S}) and recalling that 
$\hat S_I  $ acts 
also on the nucleon states in Eq.~(\ref{AMP}), we obtain
\begin{eqnarray}
{\cal A}^{p  \to n}_{(i)} 
&=&
G_F \int_{-\infty}^{\infty} dt \int_{-\infty}^{\infty} dz 
\frac{ e^{i  \Delta m  \tau} 
\delta(z-\sqrt{t^2+a^{-2}})}{az}
\nonumber \\
&\times&
{u_{\mu}} \!\langle e^+_{k_e \sigma_e} , \nu_{k_\nu \sigma_\nu}\,{}_{\rm M} \vert\;
\hat{\bar \Psi}_\nu \gamma^\mu \hat \Psi_e\;
\vert 0_{\rm M} \rangle,
\label{AMPI}
\end{eqnarray}
where 
$
\Delta m \equiv m_n - m_{p }
$,  
$
\tau = a^{-1} \sinh^{-1} (at)
$
is the proton-neutron proper time and we recall that in Minkowski 
coordinates the two-velocity is $u^\mu = (\sqrt{a^2 t^2 + 1} , at)$
[see below Eq.~(\ref{C})]. 
By using the fermionic field 
(\ref{FF}) in Eq.~(\ref{AMPI}) and carrying out the integral over $z$,
we obtain  
\begin{widetext}
\begin{eqnarray}
{\cal A}^{p  \to n}_{(i)} 
& = &
\frac{- \; (G_F /4\pi )\;\; \delta_{\sigma_e,-\sigma_\nu}}
      { \sqrt{\omega_\nu \omega_e 
      (\omega_\nu + m_\nu) (\omega_e - m_e)}}
\int_{-\infty}^{\infty} d\tau 
e^{i( \Delta m \tau 
        + a^{-1} (\omega_e + \omega_\nu) \sinh a\tau 
        - a^{-1} (k_e+k_\nu)             \cosh a\tau)
   }
\nonumber
\\
&\times &
\{ [(\omega_\nu + m_\nu) (\omega_e - m_e) + k_\nu k_e] \cosh a\tau
-  [(\omega_\nu + m_\nu) k_e + (\omega_e - m_e) k_\nu] \sinh a\tau
\}.
\nonumber
\end{eqnarray}
Thus, the  differential transition rate  
$
{d^2 {\cal P}^{p  \to n}_{\rm in}}/{dk_e \; dk_\nu} =
\sum_{\sigma_e=\pm} \sum_{\sigma_\nu=\pm} 
| {\cal A}_{(i)}^{p \to n} |^2
$
calculated in the inertial frame is
\begin{eqnarray}
\frac{d^2 {\cal P}^{p  \to n}_{\rm in}}{dk_e \; dk_\nu}
& = &
\frac{G_F^2}{4 \pi^2 \omega_\nu \omega_e} 
\int_{-\infty}^{\infty} d s  
\int_{-\infty}^{\infty} d \xi \;
e^{ i (  
        \Delta m \xi + 2 a^{-1} \sinh(a \xi/2) 
        [ 
        (\omega_\nu + \omega_e) \cosh as 
        -(k_\nu + k_e) \sinh as 
        ]
       ) 
  }
\nonumber
\\
& &\times
[(\omega_\nu \omega_e + k_\nu k_e) \cosh 2as -
 (\omega_e k_\nu + \omega_\nu k_e) \sinh 2as -
  m_\nu m_e \cosh a\xi].
\label{TS}
\end{eqnarray}
\end{widetext}
Next, by defining 
$$
k_{e(\nu)} \to k'_{e(\nu)}= 
- \omega_{e(\nu)} \sinh (as)+k_{e(\nu)} \cosh (as),
$$
we are able to perform the integral in the $s$ variable, and
the differential transition rate (\ref{TS}) can be cast 
in the form
\begin{eqnarray}
\frac{1}{T} 
\frac{d^2 {\cal P}^{p \to n}_{\rm in}}{d k'_e  d k'_\nu} 
\!\!\!\!\!\!
&& \!=\!
\frac{G_F^2}{4 \pi^2 {\omega}'_e {\omega}'_\nu} 
\int_{-\infty}^{\infty} \!\!\! d\xi 
e^{ i \Delta m \xi + i 2 a^{-1} 
   ({\omega}'_e + {\omega}'_\nu) \sinh (a\xi/2) } 
\nonumber
\\      
&& \times   
     ({\omega}'_\nu {\omega}'_e  + k'_\nu k'_e 
       - m_\nu m_e\cosh a\xi),
\label{DP2}
\end{eqnarray}
where $ T \equiv \int_{-\infty}^{\infty} ds $ is the total  proper 
time and 
$ 
{\omega}'_{e(\nu)}\equiv \sqrt{{k'}^2_{e(\nu)} + m^2_{e(\nu)}}
$. 

The total transition rate 
$
{\Gamma}^{p  \to n}_{\rm in}={\cal P}^{p \to n}_{\rm in}/T
$
is obtained after integrating Eq.~(\ref{DP2}) over both momentum variables. 
For this purpose it is useful to make the change of variables
$
k'_{e(\nu)} \to {{\tilde{k}}}_{e(\nu)} \equiv k'_{e(\nu)}/a
$
and
$
\xi \to \lambda \equiv e^{a \xi /2}.
$
(Note that ${{\tilde{k}}}_{e(\nu)}$ is dimensionless.)
Thus, we obtain
\begin{eqnarray}
 {\Gamma}^{p  \to n}_{\rm in} 
& = & 
\frac{G_F^2 a}{2 \pi^2}
\int_{-\infty}^{\infty} \frac{d {\tilde{k}}_e }{\tilde\omega_e}
\int_{-\infty}^{\infty} \frac{d {\tilde{k}}_\nu }{\tilde\omega_\nu}
\int_{0}^{\infty} \frac{d\lambda }{\lambda^{1-2i\Delta m/a}}  
\nonumber
\\      
& \times &  
[
\tilde\omega_\nu  \tilde\omega_e  + {\tilde{k}}_\nu {\tilde{k}}_e 
- m_\nu m_e (\lambda^2+\lambda^{-2})/(2 a^2) 
] 
\nonumber
\\ 
& \times &  
\exp[i (\tilde\omega_e + \tilde\omega_\nu) (\lambda-\lambda^{-1})] 
\label{RI}
\end{eqnarray}
with 
$\tilde\omega_{e(\nu)}\equiv ({\tilde{k}}^2_{e(\nu)}+m^2_{e(\nu)}/a^2)^{1/2}$.
Let us assume at this point $m_\nu \to 0$. In this case, using (3.871.3-4) 
in \textcite{Gradshteynbook}, we perform the integration over $\lambda$ and obtain
the following final expression for the proton decay rate:
\begin{eqnarray}
{\Gamma}^{p  \to n}_{\rm in} 
&  = & 
\frac{G_F^2 \tilde m_e a}{2 \pi^{3/2} e^{\pi \widetilde{\Delta m} } }
\nonumber \\
& \times &
G_{1\;3}^{3\;0} 
\left( \tilde m_e^2 \left|
\begin{array}{l}
\;\;\;1\\ 
-{1}/{2}, {1}/{2}+i \widetilde{\Delta m}, {1}/{2}-i 
\widetilde{\Delta m} \!
\end{array}
\right.
\right),
\nonumber \\
\label{RIF}
\end{eqnarray}
where 
$G_{p\; q}^{m n}$ is the Meijer function \cite{Gradshteynbook},
$ \widetilde{\Delta m} \equiv \Delta m /a$ and
$\tilde m_e \equiv m_e/a$.
In Fig.~\ref{proton} the proton mean proper lifetime 
$\tau_p = 1/\Gamma_{\rm in}^{p \to n}$ in this model is plotted. 

\subsubsection{Rindler observer perspective}
\label{subsubsection:DnipRop}

In order to re-analyze the proton decay from the point of view of
Rindler observers, it is useful to review  the 
quantization of the fermionic field in the Rindler wedge
\cite{Candelasetal78,Soffeletal80,Jareguietal91,Bautista93}. 

The line element of the Rindler wedge in terms of the Rindler 
coordinates $(\rho,\eta)$ given in Eq.~(\ref{RCp}) is written as 
[see Eq.~(\ref{sec2:RindlerRL})]
\begin{equation}
ds^2 = \rho^2 d\eta^2 - d\rho^2  .
\label{LEp}
\end{equation}
Now, the Dirac equation in a general spacetime covered 
with arbitrary coordinates is written as 
$ (i\gamma_R^\mu \tilde \nabla_\mu -m) \psi_{\omega \sigma}=0$,
where
$\gamma_R^\mu \equiv (e_\alpha)^\mu \gamma^\alpha$
are the Dirac matrices in curved spacetime,
$\tilde \nabla_\mu \equiv \partial_\mu +\Gamma_\mu $
and 
$ \Gamma_\mu
= 
\frac{1}{8} [\gamma^\alpha,\gamma^\beta] (e_\alpha)^\lambda 
\nabla_\mu (e_\beta)_\lambda
$
are the 
Fock-Kondratenko connections. (The $\gamma^\mu$ are
the usual flat-spacetime Dirac matrices.) In the
Rindler wedge the relevant  tetrads are
$ (e_0)^{\mu}= \rho^{-1} \delta^\mu_0$, 
$ (e_i)^\mu = \delta^\mu_i$. 
As a consequence, the Dirac equation takes the form
\begin{equation}
(i {\partial }/{\partial \eta})\psi_{\omega \sigma}=
( 
\gamma^0 m\rho - {i \alpha_3}/{2} - 
i \rho \alpha_3 {\partial}/{\partial \rho}
)
\psi_{\omega \sigma},
\label{DEQ}
\end{equation}
where 
$\alpha_i \equiv \gamma^0 \gamma^i$. 
\begin{figure}
\begin{center}
\mbox{\epsfig{file=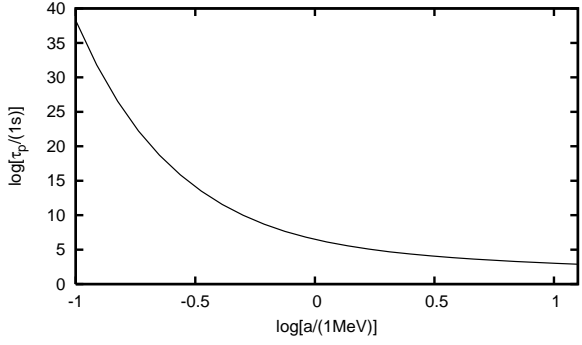,width=0.45 \textwidth,angle=0}}
\end{center}
\caption{ 
The proton mean proper lifetime is plotted 
as a function of the proper acceleration $a$
for $G_F = 10^{-12}$. (This value of $G_F$
approximately reproduces the inertial neutron proper 
lifetime of $900$ seconds in this model.) 
Note that $\tau_p \to +\infty$ for inertial protons ($a \to 0$). 
}
\label{proton}
\end{figure}

We shall express the fermionic field as
\begin{equation}
\hat \Psi(\eta,\rho)\!= \! \sum_{\sigma = \pm } 
\int_{0}^{\infty} \! d\omega
\left( \hat b_{\omega \sigma} \psi_{\omega \sigma}(\eta,\rho)
       + \hat d^\dagger_{\omega \sigma} \psi_{-\omega -\sigma}(\eta,\rho) 
\right), 
\label{PSA}
\end{equation}
where $\psi_{\omega \sigma}=f_{\omega \sigma}(\rho) e^{-i \omega \eta/a}$
are positive-energy solutions for $\omega >0$  and negative-energy
solutions for $\omega <0$ with respect to the boost Killing field 
$\partial / \partial \eta$ with polarization $\sigma = \pm$.
{}From Eq.~(\ref{DEQ}) we obtain
\begin{equation}
\hat H_\rho f_{\omega \sigma} = \omega f_{\omega \sigma},
\label{ENA}
\end{equation}
where
$
\hat H_\rho \equiv  a 
( m \rho \gamma^0 - {i \alpha_3}/{2} - 
i \rho \alpha_3 {\partial}/{\partial \rho})
$.
Now, by ``squaring'' Eq.~(\ref{ENA}) and defining two-component 
spinors $\chi_{j}$ ($j=1,2$) through 
\begin{equation}
 f_{\omega \sigma}(\rho) \equiv 
\left(
\begin{array}{c}
\chi_1 (\rho) \\
\chi_2 (\rho) 
\end{array}
\right),
\label{(4.5)}
\end{equation}
we obtain 
\begin{equation}
\left( 
\rho \frac{d}{d\rho} \rho\frac{d}{d\rho} 
\right) \chi_1 
 = 
\left( 
m^2 \rho^2 + \frac{1}{4} - \frac{\omega^2}{a^2} 
\right) \chi_1
-\frac{i \omega}{a}\sigma_3 \chi_2,
\label{XIa}
\end{equation}
\begin{equation}
\left( 
\rho \frac{d}{d\rho} \rho\frac{d}{d\rho} 
\right)  \chi_2
 = 
\left( 
m^2 \rho^2 + \frac{1}{4} - \frac{\omega^2}{a^2} 
\right) \chi_2
-\frac{i \omega}{a}\sigma_3 \chi_1  .
\label{XIb}
\end{equation} 
Next, we introduce the definition  $\phi^\pm \equiv \chi_1 \mp \chi_2$
and define $\xi^\pm $ and $\zeta^\pm $ through 
\begin{equation}
\phi^\pm \equiv 
\left(
\begin{array}{c}
\xi^\pm (\rho) \\
\zeta^\pm (\rho) 
\end{array}
\right) .
\label{(4.7)}
\end{equation}
In terms of these variables Eqs.~(\ref{XIa})-(\ref{XIb}) become 
\begin{equation}
\left( \rho\frac{d}{d\rho}\rho\frac{d}{d\rho} \right) \xi^\pm 
= 
(m^2\rho^2+(i\omega/a \pm 1/2)^2) \xi^\pm,
\label{(4.8)}
\end{equation}
\begin{equation}
\left( \rho\frac{d}{d\rho}\rho\frac{d}{d\rho} \right) \zeta^\pm 
= 
(m^2\rho^2+(i\omega/a \mp 1/2)^2) \zeta^\pm.
\label{(4.82)}
\end{equation}
The solutions of these differential equations can be written
in terms of Hankel functions 
$ H^{(j)}_{i\omega/a \pm 1/2}(im\rho)$, $j=1,2$,  
or modified Bessel functions 
$ 
K_{i \omega/a \pm 1/2}(m\rho)
$, 
$
I_{i \omega/a \pm 1/2}(m\rho)
$.
Hence, by using Eqs.~(\ref{(4.5)}) and (\ref{(4.7)}), 
and requiring that the solutions  satisfy the 
first-order equations~(\ref{ENA}), we obtain 
$$
f_{\omega +} (\rho) = 
A_+ 
\left(
\begin{array}{c}
   K_{i \omega/a + 1/2}(m\rho) +
   i K_{i \omega/a - 1/2}(m\rho) \\
0 \\
   K_{i \omega/a - 1/2}(m\rho) -
   i K_{i \omega/a + 1/2}(m\rho) \\
0
\end{array}
\right),
$$
$$
f_{\omega -} (\rho) =  
A_- 
\left(
\begin{array}{c}
0\\
   K_{i \omega/a + 1/2}(m\rho) +
   i K_{i \omega/a - 1/2}(m\rho) \\
0\\
   K_{i \omega/a + 1/2}(m\rho) -
   i K_{i \omega/a - 1/2}(m\rho)
\end{array}
\right).
$$
Note that solutions involving $I_{i \omega/a \pm 1/2}$ turn out to be 
non-normalizable and thus must be discarded.
In order to find the normalization constants 
\begin{equation}
A_+=A_-= [{m \cosh (\pi \omega/a)}/{(2\pi^2 a)}]^{1/2},
\label{CONST}
\end{equation} 
we  have used \cite{Birrelletal82} 
\begin{equation}
(
\psi_{\omega \sigma} , \psi_{\omega' \sigma'} 
) 
\equiv 
\int_\Sigma d\Sigma_\mu 
\bar \psi_{\omega \sigma} \gamma^\mu_R 
\psi_{\omega' \sigma'}
=
\delta(\omega-\omega') \delta_{\sigma \sigma'},
\label{IP2}
\end{equation}
[see also Eq.~(\ref{IP})], where 
$\bar \psi \equiv \psi^\dagger \gamma^0$ 
and  $\Sigma$ is set to be the line $\eta={\rm const}$.
Thus, the normal modes of the fermionic field~(\ref{PSA}) are 
\begin{eqnarray}
\psi_{\omega +} 
&=&  
[{m \cosh (\pi \omega/a)}/{(2 \pi^2 a)}]^{1/2}
e^{-i\omega \eta/a}
\nonumber \\
&\times& 
\left(
\begin{array}{c}
K_{i \omega/a + 1/2}(m\rho) +
i K_{i \omega/a - 1/2}(m\rho) \\
0\\
-K_{i \omega/a + 1/2}(m\rho) +
i K_{i \omega/a - 1/2}(m\rho) \\
0
\end{array}
\right)
\nonumber \\
\label{RNM1}
\end{eqnarray}
and
\begin{eqnarray}
\psi_{\omega -}
&=&
[{m \cosh (\pi \omega/a)}/{(2 \pi^2 a)}]^{1/2}
e^{-i\omega \eta/a}
\nonumber \\
&\times & 
\left(
\begin{array}{c}
0\\
K_{i \omega/a + 1/2}(m\rho) +
i K_{i \omega/a - 1/2}(m\rho) \\
0\\
K_{i \omega/a + 1/2}(m\rho) -
i K_{i \omega/a - 1/2}(m\rho) 
\end{array}
\right).
\nonumber \\
\label{RNM2}
\end{eqnarray}
As a consequence, the 
canonical anticommutation relations for the fields and
conjugate momenta imply that the
 annihilation and creation operators satisfy
the following anticommutation relations:
\begin{equation}
\{\hat b_{\omega \sigma},\hat b^\dagger_{\omega' \sigma'}\}=
\{\hat d_{\omega \sigma},\hat d^\dagger_{\omega' \sigma'}\}=
\delta(\omega-\omega') \; \delta_{\sigma \sigma'},
\label{ACR2}
\end{equation}
\begin{equation}
\{\hat b_{\omega \sigma},\hat b_{\omega' \sigma'}\} \!\! = \!\!
\{\hat d_{\omega \sigma},\hat d_{\omega' \sigma'}\} \!\! = \!\!
\{\hat b_{\omega \sigma},\hat d_{\omega' \sigma'}\} \!\! = \!\!
\{\hat b_{\omega \sigma},\hat d^\dagger_{\omega' \sigma'}\} \!\! = \!
0.
\end{equation} 

Now, we are in the position to turn our attention to the inverse 
$\beta$-decay of accelerated protons from the point of view of 
Rindler observers. In particular, the mean proper 
lifetime must be the same as the one obtained 
in Sec.~\ref{subsubsection:DnipIop}, 
but the corresponding particle interpretation changes 
significantly. As will
be shown, the proton decay, which is represented in the inertial frame 
in terms of Minkowski particles by process~$(i)$,
will be represented in the uniformly accelerated frame
as the combination of the 
processes~$(ii)$, $(iii)$ and $(iv)$ in terms of Rindler particles 
[see above Eq.~(\ref{RCp})].
These processes are characterized by the
conversion of protons to neutrons due to the absorption 
of $ e^- $ and/or $\bar\nu$ and emission of $\nu$, $e^+$ or no particle, 
from and to the Unruh thermal bath.  
Note that process $(i)$ in terms of Rindler 
particles is forbidden because the proton is static in the Rindler frame.

Let us calculate first the transition amplitude for 
process $(ii)$:
\begin{equation}
{\cal A}^{p  \to n}_{(ii)} = 
\; \langle  n \vert \otimes 
\langle \nu_{\omega_\nu \sigma_\nu}\,{}_{\rm R} \vert \;
\hat S_I \;
\vert e^-_{\omega_{e^-} \sigma_{e^-}}\,{}_{\rm R} 
\rangle \otimes \vert p  \rangle,
\label{ACA}
\end{equation}
where $\hat S_I$ is given by Eq.~(\ref{S}) with $\gamma^\mu$ replaced by 
$\gamma^\mu_R$ and our current is given by
Eq.~(\ref{CI}). 
Thus, we obtain [we recall that in Rindler coordinates $u^\mu=(a,0)$]
\begin{eqnarray}
{\cal A}^{p  \to n}_{(ii)} 
& = &
\frac{G_F}{a} \int_{-\infty}^{\infty} d\eta \exp (i \Delta m \eta/a) 
\nonumber \\
& \times &
\langle \nu_{\omega_\nu \sigma_\nu \,{}{\rm R}} \vert
\hat\Psi^\dagger_\nu  (\eta,a^{-1}) \hat \Psi_e (\eta,a^{-1})
\vert e^-_{\omega_{e^-} \sigma_{e^-}}\,{}_{\rm R} \rangle,
\nonumber \\
\label{V}
\end{eqnarray}
where we note that the second term in the 
parentheses of Eq.~(\ref{S}) does not contribute.
Next, by using Eq.~(\ref{PSA}), we obtain
\begin{eqnarray}
{\cal A}^{p  \to n}_{(ii)} 
&=&
\frac{G_F }{a}
\int_{-\infty}^{\infty} d\eta  \exp (i \Delta m \eta/a)
\nonumber \\
&\times&
\delta_{\sigma_{e^-},\sigma_\nu}
\psi^\dagger_{\omega_\nu\sigma_\nu} (\eta,a^{-1}) 
\;
\psi_{\omega_{e^-}\sigma_{e^-}}(\eta,a^{-1}).
\nonumber \\
\label{V1}
\end{eqnarray}
Using now Eq.~(\ref{RNM1}) and Eq.~(\ref{RNM2}) and 
performing the integral, we obtain
\begin{eqnarray}
{\cal A}^{p  \to n}_{(ii)} & = & 
\frac{4G_F}{\pi a}\sqrt{m_em_\nu \cosh (\pi
\omega_{e^-}/a)\cosh (\pi \omega_\nu/a)} \nonumber \\
& \times & 
Re[ K_{i\omega_\nu/a-1/2}(m_\nu /a) K_{i\omega_{e^-}/a +1/2} (m_e /a)]
\nonumber \\
& \times & 
\delta_{\sigma_{e^-}, \sigma_\nu} 
\delta( \omega_{e^-}-\omega_\nu-\Delta m) .
\label{V2}
\end{eqnarray}
The corresponding differential transition rate
per absorbed and emitted particle energies 
is given by
\begin{equation}
\frac{1}{T}\frac{d^2 {\cal P}^{p \to n}_{(ii)}}
{d\omega_{e^-}  d\omega_\nu} 
=\!\!
\frac{1}{T} \!\!\! \sum_{\sigma_{e^-}=\pm} \sum_{\sigma_{\nu}=\pm}
\!\!\! |{\cal A}^{p  \to n}_{(ii)}|^2
n_F(\omega_{e^-}) [1-n_F(\omega_\nu)],
\label{AP1}
\end{equation}
where 
$
n_F(\omega) \equiv 1/(1+e^{2\pi \omega/a})
$
is the fermionic thermal factor associated with the Unruh thermal bath
and $T=2\pi\delta (0)$ is the total nucleon proper time.
By using Eq.~(\ref{V2}) in Eq.~(\ref{AP1}), we obtain 
\begin{eqnarray}
& &  
\frac{1}{T} \frac{d^2 {\cal P}^{p \to n}_{(ii)}}{d\omega_{e^-}  
d\omega_\nu} 
\!\! = \!\!
\frac{4G_F^2 m_e m_\nu}{\pi^3 a^2}  e^{-\pi\Delta m/a} 
\delta (\omega_{e^-}-\omega_\nu-\Delta m)
\nonumber
\\
&& \times 
\{
Re [ K_{i\omega_\nu/a -1/2} (m_\nu/a) K_{i\omega_{e^-}/a+1/2}(m_e/a)] 
\}^2.
\nonumber \\
\label{AP4}
\end{eqnarray}
By integrating Eq.~(\ref{AP4}) over $\omega_{\nu}$, we obtain the 
following transition rate associated with process (ii):
\begin{eqnarray}
&& 
\Gamma_{(ii)}^{p\to n} =  
\frac{4G_F^2m_e m_\nu }{\pi^3 a^2 e^{\pi\Delta m/a}}  
\int_{\Delta m}^{\infty} d\omega_{e^-}
\nonumber \\
&& \times
\{ 
Re[K_{i(\omega_{e^-}-\Delta m)/a -1/2} (m_\nu/a)
   K_{i\omega_{e^-}+1/2}(m_e/a) ] 
\}^2.
\nonumber
\end{eqnarray}
We recall that Rindler frequencies may take 
arbitrary positive real values (see 
Sec.~\ref{subsubsection:E<mc2}).
Analogous calculations lead to the following transition rates for 
processes~$(iii)$ and~$(iv)$: 
\begin{eqnarray}
&&
\Gamma_{(iii)}^{p\to n}=  
\frac{4G_F^2m_e m_\nu }{\pi^3 a^2 e^{\pi\Delta m/a}}  
\int_{0}^{\infty} d\omega_{e^+}
\nonumber \\
&&
\times 
\{Re
[ K_{i(\omega_{e^+}+\Delta m)/a +1/2} (m_\nu/a)
  K_{i\omega_{e^+}-1/2}(m_e/a)
] 
\}^2,
\nonumber
\\
&&
\Gamma_{(iv)}^{p\to n}= 
\frac{4G_F^2m_e m_\nu }{\pi^3 a^2e^{\pi\Delta m/a}}
\int_{0}^{\Delta m} d\omega_{e^-}
\nonumber \\
&& 
\times 
\{ 
Re[K_{i(\omega_{e^-}-\Delta m)/a -1/2} (m_\nu/a)
   K_{i\omega_{e^-}+1/2}(m_e/a) ] 
\}^2.
\nonumber
\end{eqnarray}
\begin{figure}
\begin{center}
\mbox{\epsfig{file=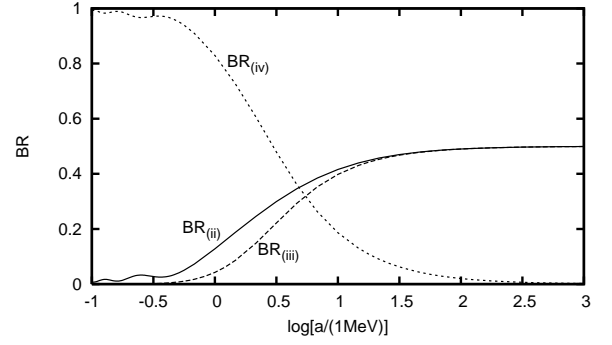,width=0.45 \textwidth,angle=0}}
\end{center}
\caption{Branching ratios 
$BR_{(ii)}$, $BR_{(iii)}$ and $BR_{(iv)}$ are plotted. 
Process $(iv)$ dominates over processes 
$(ii)$ and $(iii)$ for small accelerations, while  
processes $(ii)$ and $(iii)$ dominate over  process $(iv)$
for high accelerations.
}
\label{BR}
\end{figure}
The proton decay rate is given by adding up all contributions: 
$
\Gamma_{\rm acc}^{p\to n}= 
 \Gamma_{(ii)}^{p\to n} 
+\Gamma_{(iii)}^{p\to n}
+\Gamma_{(iv)}^{p\to n},
$
namely
\begin{eqnarray}
&&
\Gamma_{\rm acc}^{p\to n} 
= \frac{4G_F^2m_e m_\nu \exp (-\pi\Delta m/a)}{\pi^3 a^2 }  
\int_{-\infty}^{\infty} d\omega 
\nonumber \\
&& 
\times 
\{ 
Re 
  [K_{i(\omega-\Delta m)/a -1/2} (m_\nu/a)
   K_{i\omega/a+1/2}(m_e/a)] 
\}^2.
\nonumber
\end{eqnarray}
It is interesting to note that although 
transition rates  have fairly distinct interpretations in the inertial
and accelerated frames, mean proper lifetimes are scalars and must be the same
in both frames. {\em Indeed, by taking the limit
$m_\nu \to 0$ and plotting
$\tau_p(a)=1/\Gamma_{\rm acc}^{p\to n}$
as a function of acceleration, we do reproduce} 
Fig.~\ref{proton}.\footnote{At first, the decay rates calculated from the
point of view of inertial and uniformly accelerated observers
were shown to be equal only numerically and the equality was 
limited by the machine precision \cite{Vanzellaetal01}. The precise 
analytic equivalence was derived soon afterward
by \textcite{Suzukietal03}.}
In Fig.~\ref{BR} we plot the branching ratios 
$$
BR_{(ii)} \equiv\frac {\Gamma_{(ii)}^{p\to n}}{\Gamma_{\rm acc}^{p\to n}},\;\;
BR_{(iii)} \equiv \frac{\Gamma_{(iii)}^{p\to n}}{\Gamma_{\rm acc}^{p\to n}},\;\;
BR_{(iv)} \equiv \frac{\Gamma_{(iv)}^{p\to n}}{\Gamma_{\rm acc}^{p\to n}} .
$$
We note that for small accelerations where ``few'' high-energy particles
are available in the Unruh thermal bath, process $(iv)$ dominates over  
processes $(ii)$ and $(iii)$, while for high accelerations processes 
$(ii)$ and $(iii)$ dominate over  process $(iv)$. 
This is a interesting 
example of how inertial and Rindler observers may differ 
in the a phenomenon description, although they must agree on the output 
of the experiments associated with scalar observables.

\subsection{Bremsstrahlung}
\label{subsection:Bremsstrahlung}

In our next example we use the Unruh effect to discuss how the
bremsstrahlung from a uniformly accelerated charge is described in the
Rindler frame,
addressing also the
celebrated question whether or not uniformly accelerated electric
charges radiate with respect to coaccelerated 
observers.\footnote{For a series of recent papers which 
include a critical historical description on hyperbolically 
moving charges in the context of classical electrodynamics see
\textcite{Eriksenetal00a,Eriksenetal00b,Eriksenetal00c,Eriksenetal02,
Eriksenetal04}.}  
It will turn out that the Rindler photons with \emph{``zero energy"},
which are characterized by their transverse momenta, play a central role.
Our discussion closely follows \textcite{Higuchietal92,Higuchietal92b}
and does not assume that the reader is familiar with quantization of the
electromagnetic field.

A point charge uniformly accelerated along the $z$-axis in the Cartesian
coordinate system can be represented in the Rindler
coordinates~(\ref{sec2:rightcoords})
by $\xi = x = y = 0$. The corresponding
conserved current ($\nabla_{\mu} j^{\mu}=0$) is, then, given 
by 
\begin{equation}
j^\tau = q \delta (\xi ) \delta (x) \delta (y), 
\; \; j^\xi = j^x = j^y = 0.
\label{3}
\end{equation}
Let us analyze the emission of photons with fixed transverse 
momentum ${\bf k}_\bot = (k_{x}, k_{y})$. The fact that ${\bf k}_\bot$ 
is invariant under boosts in the $z$-direction allows us to directly 
compare the emission and absorption rates corresponding to Minkowski 
and Rindler photons with the same transverse momentum.  

We shall quantize the electromagnetic field defined by the Lagrangian 
density 
\begin{equation}
{\cal L}=-\sqrt {-g} [ ({1}/{4}) F_{\mu \nu} F^{\mu \nu}
+ (2\alpha)^{-1} (\nabla^\mu A_\mu)^2 ]
\label{8}
\end{equation}
with the corresponding field equations  in the Feynman gauge
$(\alpha = 1)$ being
\begin{equation}
\nabla^\mu F_{\mu\nu} 
+ \nabla_\nu(\nabla^\mu A_\mu) = \nabla_\mu \nabla^\mu A_\nu = 0,
\label{9}
\end{equation}
and calculate the response rate of the charge with respect to both
inertial and Rindler observers. 

\subsubsection{Inertial observer perspective}
\label{subsubsection:iopbremsstrahlung}

According to inertial observers, 
we write the quantized electromagnetic field 
$\hat A_\mu (x) $ as 
\begin{equation}
\hat A_\mu (x) = \int \frac{d^3 {\bf k}}{2 (2 \pi)^3 k_0} 
\sum_{\lambda = 0}^3 \left[ a^{(\lambda )} ({\bf k}) 
\epsilon_\mu^{(\lambda )} ({\bf k}) e^{-i k_\nu x^\nu } +{\rm H.c.} \right]
\label{45.5}
\end{equation}
with $k_0  \equiv \sqrt{k_z^2 + k_\perp^2}$, 
where $\lambda$ labels the mode polarization. We shall 
adopt here the  notation used in  \textcite{IZbook}.

We assign $\lambda$ the value $0$ for what we call
the non-physical modes, $1$ or $2$ for the physical
modes and $3$ for the pure-gauge modes.
The pure-gauge modes are those which can be written as
$A_{\mu}^{({\rm 3},{\bf k} )} = \nabla_{\mu} \Phi$ for some  
scalar field $\Phi (x)$ and satisfies the Lorenz condition 
\begin{equation}
\nabla_\mu A^\mu = 0. 
\label{11}
\end{equation}
The physical modes satisfy 
the Lorenz condition and are not pure gauge.
Finally the non-physical modes do not satisfy
the Lorenz condition.
Accordingly, we choose the polarization vectors 
$\epsilon_\mu^{(\lambda )}$ as
\begin{eqnarray}
& \epsilon^{(0)\mu} & = (-1 , 0, 0, 1)/\sqrt {2},
\label{47} \\
& \epsilon^{(1)\mu} & = (0 , 1 , 0, 0),
\label{48} \\
& \epsilon^{(2)\mu} & = (0 , 0, 1, 0),
\label{49} \\
& \epsilon^{(3)\mu} & = (1 , 0, 0, 1)/\sqrt{2},
\label{50}
\end{eqnarray}
in the Cartesian frame chosen such that
$k^{\mu}=(\vert {\bf k}\vert, 0,0,\vert {\bf k}\vert )$ (where
the first component is the time component).

The amplitude of emission of a photon with momentum ${\bf k}$ and
polarization $\lambda$ by the accelerated charge in the Minkowski
vacuum is
\begin{eqnarray}
{\cal A}^{(\lambda ,{\bf k} )} &=&
\; \!\langle  {\bf k}, \lambda\,{}_{\rm M} \vert 
i \int d^4x j^\mu (x) \hat A_\mu (x)
\vert 0_{\rm M} \rangle
\nonumber \\ 
&=& i \int d^4x j^\mu (x)
\epsilon_{\mu}^{(\lambda)}({\bf k})
e^{i(\omega t-{\bf k}\cdot{\bf x})}.
\label{46}
\end{eqnarray}
The Cartesian components of the current~(\ref{3}) can be 
written as
\begin{equation}
j^{\mu} = q a \,(z,0,0,t)\, 
\delta (\xi) \delta (x) \delta (y),
\label{44}
\end{equation}
where 
$
\delta (\xi ) = \delta [ z -  (t^2 + a^{-2})^{1/2} ] / (a z).
$

Next, we can express the total probability of emission of photons 
with fixed transverse momentum ${\bf k}_\bot$, divided by the 
total proper time $T = 2 \pi \delta (0)$ of the accelerated 
charge during which the interaction remains turned on, as 
\begin{equation}
^{\rm in}\!R^{\rm tot}_{k_\bot} = \sum_{\lambda = 1}^2
\int_{-\infty}^{\infty} d\tilde k_z \vert {\cal A}^{(\lambda ,{\bf k})} 
\vert^2 /T,
\label{51}
\end{equation}
where $d\tilde k_z \equiv 
dk_z / [( 2 \pi )^3 2k_0]$, and the sum runs only over the physical
polarizations $\lambda = 1, 2$. Using Eq.~(\ref{46}) in 
Eq.~(\ref{51}), one has
\begin{eqnarray}
^{\rm in}\!R_{k_\bot}^{\rm tot} 
& = &
\int_{-\infty}^{\infty}d\tilde{k}_z 
\int d^{4}x \int d^{4}x'
e^{i \omega(t-t')-i{\bf k}\cdot({\bf x}-{\bf x'})}
\nonumber \\
& & \times 
j^{\mu}(x)j^{\nu}(x')
\sum_{\lambda=1}^{2}
\epsilon_{\mu}^{(\lambda)} ({\bf k})
\epsilon_{\nu}^{(\lambda)} ({\bf k}).
\label{51.5}
\end{eqnarray}
Now, we note the identity
\begin{equation}
\sum_{\lambda = 1}^2 \epsilon _\mu^{(\lambda)} \epsilon _\nu^{(\lambda )}
= - \epsilon _\mu^{(0)} \epsilon _\nu^{(3)}
- \epsilon _\mu^{(3)} \epsilon _\nu^{(0)} - \eta _{\mu \nu},
\label{52}
\end{equation}
where $\eta_{\mu\nu}$ is the metric of Minkowski spacetime.
Because of the 
current conservation $\partial_{\mu}j^{\mu} = 0$, and due to
the fact that $\epsilon_{\mu}^{(3)}$ is proportional to $k_{\mu}$, the
first two terms in Eq.~(\ref{52}) do not contribute when one substitutes it
in Eq.~(\ref{51.5}).  Hence, as is well known, 
\begin{eqnarray}
^{\rm in}\!R^{\rm tot}_{k_\bot} 
& = &
-\frac{1}{T} \int d\tilde k_z \int d^4 x \int d^4 x'
j^{\mu}(x)j_{\mu}(x')
\nonumber \\
& & \times 
\exp [i\omega(t-t')-i {\bf k} \cdot ({\bf x} - {\bf x}')] .
\label{53}
\end{eqnarray}
Next, by substituting the current~(\ref{44}) in this formula, 
we obtain  
\begin{eqnarray}
^{\rm in}\!R^{\rm tot}_{k_\bot} &  =  & -\frac{q^2}{T} 
\int_{-\infty}^{\infty} \!\! d\tau '  \!\!
\int_{-\infty}^{\infty} \!\! d\tau '' 
\cosh a (\tau ' -\tau '' ) \! \!
\nonumber \\
& & \times  
\int_{-\infty}^{\infty} \!\! d\tilde k_z 
\exp [- (i k_z/a) (\cosh a\tau ' - \cosh a\tau '' )] 
\nonumber \\
& & \times  
\exp [  (i k_0/a) (\sinh a\tau ' - \sinh a\tau '' )],
\nonumber
\end{eqnarray}
where we have made the coordinate transformation 
$t=a^{-1} \sinh a\tau $. Now, it is necessary again
to factor out the total proper time 
$T=\int_{-\infty}^{\infty}d\tau$, where $\tau=(\tau'+\tau'')/2$. 
To this end, we use the momentum transformation~(\ref{k->k'})
with $m=0$. Then, we find 
\begin{eqnarray}
^{\rm in}\!R_{k_\bot}^{\rm tot} 
& = & 
- q^{2}\int_{-\infty}^{\infty}
d\tilde k'_z \int_{-\infty}^{\infty}d\sigma\cosh a\sigma
\nonumber \\ 
& & \times
\exp\left[\frac{2i k'_{0}}{a}\sinh\frac{a\sigma}{2}\right],
\label{54.3}
\end{eqnarray}
where $d\tilde{k}'_z \equiv dk'_z /[(2\pi)^{3}2k'_0]$
and $\sigma \equiv \tau'-\tau''$.  To evaluate this integral we cut off
the contribution from large $\vert\sigma\vert$ smoothly by letting
$\sigma \rightarrow \sigma + 2i\epsilon$ (where $\epsilon$ is an 
infinitesimal positive number) in the exponent, and taking the limit
$\epsilon\rightarrow 0$ in the end. (Otherwise this integral would
be indefinite.) Then, by introducing another change of variables as
$
s_{\pm} = [( k'_0 + k'_z )/ k_{\perp}] e^{\pm a\sigma/2}
$,
and using the formula \cite{Gradshteynbook}
\begin{equation}
\int_{0}^{\infty}dx\,x^{\nu-1}\exp\left[
\frac{i\mu}{2}\left( x - \frac{\beta^{2}}{x}\right)\right] 
= 2\beta^{\nu}e^{i\nu\pi/2}K_{\nu}(\beta\mu),
\end{equation}
where Im$\;\mu> 0$, Im($\beta^2\mu) < 0$, we obtain
\begin{equation}
^{\rm in}\!R^{\rm tot}_{k_\bot} d^2{\bf k}_\bot =
\frac{q^2}{4\pi ^3 a} \vert K_1(k_{\perp }/a) \vert^2  d^2{\bf k}_\bot. 
\label{55}
\end{equation}
This is the total emission rate of Minkowski particles associated with a 
uniformly accelerated charge. A similar discussion as the one commented 
on in Sec.~\ref{subsubsection:controversy1} concerning whether or not 
uniformly accelerated electric charges radiate can be traced back up to 
about half a century ago 
[see \textcite{Fultonetal60} and references therein]. We recall
that the radiation reaction force on a uniformly accelerated electric charge 
vanishes. The classical radiation reaction force is known to have some unusual 
features \cite{Barutbook80} but it was recently shown to be in agreement with 
quantum field theory \cite{Krivitskii91, Higuchi02, Higuchietal04, 
Higuchietal05}. Clearly, no problems arise when one deals with physical
situations where electric charges are accelerated for a finite time interval 
\cite{Jacksonbook99}. Accordingly, Eq.~(\ref{55}) should be seen as 
approximating the one  obtained when an electric charge is uniformly 
accelerated for long enough.

\subsubsection{Rindler observer perspective}

Now, we shall evaluate the response rate associated with the 
current~(\ref{3})
according to Rindler observers by considering the Unruh thermal bath. 
The response will consist 
of emission and absorption of photons to and from the Unruh thermal
bath. It is clear that the rate of spontaneous emission is zero
because the current~(\ref{3}) is static. However, it does not
imply that the rates of induced emission and absorption vanish as well. 
This is
because these rates are proportional to the number of photons
present in the thermal bath which couple to the current~(\ref{3}). 
Since the number of {\em zero-energy} (Rindler) photons in the 
(Unruh) thermal bath, which are the relevant ones in this case, 
is infinite, the rates of induced emission and absorption are 
{\em indefinite}. Hence one needs to {\em regularize} the current~(\ref{3})
to make both its strength of coupling to the field and the relevant photon
number finite. (The regulator is removed in the end.) 

Let us discuss our regularization procedure in two steps.  First
we modify Eq.~(\ref{3}) by considering a charge
oscillating with frequency $E$
\begin{equation}
j^\tau = \sqrt 2 q \cos E \tau \delta (\xi ) \delta (x) \delta (y), \; 
j^\xi = j^x = j^y = 0,
\label{4}
\end{equation}
and take the limit $E \to 0$ in the end \cite{Kolbenstvedt88}. The 
factor $\sqrt 2$ 
appears because the radiation rate, in first order of perturbation,
is proportional to the square of the charge. When the oscillation is
slow,  i.e.~when $E \ll a, k_{\perp}$, 
the charge is expected to interact with
the field as if it were a constant charge at each $\tau$.  (We assume
continuity of the rate in the limit $E\to 0$.) Hence, 
the $\tau$-average of the square of the charge must be set equal to
$q^2$ and, therefore, the factor $\sqrt{2}$ is necessary. 

Now, the current~(\ref{4}) 
does not satisfy electromagnetic charge conservation. For this reason
we replace this
current by an oscillating dipole arrangement with a charge at $\xi=0$
and the other one at infinity (which is omitted here because it does
not affect the final result) described by
\begin{eqnarray}
j^\tau & = & \sqrt 2 q \cos (E\tau)\delta (\xi )\delta^2({\bf x}_\bot),
\label{5}\\
j^\xi & = & 
\sqrt 2 q E \sin ( E\tau) e^{-2a\xi}
\theta (\xi )\delta^2({\bf x}_\bot),
\label{6}\\
j^x& = & j^y=0 .
\label{7}
\end{eqnarray}
The current $j^\xi$ flowing to $\xi =\infty$ 
will not contribute to the final 
results. Its only importance is to keep the condition 
$\nabla^\mu j_\mu = 0$
valid and make the computation gauge independent even 
before taking the limit
$E\to 0$.

Next, we analyze the interaction of the source~(\ref{5})--(\ref{7})
with the Maxwell field in the Rindler wedge.
For this purpose we need to expand the electromagnetic field 
by the positive- and negative-frequency modes 
defined with respect to the Rindler 
time $\tau$. We again deal with the Lagrangian 
density for the electromagnetic field in the Feynman gauge 
given by Eq.~(\ref{8}) with $\alpha=1$,
and the field equations in the Feynman gauge given by Eq.~(\ref{9})
considered now in the Rindler wedge.
The presence of
$\partial_\tau$, $\partial_x$ and
$\partial_y$ as Killing fields makes it sufficient to look for solutions of
Eq.~(\ref{9}) of the form
\begin{equation}
A_\mu^{(\lambda, \omega, {\bf k}_\bot)} (x ) = 
f_\mu^{(\lambda , \omega, k_\bot)} (\xi ) 
e^{i({\bf k}_\bot\cdot{\bf x}_\bot - \omega \tau )} .
\label{10r}
\end{equation}
Then, 
we expand the electromagnetic quantum field in terms of annihilation
and creation operators as
\begin{eqnarray}
\hat A_\mu (x ) 
& = & 
\int d^2{\bf k}_\bot
\int_{0}^{\infty} d\omega
\nonumber \\ 
&& \times 
\sum_{\lambda = 0}^{3}
\left\{ \hat{a}_{(\lambda , \omega,{\bf k}_\bot)} 
       A_\mu^{(\lambda, \omega,{\bf k}_\bot)}
       (x) + {\rm H.c.} \right\},
\nonumber \\
\label{12}
\end{eqnarray}
where $A^{(\lambda ,\omega ,{\bf k}_\bot)}_\mu (x)$ are solutions
of the form given in Eq.~(\ref{10r}). These modes are conveniently
expressed in terms of the solutions of the scalar field equation
$\Box \phi = 0 $
[see \textcite{Candelasetal77}]. 
For each set of quantum numbers the solution,
which does not diverge as $\xi \to +\infty$, is 
obtained by letting $m=0$ in Eq.~(\ref{sec2:normalizedvR}):
\begin{eqnarray}
\phi^{(\omega , {\bf k}_\bot )} & = & 
\left[\frac{\sinh(\pi\omega/a)}{4\pi^4 a}\right]^{1/2}
K_{i\omega/ a} 
[ ({k_\perp}/{a}) e^{a\xi } ]\nonumber \\
&& \times e^{i{\bf k}_\bot\cdot{\bf x}_\bot - i \omega \tau}.
\label{16.5}
\end{eqnarray}

One can choose a set of independent normal modes as 
\begin{eqnarray}
A_\mu^{(0, \omega,{\bf k}_\bot)} & = & 
C^{(0, \omega, {\bf k}_\bot )} ( 0, 0, k_x \phi , k_y \phi ),
\label{16}\\
A_\mu^{(1, \omega,{\bf k}_\bot)}  & = &
C^{(1, \omega,{\bf k}_\bot)} (0 ,0, k_y \phi, -k_x \phi),
\label{13} \\
A_\mu^{(2, \omega,{\bf k}_\bot)}  & = & 
C^{(2, \omega, {\bf k}_\bot )} (\partial_\xi \phi , -i\omega \phi, 0, 0),
\label{14} \\
A_\mu^{(3, \omega,{\bf k}_\bot)}  & = & 
C^{(3, \omega, {\bf k}_\bot )} (-i\omega \phi , 
\partial _\xi \phi, i k_x \phi, i k_y \phi ),\nonumber \\
\label{15}
\end{eqnarray}
where $A_{\mu} = (A_{\tau}, A_{\xi}, A_x, A_y)$, 
$C^{(\lambda,\omega,{\bf k}_\bot)}$ are
normalization constants, and $\phi \equiv \phi^{(\omega ,{\bf k}_\bot)}$.
The modes $A_\mu^{(0,\omega,{\bf k}_\bot)}$ are the non-physical modes
because $\nabla^\mu A_\mu^{(0,\omega,{\bf k}_\bot)} \neq 0$.  It can
readily be shown that the modes 
$A_\mu^{(\lambda,\omega,{\bf k}_\bot)}$ with $\lambda = 1,2$ satisfy the
Lorenz condition $\nabla^\mu A_\mu^{(\lambda,\omega,{\bf k}_\bot)} = 0$.
Thus, these are the physical modes.  The modes
$A_\mu^{(3,\omega,{\bf k}_\bot)} \propto \nabla_\mu\phi^{(\omega,{\bf
k}_\bot)}$ are the pure-gauge modes.

The normalization constants $C^{(i)}$ can be determined from the canonical
commutation relations of the fields by 
requiring suitable commutation relations for the operators $a_{(i)}$
and $a_{(i)}^\dagger$. [Here the label $(i)$ represents 
$(\lambda, \omega,{\bf k}_\bot)$.] 
In this context, it is convenient to 
introduce the generalized Klein-Gordon inner product  
\begin{equation}
(A^{(i)}, A^{(j)})_{\rm KG} \equiv 
\int_{\Sigma} d\Sigma_\mu W^\mu [ A^{(i)}, A^{(j)} ]
\label{17}
\end{equation}
between any two modes
$A^{(i)}_\mu$ and $A^{(j)}_\mu$, where
the integration is performed on some Cauchy surface
$\Sigma$ for the Rindler wedge, e.g., any hypersurface $\tau =$ const,
and where
\begin{equation}
W^\mu [ A^{(i)}, A^{(j)} ] \equiv
\frac {i}{\sqrt{-g}} 
( A^{(i) *}_{\nu} \pi^{(j)\mu \nu } - 
A^{(j)}_{\nu} \pi^{(i) \mu \nu *})
\label{18}
\end{equation}
with $\pi^{(i)\mu \nu } \equiv 
\partial {\cal L} / \partial_\mu A_\nu \vert_{A_\mu = A_{(i) \mu }} $. 
The $\pi^{(i)\mu \nu}$ are calculated in the Feynman gauge to be
\begin{equation}
\pi^{(i)\mu \nu} = \sqrt{- g} [ 
\nabla^\nu A^{(i) \mu} - \nabla^\mu A^{(i) \nu} - g^{\mu \nu}
\nabla_\alpha A^{(i) \alpha} ] .
\label{18.5}
\end{equation}
It can be seen [see, e.g., \textcite{Friedman78}] 
that the field equations ensure conservation
of the current~(\ref{18}), and thus the inner product~(\ref{17}) is 
independent of the choice of the Cauchy surface $\Sigma$.

{}From 
the canonical commutation relations one finds
\begin{equation}
[ (A^{(i)} , \hat A )_{\rm KG}, (\hat A , A^{(j)})_{\rm KG} ] =
(A^{(i)}, A^{(j)})_{\rm KG}.
\label{19.3}
\end{equation}
[See Eq.~(\ref{sec2:KGfg}).]
This equation and Eq.~(\ref{12}) imply that 
\begin{equation}
(A^{(i)} , A^{(l)} )_{\rm KG} [ \hat{a}_{(l)} , \hat{a}_{(l')}^\dagger ]
( A^{(l')} , A^{(j)})_{\rm KG} =
(A^{(i)}, A^{(j)})_{\rm KG} ,
\label{19.4}
\end{equation}
where we have used the fact that positive- and negative-frequency
modes can be shown to be orthogonal to each other. 
The schematic summation over
 $l$ represents integrations over $\omega$ and ${\bf k}_\bot$
as well as the summation over $\lambda$. Next, define the matrix
$
M^{(i)(j)} \equiv (A^{(i)} , A^{(j)} )_{\rm KG}
$.
Then,  Eq.~(\ref{19.4}) implies [see. e.g., \textcite{Higuchi89}]
\begin{equation} 
[ \hat{a}_{(i)}, \hat{a}_{(j)}^{\dagger} ] = (M^{-1})_{(i)(j)} ,
\label{21}
\end{equation}
where $(M^{-1})_{(i)(j)}$ is defined by
\begin{equation}
(M^{-1})_{(i)(l)} M^{(l) (j)} = \delta^{\lambda \lambda'}
\delta (\omega - \omega') \delta^2({\bf k}_\bot -{\bf k}'_\bot)
\label{21.2}
\end{equation}
with $(i) = (\lambda,\omega,{\bf k}_\bot)$ and 
$(j)=(\lambda',\omega',{\bf k}_\bot')$.

Now, by using the inner product~(\ref{17}) for the normal 
modes~(\ref{16})--(\ref{15}),
we can verify the following orthogonality properties:
\begin{equation}
( A^{(\lambda ,\omega ,{\bf k}_\bot )} , 
A^{(\lambda ',\omega ',{\bf k}_\bot' )} )_{\rm KG} = 0,\ \ 
\lambda = 1,2,\ \lambda'=0,3.
\label{19}
\end{equation}
In other words, the physical modes are orthogonal 
to the {\em pure gauge} mode $\lambda = 3$ and to the {\em Lorenz condition
violating} non-physical mode $\lambda = 0$ and to each other. 
Hence, it is sufficient to know the restriction of the matrix
$M^{(i)(j)}$ to the physical subspace (i.e.~to $\lambda = 1,2$) in order to
derive the commutators among the physical annihilation and creation operators
according to Eq.~(\ref{21}).  Thus, 
by requiring the commutators of annihilation and creation operators associated
with the physical modes (i.e.~with $\lambda$ and $\lambda'$ being $1$ or
$2$) to be
\begin{equation}
[ \hat{a}_{(\lambda ,\omega,{\bf k}_\bot )},
\hat{a}_{(\lambda ' ,\omega ',{\bf k}_\bot' )}^{\dagger} ] =
\delta _{\lambda \lambda '}
\delta (\omega -\omega ')
\delta^2({\bf k}_\bot -{\bf k}_\bot'),
\label{22}
\end{equation}
we find the normalization condition
\begin{equation} 
(A^{(\lambda,\omega,{\bf k}_\bot)} , 
A^{(\lambda',\omega', {\bf k}_\bot')})_{\rm KG} = 
\delta^{\lambda \lambda'}\delta(\omega -\omega')\delta^2({\bf k}_\bot
-{\bf k}_\bot')
\label{23}
\end{equation}
for $\lambda,\lambda'=1,2$.  For these modes we find
\begin{eqnarray}
(A^{(\lambda,\omega,{\bf k}_\bot)},
A^{(\lambda',\omega',{\bf k}_\bot')})_{\rm KG}
& = & \delta^{\lambda\lambda'}
k_\bot^2|C^{(\lambda,\omega,{\bf k}_\bot)}|^2\nonumber \\
&& \times
(\phi^{(\omega,{\bf k}_\bot)},\phi^{(\omega',{\bf k}'_\bot)})_{\rm
KG},\nonumber \\
\label{26}
\end{eqnarray}
where
\begin{equation}
(\phi^{(\omega,{\bf k}_\bot)},\phi^{(\omega',{\bf k}'_\bot)})_{\rm KG}  = 
i\int_\Sigma d\xi d^2{\bf x}_\bot \phi^{(\omega,{\bf k}_\bot) *} 
                        \stackrel{\leftrightarrow}{\partial_\tau}
                        \phi^{(\omega',{\bf k}'_\bot)}
\end{equation}
is the Klein-Gordon inner product for the scalar field defined
by Eq.~(\ref{sec2:KGfg}) and where
$\phi^{(\omega ,{\bf k}_\bot )}$ is given by Eq.~(\ref{16.5}).
Since the solutions $\phi^{(\omega,{\bf k}_\bot)}$ are
normalized as scalar mode functions, we have
\begin{equation}
(\phi^{(\omega,{\bf k}_\bot)},\phi^{(\omega',{\bf k}_\bot')})_{\rm KG} 
= \delta(\omega-\omega')\delta^2({\bf k}_\bot -{\bf k}_\bot').
\end{equation}
Substituting this equation in Eq.~(\ref{26}) and comparing the result
with Eq.~(\ref{23}), we find
$|C^{(\lambda,\omega,{\bf k}_\bot)}| = k_\bot^{-1}$ for $\lambda=1,2$.
Thus, the physical modes with $\lambda=2$ [see Eq.~(\ref{14})] are
\begin{equation}
A_\mu^{(2, \omega,{\bf k}_\bot)} = 
\frac{1}{2 \pi^2 k_\perp} \left[\frac{\sinh (\pi \omega/a)}{a}\right]^{1/2}
(\partial_\xi \phi , -i\omega \phi, 0, 0),
\label{33}
\end{equation}
up to a constant phase factor.
In fact we only need these modes because the
current~(\ref{5})--(\ref{7}) will excite neither the physical modes with
$\lambda=1$ 
[see Eq.~(\ref{13})] nor the modes with $\lambda = 0$ 
or $\lambda = 3$ via the interaction Lagrangian density 
\begin{equation}
{\cal L}_{int} = \sqrt{-g} j^\mu \hat A_\mu .
\label{24}
\end{equation}
Now, to lowest order in perturbation, the amplitude 
${\cal A}^{\rm em}_{(\lambda,\omega,{\bf k}_\bot)}$ 
of emission of a Rindler photon with quantum numbers
$ (\lambda,\omega,{\bf k}_\bot)$
into the Rindler vacuum 
state $\vert 0_{\rm R}\rangle$,
which is defined by
$
\hat{a}_{(\lambda,\omega,{\bf k}_\bot)}\vert 0_{\rm R}\rangle = 0
$
for all
$ (\lambda,\omega,{\bf k}_\bot)$,
is given by
\begin{equation}
{\cal A}^{\rm em}_{(\lambda,\omega,{\bf k}_\bot )} = 
\! \langle \lambda, \omega , {\bf k}_\bot\, {}_{\rm R}\vert 
i \int d^4x \sqrt{-g} 
j^\mu(x) \hat A_\mu (x) \vert 0_{\rm R} \rangle,
\label{34}
\end{equation}
where
$
\vert \lambda,\omega,{\bf k}_\bot\,{}_{\rm R}\rangle \equiv
\hat{a}_{(\lambda,\omega,{\bf k}_\bot)}^{\dagger}\vert 0_{\rm R}\rangle 
$.
It is straightforward to compute 
${\cal A}^{\rm em}_{(2,\omega , k_\bot )}$, which is the only
non-vanishing amplitude,
for the current~(\ref{5})--(\ref{7}) using Eqs.~(\ref{12}) and (\ref{22})
with Eq.~(\ref{33}).  
We obtain
\begin{eqnarray}
 {\cal A}^{\rm em}_{(2,\omega, k_\bot )} 
&= & 
i q \left[\frac{\sinh (\pi E/a)}{a}\right]^{1/2} \delta ( E - \omega ) 
\nonumber \\
& & \times 
\left\lbrace K'_{{i E}/{a}}(k_\perp /a)  \right. 
\nonumber \\
&  & - 
\left. 
\frac{E^2}{ak_\perp } \int_{{k_\perp}/{a}}^{\infty}
\frac{dz}{z}
K_{{i E}/{a}}(z) \right\rbrace , 
\label{35} 
\end{eqnarray} 
where the derivative with respect to the argument is denoted by a prime.

We are interested in the differential 
probability of emission per unit time and unit area in the 
transverse-momentum space given by 
\begin{equation}
dW_0^{\rm em} (\omega, k_\bot ) =  \sum_{\lambda = 1}^2 
\vert {\cal A}_{(\lambda,\omega,k_\bot)}^{\rm em} \vert^2 d\omega /T,
\end{equation}
where $T$ is the time interval while the interaction remains turned on.
We thus obtain
\begin{eqnarray}
&& dW^{\rm em}_0 (\omega, k_\bot ) \nonumber \\
&& = 
\frac{ q^2 }{4 \pi^3 a} \sinh (\pi E/a) \delta(E-\omega)
\nonumber \\
&& \ \ \ \ \times 
\left| K'_{{i E}/{a}}(k_{\perp }/a) -\frac{E^{2}}{a k_{\perp}}
\int_{k_{\perp}/a}^{\infty} \!\! \frac{dz}{z}K_{{iE}/{a}}(z)\right|^2
d\omega, 
\nonumber\\
\label{36}
\end{eqnarray}
where we have used $\delta (0) = T / 2\pi$. 

The total differential rate (per unit area in the transverse-momentum space) 
of emission of photons with given transverse momentum ${\bf k}_\bot$ 
into the thermal bath can be written as [see Eq.~(\ref{PMEM})]
\begin{equation}
R^{\rm em}_{k_\bot} = 
\int_0^{\infty} dW_0^{\rm em} (\omega, k_\bot ) 
\left( \frac{1}{e^{2 \pi \omega /a} -1} + 1 \right) . 
\label{37}
\end{equation}
The two terms inside the parentheses are associated with the 
induced and spontaneous emissions, respectively. Evaluating
the integral in Eq.~(\ref{37}) and taking the limit $E\rightarrow 0$ 
(thus removing the regulator), we obtain
\begin{equation}
R^{\rm em}_{k_\bot} d^2{\bf k}_\bot =
\frac{q^2}{8\pi ^3 a} \vert K_1 (k_{\perp }/a) \vert^2  d^2{\bf k}_\bot .
\label{38}
\end{equation}
Similarly, the total absorption rate of photons per unit area in the
transverse-momentum space is
\begin{equation}
R^{\rm abs}_{k_\bot} = 
\int_0^{\infty} dW_0^{\rm abs} (\omega, k_\bot )
\frac{1}{e^{2 \pi \omega/a} -1} .
\label{39}
\end{equation}
On unitarity  grounds we have
\begin{equation}
 dW^{\rm abs}_0 (\omega, k_\bot) = 
dW_0^{\rm em} (\omega, k_\bot),
\label{40}
\end{equation} 
and one can evaluate Eq.~(\ref{39}) using Eq.~(\ref{36}). We obtain
in the limit $E \to 0$ 
\begin{equation}
R^{\rm abs}_{k_\bot} d^2{\bf k}_\bot =
\frac{q^2}{8\pi ^3 a} \vert K_1(k_{\perp }/a) \vert^2  d^2{\bf k}_\bot .
\label{41}
\end{equation}
The reason for the equality of $R^{\rm em}_{k_\bot}$ and 
$R^{\rm abs}_{k_\bot}$
is that the spontaneous emission becomes negligible in comparison to the
induced emission as $E$ approaches zero.
It is also interesting to note that it is the 
existence of an infinite number of zero-energy Rindler photons
in the thermal bath that prevents
$R^{\rm em}_{k_\bot}$ and  $R^{\rm abs}_{k_\bot}$ 
from vanishing.
In the absence of the thermal bath, the 
emission and absorption rates would vanish.

Next, we recall that since there is no interference between the processes 
of emission and absorption of  Rindler photons at the
tree level, the total response rate will
be given by adding Eqs.~(\ref{38}) and (\ref{41}). We find, thus,
\begin{equation}
^{\rm ac}\!R^{\rm tot}_{k_\bot} d^2{\bf k}_\bot =
\frac{q^2}{4\pi^3 a} \vert K_1(k_{\perp }/a) \vert^2  d^2{\bf k}_\bot.
\label{42}
\end{equation}
By comparing this equation and Eq.~(\ref{55}) we find 
\begin{equation}
^{\rm ac}\!R_{k_\bot}^{\rm tot}=
\;^{\rm in}\!R_{k_\bot}^{\rm tot}.
\label{42a}
\end{equation}
Thus, we have established by explicit calculations that the rate of
photon emission from a uniformly accelerated charge can be reproduced by
summing the rates of emission and absorption of zero-energy Rindler photons
in the Unruh thermal bath.\footnote{
The analogous situation where the electric charge coupled to
the Maxwell field is replaced by a scalar source coupled to a 
massless Klein-Gordon field has also been investigated in 
free space \cite{Renetal94,Diazetal02} and in the presence of 
boundaries \cite{Alvesetal04}. An equality analogous 
to Eq.~(\ref{42a}) is satisfied in these cases as well.
                                   }
This also answers one of a series of questions concerning the Equivalence
Principle and the radiation concept [see \textcite{Rohrlich61,Ginzburg69} 
and references therein]. {}From our discussion in 
Sec.~\ref{subsubsection:E<mc2} it should be clear that zero-energy 
Rindler photons are not detectable since they concentrate on the horizon.
As a consequence, Rindler observers do not find emission of
classical radiation from uniformly accelerated charges although inertial 
observers do.\footnote{A discussion 
on how one can account for the change in the energy-momentum content 
of the radiation field in spite of the fact that uniformly accelerated 
charges are in {\em equilibrium} with the undetectable zero-energy 
Rindler photons of the Unruh thermal bath can be found in 
\textcite{Penaetal05}.} 
This is in agreement with conclusions obtained in the context of 
classical electromagnetism \cite{Boulware80,Eriksenetal04}.

A related question raised in this context is whether or not static
charges in gravitational fields should radiate. The quantization of the 
electromagnetic field outside black holes can be found 
in \textcite{Cognolaetal98} and \textcite{Crispinoetal01}. This was used to analyze 
the response of static charges coupled to the Hawking radiation
\cite{Crispinoetal98}. Because these charges lie at rest with respect
to the observers following the integral curve of the Killing vector
generating the global timelike isometry with 
respect to whom the particle content of the field theory is extracted,
the response is solely associated with the emission and absorption 
of zero-energy Boulware photons \cite{Boulware75,Boulware75b}. 
As a result, no classical radiation is emitted by the static 
charges as seen by the static observer\cite{Eriksenetal04}.\footnote{ 
A surprising coincidence appears as one considers the response of
static scalar sources interacting with a massless Klein-Gordon field 
outside a Schwarzschild black hole with the Unruh vacuum. Such a source 
behaves as if it were moving with the same proper acceleration in the 
inertial vacuum of Minkowski spacetime \cite{Higuchietal97}. 
This equivalence was expected when the source is close to the 
horizon \cite{Grishchuketal87} but not everywhere. Indeed, by considering 
other vacua \cite{Higuchietal98} as in \textcite{Hartleetal76}, 
other spacetimes \cite{Castineirasetal00}, fields 
\cite{Crispinoetal98,Castineirasetal03} 
or spacetime dimensionalities \cite{Crispinoetal04} 
the equivalence is broken.
                                      }

\section{Experimental proposals}
\label{section:Experimentalproposals}

This section will be mainly concerned with reviewing two complementary aspects, namely,
``proposed experimental tests of the Unruh effect" and ``the possible contributions of 
the Unruh effect for the explanation of experimental data". 
[For an extensive reference 
list of experiments related to the Unruh and Hawking effects 
see \textcite{Rosu01}.] 
We have already stressed that the Unruh effect 
does not need experimental confirmation 
any more than free Quantum Field Theory does. This fact 
does not invalidate, however, 
explanations of laboratory phenomena from the point of view of Rindler observers in 
terms of the Unruh effect. On the contrary, such explanations are 
interesting, and looking at 
some problems from the point of view of Rindler observers also 
can bring new insights.
This is how we shall understand here the experimental proposals of ``testing" the 
Unruh effect.

\subsection{Electrons in particle accelerators}
\label{subsection:polarization}

Among the first attempts to explain experimental data in terms of the 
Unruh effect is the one due to \textcite{Belletal83}. The fact that the 
transverse polarization of electrons and positrons in particle storage 
rings is not perfect has been observed for some time. The ``transverse 
polarization" here means the polarization perpendicular to both space 
velocity and acceleration, i.e.~along the direction of the magnetic
field responsible for the bending. Positrons and electrons are polarized 
in the directions 
parallel and antiparallel to the magnetic field, respectively. 
\textcite{Sokolovetal63} studied the electron-positron
polarization in storage rings first assuming a homogeneous magnetic 
field. Next, \textcite{Baieretal67}
generalized this result to inhomogeneous magnetic fields.
A more comprehensive analysis was performed further by
\textcite{Derbenevetal73}. 
The polarization is built up gradually in time according to the formula 
$
P(t) = P_0 (1-e^{-t/t_{{\rm build}\;{\rm up}}})
$, 
where the maximum polarization achieved in ideal conditions is 
$
P_0 = 8/(5\sqrt{3}) \approx 92.38 \%
$
and the characteristic build-up time in the laboratory frame is
$$
t_{{\rm build}\;{\rm up}} = \frac{8 }{ 5 \sqrt{3} } 
                             \frac{m_e^2 \rho^3 }{e^2 \gamma^5}.
$$
Here $m_e$ and $e$ are the electron mass and charge, respectively,
$\rho$ is the curvature radius of the orbit and
$\gamma = 1/\sqrt{1-v^2}$ is the usual relativistic factor.
For circular orbits $\rho$ coincides with the circle radius;
otherwise
$$
\rho^{-3} = \oint \rho(s)^{-3} ds \bigg/ \oint ds,
$$
where $\rho (s)$ is the radius of curvature at each
point on the orbit and $ s $ is the spatial distance 
of the corresponding point from some 
arbitrary origin defined on the orbit.
The photon power radiated due to the spin flip, 
$ I_{{\rm spin}\; {\rm flip}}$, can be compared with
the one due to synchrotron radiation, $I_{\rm synchrotron}$,
by
$$
\frac{I_{{\rm spin}\; {\rm flip}}}{I_{\rm synchrotron}}
= 3 \left(\frac{\gamma^2}{m_e \rho} \right)^2 
    \left(1\pm \frac{35 \sqrt{3}}{64}\right)^2,
$$
where the positive and negative signs 
should be used when initially the spin state 
is excited and deexcited, respectively [see \textcite{Jackson76} 
and \textcite{Montage84} for comprehensive reviews on the spin-flip 
synchrotron radiation and the 
polarization of electrons in storage rings]. 

Although theoretical 
investigations adapted to {\em inertial observers} 
were already performed, Bell and Leinaas posed the question 
whether or not it would 
be possible to use the spin as a sensitive thermometer and
interpret the depolarization of accelerated electrons from 
the point of view  of comoving observers through the Unruh effect. 
The coupling between the electron spin and a background magnetic 
field induces an energy gap $\Delta E$ between the ``spin up" and 
``spin down" states, making it a two-level system. If the distribution 
of spin-up and spin-down states of the accelerated electrons satisfied
the detailed balance relation, one could easily
interpret  the observed depolarization in terms of the Unruh effect 
(see Sec.~\ref{subsection:detector}).  If this was the case, the 
polarization 
\begin{equation}
P \equiv
\frac{^{\rm deexc}\! R - 
      ^{\rm exc}\! R}
     {^{\rm deexc}\! R + 
      ^{\rm exc}\! R}
\label{polarizationa}
\end{equation}
would be given by
\begin{equation}
P = 
\frac{1 - e^{-\beta \Delta E}}{1 + e^{-\beta \Delta E}},
\label{polarizationb}
\end{equation}
where we have used Eq.~(\ref{ratio}).

For {\em linear accelerators}, Bell and Leinaas obtained for
the excitation and deexcitation transition 
rates, $ ^{\rm exc}\! R$   and $ ^{\rm deexc}\! R$,  
(here denoted by $\Gamma_{+}$ and $\Gamma_{-}$, respectively)
$$
\Gamma_{\pm}= 
\mp \frac{8 \mu^2 }{3}\frac{\Delta E (\Delta E^2 + a^2)}{1- \exp(\pm 2\pi \Delta E/a)},
$$
where it is assumed for these machines that the magnetic field points to the 
acceleration direction, 
$\mu = g e / (4 m_e )$ is the magnetic moment and 
$g \approx 2.0023 $ is the electron gyromagnetic factor.
As a result, in this case the electron polarization~(\ref{polarizationa}) 
would indeed lead to Eq.~(\ref{polarizationb})
{\em if the actual machine specifications did not impose technical impediments}. 
At the Stanford linac with an accelerating field of $10 \, {\rm MV/m}$, 
for example, the Unruh temperature associated with the corresponding 
proper acceleration of 
$2 \, \times \, 10^{16} \, g_{\oplus}$ ($g_{\oplus} \approx 9.8 \, {\rm m/s}^2 $) 
would be about  $\beta^{-1}=0.7 \, \times \, 10^{-3}\, {\rm K}$. The fact that this 
temperature is much smaller than the ordinary background temperature of about 
$300 \, {\rm K}$ does not cause 
a substantial problem since the influence of the background thermal
bath is damped for relativistic electrons \cite{Costaetal95,Guimaraesetal98}. 
This is so because the background photons
are Doppler shifted in the electron proper frame and so most of them
are pushed away from the absorbable band. The main problem here is 
related with the ``thermalization" time. 
For instance,  the polarization build-up time at the Stanford linac is
much larger than the flight time (actually much larger than the lifetime
of the Universe). 
As a result, no equilibrium polarization would be built 
up in linear accelerators.

In order to decrease the polarization build-up time, larger accelerations
are necessary. Large enough accelerations are
indeed achieved in storage rings \cite{Barber99}. 
For instance, 
at the LEP/CERN, HERA/DESY and SPEAR/Stanford conditions, polarization
equilibrium states could be achieved in a couple of hours, half an hour 
and 
10 minutes, 
respectively.\footnote{Experiments at SPEAR measuring the maximum 
polarization and corresponding build-up time have been reported 
by \textcite{Camerinietal75} and \textcite{Johnsonetal83}, 
while spin polarization at the HERA, and LEP electron storage-rings
have been reported by \textcite{Barberetal94} and 
\textcite{Knudsenetal91,Assmannetal95}, 
respectively.                  } 
However, some cautionary remarks are in order. 
Firstly, the {\em Thomas precession} plays a major role when electrons 
are in circular motion, in contrast to 
the case of linear acceleration, and cannot 
be disregarded. Secondly, if the electrons are not linearly (and
uniformly) accelerated, the results concerning the Unruh effect are not
guaranteed to be applicable to them.
Thus, there is no compelling reason to expect that the detailed balance 
relation~(\ref{ratio}) and consequently Eq.~(\ref{polarizationb}) should
hold here. The intrinsic difficulties in the attempt to 
derive a variant of the Unruh effect for circularly moving detectors 
was already 
discussed in Sec.~\ref{subsubsection:circular}. Nevertheless, 
\textcite{Letawetal80}, \textcite{Belletal83} 
and \textcite{Takagi86} have argued that the response 
of ultra-relativistic Unruh-DeWitt detectors 
in uniform circular motion [see Eq.~(\ref{RMcircultrarel})] 
and that for those 
linearly accelerated [see Eq.~(\ref{RPlanck})] have some 
resemblance.\footnote{Actually, 
                                \textcite{Takagi84} found that the response 
                                of a Unruh-DeWitt detector with 
                                uniform circular motion and speed 
                                $v$ is better approximated by the 
                                one of an Unruh-DeWitt detector 
                                with a combined motion made of
                                linear acceleration and uniform 
                                translation with speed $v$ in the
                                direction perpendicular to the acceleration.}
Indeed, by calculating the excitation-to-deexcitation ratio 
$^{\rm exc}\! R^{m=0}_{\rm circ}/ ^{\rm deexc}\! R^{m=0}_{\rm circ}$
for circular motions, where $^{\rm exc}\! R^{m=0}_{\rm circ}$
is given in Eq.~(\ref{RMcircultrarel}) and  
$$
^{\rm deexc}\! R^{m=0}_{{\rm circ}} = 
\frac{|q|^2}{2 \pi} 
\frac{a (e^{-\sqrt{12} \Delta E/a} + 2 \sqrt{12} \Delta E/a  )}{2\sqrt{12}},
$$
and equating 
$^{\rm exc}\! R^{m=0}_{\rm circ}/ ^{\rm deexc}\! R^{m=0}_{\rm circ}$ 
to the detailed balance relation~(\ref{ratio})
satisfied by uniformly accelerated detectors, one is led 
to define the  {\em frequency dependent}  
temperature\footnote{We recall that Eq.~(\ref{RMcircultrarel})
was calculated assuming ultra-relativistic detectors and, thus,
Eq.~(\ref{fdt}) should be seen as an approximation. See 
\textcite{Letawetal82} and \textcite{Obadiaetal07} for more details. }
\begin{equation}
\frac{\beta^{-1}}{a} = 
\frac{\Delta E/a}{\ln [1+ 4 \sqrt{3} (\Delta E/a) \exp (2\sqrt{3} \Delta E/a)]}.
\label{fdt}
\end{equation}
\begin{figure}
\begin{center}
\mbox{\epsfig{file=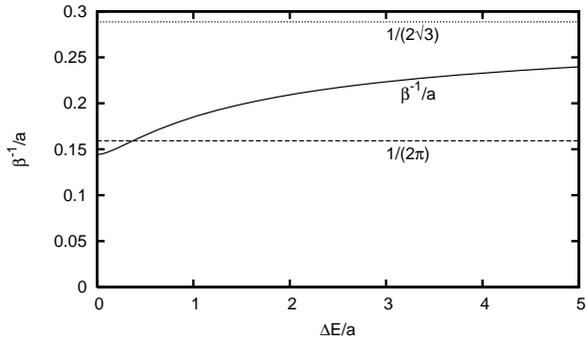,width=0.45\textwidth,angle=0}}
\end{center}
\caption{The frequency-dependent temperature $\beta^{-1}/a $ 
[see Eq.~(\ref{fdt})] is plotted as a function of $\Delta E/a$ 
and compared with the corresponding one given by the Unruh 
effect: $ 1/(2\pi)$. For $\Delta E \gg a$, one obtains 
$ \beta^{-1}/a = 1/(2 \sqrt{3})$.
}
\label{effect-temp}
\end{figure}
Note that for $\Delta E \ll a $ and $\Delta E \gg a $, one gets
$$ 
\beta^{-1}/a \approx  {1}/(2 \sqrt{12})
\;\;\;\; {\rm and} \;\;\;\;
\beta^{-1}/a \approx  {1}/(2 \sqrt{3}),
$$
respectively (see Fig.~\ref{effect-temp}). One should interpret 
$\beta^{-1}$ in Eq.~(\ref{fdt}) as an effective temperature
experienced by the detector in circular motion.\footnote{This temperature
is ``effective" because, due to the dependence of 
$\beta^{-1}$ on $\Delta E$, it cannot
be considered as the temperature of a legitimate thermal bath in 
contrast to that for the Unruh thermal bath.}

Now, at first sight it would not be unnatural to expect that 
ultra-relativistic electrons in storage rings had a polarization 
approximated by
\begin{equation}
P_1 = 
\frac{1 - e^{-\pi g}}{1 + e^{- \pi g}},
\label{polarization2}
\end{equation}
where we have used Eq.~(\ref{polarizationb}) with
$
\Delta E = 2 | \mu | | {\bf B}'_0 |
$, $\mu = ge/4m_e$,
$
\beta^{-1} = a/(2\pi) = e |{\bf B}'_0|/(2 \pi m_e)
$
and we recall that $g \approx 2.0023 $ is the electron gyromagnetic
factor.
Here ${\bf B}'_0$ is the magnetic field in the inertial frame 
instantaneously at rest with the electron.
In this case a description of the depolarization in terms of Rindler 
observers could be discussed along the same lines as the excitation
of accelerated detectors (see Sec.~\ref{subsection:detector}).
Clearly, this would be an ``indirect" connection with the Unruh effect,
since no real thermal bath of Rindler particles could be associated 
with observers comoving with the rotating electrons.

\begin{figure}
\begin{center}
\mbox{\epsfig{file=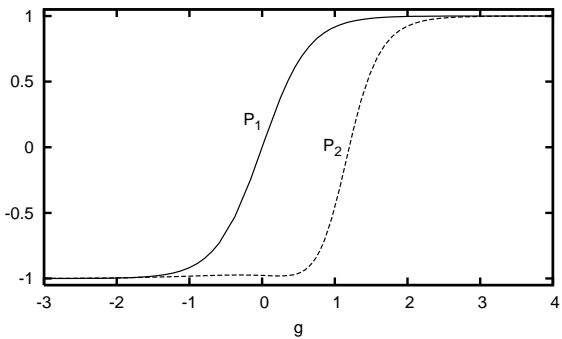,width=0.45\textwidth,angle=0}}
\end{center}
\caption{The polarizations $P_1$ (solid line) and $P_2$ (dashed line) given 
in Eqs.~(\ref{polarization2}) and (\ref{derbenevetal}), respectively, are 
plotted as functions of $g$. The two curves show substantial differences
although some similarities can be also seen.}
\label{polarizationfig}
\end{figure}

On the other hand, detailed inertial frame
calculations \cite{Derbenevetal73, Jackson76} 
show that the polarization is actually given by 
\begin{equation}
P_2= 
F_2(\tilde g )/
[F_1(\tilde g) e^{-\sqrt{12} |\tilde g|} + (\tilde g/|\tilde g|) F_2(\tilde g) ],
\label{derbenevetal}
\end{equation}
where 
$
\tilde g = (g-2)/2
$,
\begin{eqnarray}
& & 
F_1(\tilde g) 
= 
1+ \frac{41 \tilde g}{45} - \frac{23 \tilde g^2}{18} 
 - \frac{8 \tilde g^3}{15} +\frac{14 \tilde g^4}{15} 
\nonumber \\
& -&
\frac{8 \tilde g }{5 \sqrt{3} |\tilde g|} 
\left( 
1+ \frac{11 \tilde g }{12} - \frac{17 \tilde g^2}{12} 
 - \frac{13 \tilde g^3}{24} + \tilde g^4
\right),
\end{eqnarray}
and
\begin{equation}
F_2(\tilde g) =
\frac{8 }{5 \sqrt{3} }
\left( 
1+ \frac{14 \tilde g }{3} +8 \tilde g^2 
 + \frac{23 \tilde g^3}{3} + \frac{10 \tilde g^4}{3} +\frac{2 \tilde g^5}{3}
\right).
\end{equation}
The fact that the polarizations~(\ref{derbenevetal}) and~(\ref{polarization2})
show substantial differences (see Fig.~\ref{polarizationfig}) [although some 
similarities can be also pointed out \cite{Belletal83}] was discussed 
by \textcite{Unruh98, Unruh99}. 
The difficulty to understand the polarization 
in terms of electromagnetic vacuum fluctuations as experienced 
by the circularly moving observers stems from the fact that 
in this case the electron should be seen as a system composed of two 
field detectors coupled to each other. In addition 
to the spin, the vertical fluctuations in the orbit should be
considered \cite{Belletal87, Leinaas99, Leinaas02}.\footnote{ 
See also \textcite{Barberetal88} for a comparison 
of the procedures used by \textcite{Belletal87} and 
\textcite{Derbenevetal73}.                                }  
By analyzing the problem in the Fermi-Walker frame associated 
with the electron, it is concluded that the vertical oscillation 
responds in a thermal-like 
fashion with a frequency-dependent 
temperature similar (but not identical) 
to that associated with the spin alone. 
Nevertheless, the subtle coupling between the two systems makes
the joint system to have a polarization~(\ref{derbenevetal})
distinct from the simple thermal-like one~(\ref{polarization2}). 

\subsection{Electrons in Penning traps}
\label{subsection:penningtrap}

The use of the degree of freedom associated with vertical fluctuations 
of a single electron in circular motion as a detector was investigated 
further by \textcite{Rogers88}. In this paper an experimental proposal
is made in which an electron
is bound radially by the laboratory magnetic field ${\bf B}_0$ 
and axially by superimposing an electrostatic restoring force given 
by a quadrupole  potential 
\cite{Brownetal86}
$ 
V = V_0 {(z^2 - \rho^2/2)}/(2 d^2)
$,
where $V_0, d = {\rm const}$, and  $z$ and $\rho$ are the {\em axial} and
{\em radial} coordinates, respectively. Thus, the electron is in a
{\em Penning trap}. As a result, an electron constrained 
to move in a circular cyclotron orbit around the trap axis with angular 
frequency $\omega_0 = e|{\bf B}_0| / (\gamma m_e)$ can oscillate axially 
with frequency
$
\omega_z = \sqrt{e V_0/ ( \gamma m_e d^2)}
$.
The excitation of this degree of freedom could be interpreted as due to 
the vacuum fluctuation experienced by the circularly moving electron. By
surrounding the Penning trap with an electromagnetic cavity tuned to 
resonate at the electron axial oscillation frequency $\omega_z$, the 
energy of the axial motion would be transferred to the cavity 
electromagnetic field where it would be measured.
In general, electrons are captured by the trap in large orbits but their 
radii shrink rapidly due to 
the emission of synchrotron radiation. In order
to replace the energy lost, \textcite{Rogers88} suggests to
irradiate the system with circularly polarized waves of frequency 
$\omega_0$. In the proposed experiment,
an electron is assumed to have 
velocity $v=0.6$ in a background magnetic field of 
$|{\bf B}| = 1.5\, \times \, 10^{5} \,  {\rm G}$. The axial and angular 
frequencies would be 
$
\omega_z \approx 5 \, \times \, 10^{10} \, {\rm s}^{-1}
$
and
$
\omega_0 \approx 2 \, \times \, 10^{12} {\rm s}^{-1},
$
respectively, corresponding to a proper acceleration 
$a = \gamma^2 v \omega_0 \approx 6 \, \times \, 10^{19} g_\oplus$
with $a/2\pi \approx 2\,{\rm K}$.

\subsection{Atoms in microwave cavities}
\label{subsection:cavities}

\textcite{Scullyetal03} and 
\textcite{Belyaninetal06} have considered a {\em gedanken}
experiment assuming that a beam of two-level atoms are accelerated 
through a high-Q (i.e. low power loss) 
``single mode" microwave cavity. They have noted 
that even with a large acceleration frequency (defined as the
acceleration divided by the speed of light)
$\alpha \approx 10^8 \, {\rm Hz}$, 
corresponding to the proper acceleration as large as 
$ 3 \, \times \, 10^{15} \, g_{\oplus}$, for an atom 
with energy gap of 
$\Delta E \approx 10^{10}\, {\rm Hz}\, (\approx 4\times 
10^{-5}\,{\rm eV})$, the 
excitation-deexcitation ratio~(\ref{ratio}) of the atom would be 
\begin{equation}
^{\rm exc}\! R /^{\rm deexc}\! R = e^{- 2 \pi \Delta E/ \alpha} \sim 10^{-200},
\label{Unruhcavity}
\end{equation}
which is extremely small. 
The atoms are assumed to follow the world line
$$
t(\tau) = t_0 + \alpha^{-1} \sinh (\alpha \tau ),\;
z(\tau) = \alpha^{-1} (\cosh (\alpha \tau) -1),
$$
where $t_0 = t(\tau)|_{\tau=0}$ is the moment in the laboratory time
when the atoms begin to accelerate. They enter
the cavity at $\tau = \tau_i$ and exit it later at $\tau = \tau_e$ 
after staying in interaction for long enough, typically 
$\alpha (\tau_e - \tau_i) \gg 1$.  

In spite of the minute value predicted by the Unruh effect 
for the situation described above, 
in a real experiment the ratio 
$^{\rm exc}\! R/^{\rm deexc}\! R$ 
can be much larger because the sharp boundaries of the cavity 
induce  a nonadiabatic coupling of the form $g(\tau ) =\mu E'$ 
between the atom and the electromagnetic field, where
$\mu$ is the atomic dipole moment and $E'$ is the electric 
field as measured in the inertial frame instantaneously at rest with 
the atom. This may be seen as a sort of laboratory 
implementation of the finite-time 
detectors discussed in Sec.~\ref{subsubsection:otherresults}, 
where the scalar field is replaced by an electromagnetic one.
For the case where the photon frequency $\nu$ associated with 
the single-mode cavity satisfies $\nu \gg \Delta E \gg \alpha$, 
the ratio of the excitation to deexcitation rates is found to be
\begin{equation}
^{\rm exc}\! R/^{\rm deexc}\! R = \alpha /(2 \pi \Delta E),
\label{boundarycavity}
\end{equation}
which leads to a value enhanced by many orders 
of magnitude under the conditions given above: 
$^{\rm exc}\! R/^{\rm deexc}\! R  \sim 10^{-3}$.
We note that even when the atom enters the cavity with constant 
velocity, the sudden turning-on of the interaction induced by the  
sharp boundaries makes the atom excitation rate to be nonzero.
(Note that a photon is {\em emitted} in this excitation process.) 
Nevertheless, acceleration still plays a role in the ratio
$^{\rm exc}\! R/^{\rm deexc}\! R$
as seen in Eq.~(\ref{boundarycavity}). 
The authors note that Eq.~(\ref{Unruhcavity}) 
is recovered when one considers 
the limit of adiabatically slow switching-off of the interaction at 
infinitely distant  cavity  boundaries.
In this case
\begin{eqnarray}
^{\rm exc}\! R 
& \propto &
\pi \Delta E  [\alpha \sinh (\pi \Delta E / \alpha)]^{-1} e^{- \pi \Delta E/\alpha},
\nonumber \\
^{\rm deexc}\! R 
& \propto &
\pi \Delta E  [\alpha \sinh (\pi \Delta E / \alpha)]^{-1} e^{ \pi \Delta E/\alpha}.
\nonumber 
\end{eqnarray}
\textcite{Belyaninetal06} also discuss the existence of photons and
their number in the cavity as a result of the interaction 
with the beam of atoms (see discussion in 
Sec.~\ref{subsubsection:otherresults}). 
Clearly, the physical origin of the simultaneous increase of the
field energy and the internal energy of the atom 
is the work done by the external force, 
which drives the center-of-mass motion of the atom against the 
radiation reaction force.

Both the acceleration and boundary contributions to the photon 
emission from the ground state atom come from the 
{\em counterrotating term} 
$\propto \hat a^\dagger \hat \sigma^\dagger$ in the interaction
Hamiltonian, where $\hat a^\dagger$ is the photon creation operator and
where the operator
$\hat \sigma^\dagger \equiv |a \rangle  \langle b | $ converts the
ground state $|b\rangle$ of the atom to its excited state $|a\rangle$. 
In the single-mode 
cavity case, we could also {\em define} an {\em effective} 
temperature 
\begin{equation}
\beta^{-1}_{\rm eff} = \Delta E/ \ln (2\pi \Delta E/\alpha)
\label{effectivetemperature}
\end{equation}
by equating  Eqs.~(\ref{Unruhcavity}) and~(\ref{boundarycavity}).
Nevertheless, the Unruh and boundary effects should not be put on 
the same footing \cite{Obadia07}. We note that Eq.~(\ref{effectivetemperature}) 
is an effective temperature depending on the details of the atom 
through $\Delta E$ in contrast to the Unruh temperature which 
is associated with a legitimate thermal bath of Rindler particles.

Another possible physical implementation of a finite-time detector 
was also discussed recently. \textcite{Alsingetal05} have considered 
trapped ions prepared in the vibrational ground state with a laser 
cooling procedure. The internal electronic state of each ion would 
play the role of a phonon detector through the use of an external 
laser which would couple it to the ion vibrational motion. Thus, 
this phonon 
detector could be turned on and off at will by controlling the external 
laser. Another interesting feature of this detector is that it would
be easy to vary its energy gap by tuning the frequency of the external 
laser. \textcite{Alsingetal05} also consider a {\em linear} ion trap 
to analyze particle creation in condensed matter analogues of 
expanding universes. 

\subsection{Backreaction radiation in ultraintense lasers and related
topics}
\label{subsection:lasers}

A different path pursued in the literature 
concerns the use of ultraintense
lasers \cite{Chenetal99,Schutzholdetal06}. It is known that the electric 
field of laser-driven waves of plasma can accelerate particles 
in just a few meters to energies as high as the ones obtained by large 
conventional  radio-frequency accelerators [see \textcite{Mourouetal06} 
and references therein]. In {\em plasma wakefield accelerators}, a short pulse of 
laser light (or electrons) is responsible for a collective perturbation of 
the plasma confined in a cavity producing an electromagnetic wakefield 
in the laser propagation direction. This wakefield can be surfed by some
electrons which acquire very high accelerations. 
However, the direct effect of the laser field  on the electrons 
can induce even larger accelerations (and decelerations) along 
every laser cycle. Electric field pulses not too far below the Schwinger limit 
(about $10^{18} V/m$) are expected in future facilities.  
(The Schwinger limit is 
associated with the electric field above which the spontaneous creation 
of electron-positron pairs becomes favorable. This is so when 
the work done by the electric field along the electron Compton 
wavelength is at 
least about the mass of the electron-positron pair.) Electrons under the
influence of fields with this magnitude could reach proper accelerations 
as high as $ 10^{28} \, g_{\oplus}$. Here, we shall be interested in interpreting
the radiation emitted by such electrons in terms of the Unruh effect
rather than in the behavior of internal degrees of freedom.

For the sake of simplicity, Chen and Tajima consider the case where two
identical counterpropagating laser plane waves  produce a standing wave.
In this case, electrons can be treated as 
classical charges with well-defined trajectories. Let us consider linearly 
polarized lasers with angular frequency $\omega_0 $, wave number $k_0$ and 
propagation in the $\pm z$ directions:
\begin{eqnarray}
E_x &=& E_0[\cos (\omega_0 t - k_0 z) + \cos (\omega_0 t + k_0 z)],
\nonumber \\
B_y &=& E_0[\cos (\omega_0 t - k_0 z) - \cos (\omega_0 t + k_0 z)],
\nonumber
\end{eqnarray}
where $E_x$ and $B_y$ are the electric and magnetic fields in the $x$ 
and $y$ directions, respectively, as measured in the laboratory frame.
The equations of motion for an electron under the influence of this 
field can be written as
\begin{eqnarray}
dp_x/dt & = & -e (E_x - v_z B_y),
\nonumber \\
dp_z/dt & = & -e v_x B_y,
\nonumber
\end{eqnarray}
where 
$p_{x} = m_e \gamma v_{x}$
and 
$v_{x} \equiv dx / dt$.
The largest electric field is found at the nodal points
$k_0 z = 0, \pm 2\pi, \ldots $ where $E_x = 2 E_0 \cos (\omega_0 t)$ 
and $B_y = 0$. In particular, at $z=0$, Chen and Tajima find
$$
\gamma v_x = 2 a_0 \sin \omega_0 t, \;\;\; 
\gamma = \sqrt{1+4 a_0^2 \sin^2 \omega_0 t},
$$
where
$a_0 = e E_0/(m_e \omega_0)$ 
is a dimensionless parameter, which characterizes the laser strength. 
The corresponding electron proper acceleration is, thus, given by
$$
a = 2 a_0 \omega_0 \cos \omega_0 t,
$$
and the total Larmor radiation power ${dI_L}/{dt} = (2/3) e^2 a^2 $ is
\begin{equation}
{dI_L}/{dt} =  (8/3) e^2 a_0^2 \omega_0^2 \cos^2 \omega_0 t. 
\end{equation}
Hence, the total energy radiated during each laser half cycle is
$\Delta I_L = (4 \pi/3) e^2 a_0^2 \omega_0$. 

We have shown in Sec.~\ref{subsection:Bremsstrahlung} how the radiation 
emitted by {\em uniformly accelerated charges} can be 
interpreted from the point of view of coaccelerated observers 
in terms of the {\em emission} and {\em absorption} of zero-energy 
Rindler photons {\em to} and {\em from} the Unruh thermal bath, 
respectively. In this calculation, the charge was assumed not to recoil
in the emission/absorption process. This assumption is justified if the
mass of the charge is much larger than the typical (Minkowski) 
energy of the photon
emitted. In a real physical set-up, however, the electron 
backreacts to the Larmor radiation.  Chen and Tajima claim that this
backreaction triggers additional ``quivering radiation", which reflects
the Unruh effect.  They estimate
the power of this radiation, $\Delta I_U$, in comparison to that of the
Larmor radiation, $\Delta I_L$, as 
\begin{equation}
\frac{\Delta I_U}{\Delta I_L}
\approx \frac{9}{\pi^2} \frac{\hbar \omega_0}{m_e c^2} a_0 \log(a_0/\pi)
\approx 3\times 10^{-4},
\end{equation}
where $m_e$ is the electron mass,
for $\omega_0 \approx 2 \times 10^{15}\;{\rm sec}^{-1}$ and $a_0\approx
100$.

\textcite{Schutzholdetal06} note that a Rindler photon seen
to be scattered off a static charge 
by the Rindler observers should correspond to a pair of correlated 
Minkowski photons emitted from a uniformly accelerated charge 
as seen by the inertial observers. (Note that a Rindler photon with
{\em nonzero} energy can cause this process in contrast to the
Larmor radiation, i.e. the bremsstrahlung.)
They propose this two-photon
emission process as a distinct signal of the Unruh effect.
they argue that, 
as long as the acceleration is not close to the Schwinger limit, where
the Unruh temperature becomes comparable to the electron mass, 
Rindler observers can describe the electrons as pointlike (Thomson) 
scatterers of Rindler photons. (In this regime the electron
spin is not supposed to play any major role.) 
They assert that the most promising strategy to observe a signal of
this radiation above the background Larmor noise 
would be by probing the angular distribution of the photon 
emission; in contrast to the Larmor radiation, which has 
a well known blind spot along the motion direction,
this two-photon radiation dominates inside small backward and 
forward 
cones.\footnote{\textcite{Chenetal99} also propose to exploit the blind
spot of the Larmor radiation.}  They also note that 
another signal would be the direct detection 
of correlated photons.

Although the residual quivering radiation or correlated radiation 
of Minkowski photons 
could be explained  and  calculated by inertial observers using  
textbook quantum field theory\footnote{We have favored the term 
``quivering" rather than ``Unruh" radiation, as appears sometimes 
in the literature, to emphasize that this does not depend on the 
Unruh effect any more than the Larmor one does.}, it is certainly
interesting to understand these processes invoking the Unruh effect.

Finally, let us mention here some works concerning a detailed-balance
relation obeyed by the transverse momentum of a uniformly accelerated
electron. 
The differential probability of emission of a photon by an electron
in a constant electric field $E$ was obtained by 
\textcite{Nikishov70} by means of an inertial frame calculation,
where the electron and photon fields are quantized. The recoil 
causes the electron to change its momentum perpendicular to the
background electric field, which accelerates the electron.  
Let $P(p_\bot\to p'_\bot)$ be 
the differential probability associated with the photon emission
changing the modulus of the transverse momentum of the electron from
$p_\bot$ to $p'_\bot$.  \textcite{Nikishovetal88} find 
\begin{equation}
\frac{P(p_\bot \to p'_\bot)}{P(p'_\bot \to p_\bot)} 
= \exp (-\beta \Delta {\cal E}),
\label{ratio2}
\end{equation}
where
$
\beta^{-1} = a/ 2 \pi
$
with $a = e E /m_e $
and 
\begin{equation}
\Delta {\cal E} = p_\perp^{\prime 2}/2m_e - {p_\perp^2}/2m_e.  \label{non-rel}
\end{equation}
This expression for $\Delta {\cal E}$ is valid even if $p_\perp$ and
$p'_\perp$ are comparable to $m_e$.  If $p_\perp, p'_\perp \ll m_e$,
then
\begin{equation}
\Delta{\cal E} \approx \sqrt{p_\perp^{\prime 2} + m_e^2}
- \sqrt{p_\perp^2 + m_e^2}.
\end{equation} 
\textcite{Myhrvold85} calls $\sqrt{p_\perp^2 + m_e^2}$ the {\em
transverse energy} of the electron with transverse momentum $p_\perp$
and claims that the relation (\ref{ratio2}) reflects the Unruh effect
for $p_\perp,p'_\perp \ll m_e$.
Indeed Eq.~(\ref{ratio2}) is similar to 
the detailed balance 
relation~(\ref{ratio}), which was derived assuming hyperbolic 
motion of the source.\footnote{Recent investigations also considering the
backreaction on accelerated systems can be found in 
\textcite{Parentani95}, \textcite{Parentanietal97}, 
\textcite{Gabrieletal98} and \textcite{Reznik98}.
                } Although the relation (\ref{ratio2}) may well be
closely related to the Unruh effect, the former 
does not directly follow from the latter because the connection between the
Rindler energy [in Eq.~(\ref{ratio})] and the transverse energy
[in Eq.~(\ref{ratio2})] is not entirely clear.

\subsection{Thermal spectra in hadronic collisions}
\label{subsection:hadrons}

Now, let us turn our attention to insights that the Unruh effect
can bring to explain some experimental data in hadronic physics. The 
Unruh effect has been considered as possibly helpful in explaining 
the puzzling thermal-like emission spectra observed in hadron 
collisions \cite{Barshayetal78, Barshayetal80, Kharzeev06}. 
The main idea is that in the collision process, hadrons would feel
in their rest frame a large Unruh temperature, which would lead them to 
quiver and interact accordingly. It is conjectured, then, that the 
thermal-like emission of {\em Minkowski particles} observed in hadron 
processes would be a reflection of it. In some sense, it may be
that a quivering-like radiation which we have discussed in 
Sec.~\ref{subsection:lasers} in a quite different context  
becomes the explanation for this puzzling aspect of hadron
collisions. As we have said, the Unruh thermal bath is not required 
for the investigation of accelerated systems from the point of view 
of inertial observers. For these observers, ``plain" quantum field 
theory must suffice for a complete phenomenon description. 
Nevertheless, it would be certainly interesting if the Unruh effect 
could bring new insights to the understanding of this problem.

\subsection{Unruh and Moore (dynamical Casimir) effects}
\label{subsection:dynamicalcasimir}

There have also appeared some proposals of using the {\em Moore effect}, 
often called
{\em dynamical Casimir effect}, as a way to test the thermal 
bath observed by Rindler observers. Here we discuss why the connection 
between the Moore and Unruh effects is tenuous and comment briefly on some 
few selected proposals. We refer to \textcite{Rosu01} for a more 
extensive list.

\textcite{Moore70} and later \textcite{DeWitt75} found independently 
that photons can be created by 
moving mirrors in the Minkowski vacuum. An interesting connection 
between the Moore effect and the Hawking radiation was established 
by \textcite{Daviesetal77} and more recently revisited  by 
\textcite{Calogeracos02b,Calogeracos02}. These authors consider a 
massless scalar field in two-dimensional Minkowski spacetime equipped 
with a reflective boundary. At $t=0$ the boundary begins to move to 
the left following the trajectory
\begin{equation}
z(t) = - \kappa^{-1} \ln (\cosh \kappa t),
\end{equation}
where $\kappa = {\rm const}$ and $x^\mu= (t,z)$ are the usual Cartesian 
coordinates.\footnote{Notice that by identifying $t, z$ and $\kappa$
with $\tau, \xi$ and $a$, respectively, we obtain Eq.~(\ref{sec2:rightcoords})
with $z=a^{-1}$.} We note that asymptotically the corresponding world line 
becomes lightlike. (Notice that the proper acceleration of the boundary
is $\kappa \cosh \kappa t \neq \kappa$.) Eventually, the 
receding boundary induces a thermal flux of {\em Minkowski} particles 
to the right characterized by a temperature $\kappa/2 \pi$. 
(This would not be so if the boundary were uniformly 
accelerated.) The {\em energy content} associated 
with the {\em particle emission} was also investigated  
\cite{Fullingetal76} [see also \textcite{Calogeracos04}]. 
There is,  thus, a similarity between the flux of Minkowski particles, 
which are emitted from the receding boundary and the Hawking 
radiation of {\em Boulware} particles produced in a black hole 
formation process. 

Now, by approximating the line element of a black hole 
close to its horizon by that of the Rindler wedge as discussed 
in Sec.~\ref{subsubsection:E<mc2}, we can establish a correspondence
between {\em static observers outside the horizon} and {\em Rindler 
observers}, where the former and latter observers
are immersed in the Hawking 
radiation and Unruh thermal bath, respectively 
[see also \textcite{Ginzburgetal87}]. 
This leads to the following loose connection
$$
{\rm Moore} \; {\rm effect} \leftrightsquigarrow 
{\rm Hawking} \; {\rm radiation} \leftrightsquigarrow
{\rm Unruh} \; {\rm effect},
$$
where the Moore effect is seen as a flat spacetime analog of 
the Hawking effect and this is connected with the Unruh thermal
bath close to the black hole horizon. It should be stressed, however, 
that although the observation  of the Moore effect would
be very interesting, this would not constitute a experimental
verification of the Unruh effect. It is worthwhile to emphasize 
that the thermal flux associated with the Moore and Unruh effects are 
formed of Minkowski and Rindler particles, respectively, which are 
quite different. The Moore, Hawking and Unruh effects, although 
related, have features which make them distinct.

Despite the fact that the Moore and Unruh effects are only linked 
through the indirect reasoning above, we comment briefly on 
some experimental proposals, which are interesting in their own right.
\textcite{Yablonovitch89} [see also \textcite{Yablonovitchetal89}] has 
recently discussed the possibility of using media with varying index of 
refraction to observe a Moore-like effect. When a gas is suddenly 
photo-ionized, its index of refraction drops from about 1 to 0. 
This disturbs the vacuum in a way similar to what 
an accelerated mirror does.
[For a comparative discussion between these two similar effects see, e.g., 
\textcite{Johnstonetal95}.]  Likewise, sudden creation of electron-hole 
pairs in a semiconductor slab can quickly reduce the refractive index 
from about 3.5 to 0. 
Considering a general medium with time-varying index of refraction 
$n=n(t)$ which instantaneously jumps from  $n_0$ to $n$,
\textcite{Yablonovitch89}
found an expectation number of created modes with wave vector $k$ given by
$$
N_k =  \sum_{k'} |\beta_{k {k'}}|^2,
$$
where
$
 |\beta_{k {k'}}| = |n-n_0| \delta_{k {k'}}/(2 \sqrt{n n_0})
$.
The experimental prospect of the laboratory verification of 
Moore-like effects in the near future seem very promising [see, 
e.g., \cite{Kimetal06,Uhlmannetal04}].

\section{Recent developments}
\label{section:Recentdevelopments}

Recently, a number of issues connecting Quantum Mechanics,
Relativity and Information Theory have been investigated
[see \textcite{Peresetal04} for a critical review]. Here we 
comment briefly on some of these issues and other topics
that have the Unruh effect as the central theme. 
We refer the reader to our list of references for more details.

\subsection{Entanglement and Rindler observers }
\label{subsection:Entanglement}

As is well known, mixed states can be obtained from pure
states by tracing out (i.e.~ignoring) some of its degrees 
of freedom \cite{Zurek91}. However, it 
was not obvious until recently that the ``amount of mixing" 
could depend on the observer. For a spin-$1/2$ system
\textcite{Peresetal02} found that, in general, different 
inertial observers will find distinct values for the 
corresponding von Neumann quantum entropy 
$$
S= - {\rm Tr} (\rho^{\rm red} \ln \rho^{\rm red} ).
$$ 
Here $\rho^{\rm red}$ is the reduced density 
matrix associated with the spin-$1/2$ particle, which is 
obtained after the momentum degrees of freedom 
are traced out. 
Later on, for a {\em pair} of massive spin-$1/2$ 
particles \textcite{Gingrichetal02} found that by 
tracing out the momentum degrees of freedom, different inertial 
observers will assign in general distinct entanglements 
between the particle spins. A similar conclusion was reached
for the entanglement between the polarization of a pair 
of photon beams \cite{Gingrichetal03}. 

Although the entanglement between some degrees of freedom 
can be transferred to others as shown above, all inertial
observers will agree about the entanglement of the full state.
This is not the case, however, when non-inertial observers
are involved. \textcite{Fuentes-Schulleretal05} investigated 
the entanglement between two modes of a free massless scalar
field as seen by inertial and uniformly accelerated observers.
They reached the conclusion that the existence of a horizon for 
the Rindler observer leads in general to loss of information. 
The entanglement which appears to be maximal for inertial 
observers is degraded according to the Rindler ones because of 
the Unruh effect. The authors suggest that analogous conclusions 
should be valid close to black holes when inertial and Rindler 
observers are replaced by free-falling and static ones, 
respectively. A thorough investigation of such questions in 
general curved spacetimes would be very interesting.

\subsection{Decoherence of accelerated detectors }
\label{subsection:Decoherence}

Discussions on the decoherence of accelerated detectors
have been continuing since some years ago \cite{Audretschetal95}. 
\textcite{Koketal03} considered a qubit $|\psi \rangle $ 
represented by a (uniformly accelerated) 
Unruh-DeWitt detector with free Hamiltonian 
$
H_0 = \Delta E \; \hat b^\dagger \hat b
$, 
where $\Delta E$ is the energy gap between the two internal degrees of 
freedom  $|0 \rangle$ and $|1 \rangle $ of the qubit, and  
$\hat b$ and $\hat b^\dagger$ denote the lowering and raising 
operators, respectively, acting on the corresponding two-dimensional 
Hilbert space:
$$
\hat b |0 \rangle = 0, \;\; \; 
\hat b^\dagger |0 \rangle = |1 \rangle, \;\;\;
\hat b |1 \rangle = |0 \rangle, \;\;\;
\hat b^\dagger |1 \rangle = 0.
$$
The qubit  is coupled to a real scalar field 
$\hat \Phi ({\bf x},t)$ through the interaction Hamiltonian 
$$
H_I(t) = 
\epsilon (t) \int_\Sigma 
\hat \Phi ({\bf x},t) 
(\psi({\bf x}) \hat b + \bar \psi ({\bf x}) b^\dagger) \sqrt{-g\,}
d^3{\bf x},
$$
where $\psi ({\bf x})$ is a smooth function which vanishes outside
a small volume around the qubit. The integration is over the 
global spacelike Cauchy surface $\Sigma$ given by $t= {\rm const}$ 
(with $t$ being the usual Cartesian time coordinate) and $\epsilon (t)$ 
is a time dependent coupling constant, which vanishes everywhere
except within a finite time interval $\Delta t$
where $\epsilon (t) = \epsilon = {\rm const}$. Before the acceleration
takes place, the qubit is prepared in the state
$$
| \psi \rangle = (| 0 \rangle + | 1 \rangle)/ \sqrt{2\,},
$$
which is combined with the field state described by the reduced  
density matrix~(\ref{sec2:discdensity}). We recall that 
this is  obtained when the degrees of freedom of the Minkowski vacuum 
associated with one of the Rindler wedges are traced out. Then, the 
combined initial state 
\begin{equation}
\hat \rho_{\rm in} = \hat \rho_R \otimes |\psi \rangle  \langle \psi |
\label{rhoin}
\end{equation}
must be evolved through the interaction Hamiltonian leading to
$\hat \rho_{\rm out}$. Kok and Yurtsever find 
the final reduced density matrix associated with
the qubit by tracing out the field degrees of
freedom as
\begin{eqnarray}
\hat \rho_{q, {\rm out} } 
& = & {\rm Tr}_\Phi ( \hat \rho_{\rm out})
\nonumber \\
& = & \frac{1}{2}
      \left( 
                \begin{array}{cc}
                  S_0 + S_e &  S_0        \\
                      S_0   &  S_0 + S_a  
                \end{array}
       \right),
\end{eqnarray}
where
\begin{eqnarray}
S_0 & = & (1-e^{-2\pi \Delta E/a} ) \sum_n e^{-2\pi n \Delta E/a}/Q_n,
\nonumber \\
S_a & = & (1-e^{-2\pi \Delta E/a} ) |\mu |^2 \sum_n n e^{-2\pi n \Delta E/a}/Q_n,
\nonumber \\
S_e & = & (1-e^{-2\pi \Delta E/a} ) \nu^2 \sum_n (n+1) e^{-2\pi n \Delta E/a}/Q_n.
\nonumber
\end{eqnarray}
Here,
$
Q_n = 1 + n |\mu |^2/2 + (n+1) \nu^2/2,
$
where 
\begin{equation}
|\mu | \approx  \frac{\epsilon \Delta t}{\sqrt{ 2\Delta E \,}} 
            e^{-\kappa^2 \Delta E^2/2 }, \,\,\,\,\,
\nu \approx  \frac{\epsilon \Delta t}{2 (\sqrt{\pi} \kappa^3 )^{1/2}} 
\end{equation}
and 
$\kappa$ is a length scale setting the spatial range of the 
interaction. Then, they show that the {\em purity} 
${\rm Tr} (\hat \rho_{q, {\rm out} }^2)$ decreases
monotonically with the qubit proper acceleration $a$,
as expected.

\subsection{Generalized second law of thermodynamics 
and the ``self-accelerating box paradox"}
\label{subsection:entropy}

In a colloquium delivered at Princeton University  in the
early 1970's R. Geroch raised the possibility of violating the 
ordinary second law of thermodynamics with the help of
classical black holes. The idea was to bring {\em slowly}
from infinity a box with proper energy $E_{\rm b}$ over the event 
horizon and  throw it eventually inside the hole.
The cycle would be closed by lifting back the ideal rope 
characterized by an arbitrarily small mass. Because 
static asymptotic observers would assign zero energy to
the box at the event horizon, the hole would remain the 
same after engulfing it. This would challenge 
the ordinary second law of thermodynamics, since eventually
all entropy associated with the box would vanish from the
Universe with no compensating entropy increase elsewhere.

As an objection to Geroch's process, Bekenstein argued 
that quantum mechanics would constrain the size and 
energy of the box  accordingly. This constraint would make it impossible
for all parts of the box to reach 
the event horizon at once and, thus, the black 
hole would necessarily gain mass after engulfing 
the box. Then, \textcite{Bekenstein73} conjectured 
that black holes would have a nonzero entropy 
$S_{\rm bh} = A /4$ proportional to the 
event horizon area $A$  and formulated the 
{\em generalized second law} (GSL), namely, that the 
total entropy of a closed system (including that
associated with black holes) would never decrease.
This opened a new whole subject called
``black hole
thermodynamics".\footnote{Recently there have been some 
works on how the
laws of thermodynamics associated with black hole horizons can 
be extended to what \textcite{Jacobsonetal03} call {\em causal 
horizons}, i.e.~the boundary of the past 
of any timelike curve $\lambda$ of infinite proper length in 
the future direction.}
Now, because the GSL would be violated if the entropy of the box $S$ 
satisfied $S> 2 \pi E_{\rm b} R$, where 
$R$ is the proper radius of the smallest sphere circumscribing 
the box, Bekenstein conjectured in addition
the existence of a new thermodynamical law, namely, that every system
should have an entropy-to-energy ratio satisfying 
$S/E_{\rm b} \leq 2 \pi R$. 

Later, however, \textcite{UnruhWald82} 
concluded\footnote{See also \textcite{UnruhWald83}
in response to \textcite{Bekenstein83}.} 
that by taking into account the buoyancy force induced by the Hawking
radiation, the GSL would not be violated even without
the imposition of the constraint 
$S/E_{\rm b} \leq 2 \pi R$. The thermal 
ambiance outside the hole would prevent the box from descending 
beyond the point after which the energy delivered to the black 
hole would be too small to guarantee $\delta S_{\rm bh} \geq S$ 
as demanded by the GSL [see also \textcite{Matsasetal05}]. 
However, by accepting that the box floats due to the 
Hawking radiation, we are led to conclude that a box in 
the Minkowski vacuum would be able to self-accelerate   
because of the Unruh thermal bath.\footnote{This 
would be so because of the Unruh temperature gradient along 
the box in the acceleration direction.} 

The ``self-accelerating box paradox" was recently
revisited by \textcite{Marolfetal02}. They concluded that
the heat absorbed by the box walls would increase their  
masses preventing the box from floating outside the black hole 
and, thus, self-accelerating
in Minkowski spacetime. Although this would solve the self-accelerating 
box paradox, the GSL seemed to be in danger again. However,
\textcite{Marolfetal02} presented a way out to save 
the GSL without the introduction of any extra entropy-bound law 
by assuming the existence of ``box-antibox pairs" in the Hawking 
radiation. Further discussion can be found in \textcite{Marolfetal04b}
and in the next section.

\subsection{Entropy and Rindler observers}
\label{subsection:entropy2}

Even if the GSL is not violated in the thought experiment above, 
one could think of more extreme situations where objects with
fixed energy and volume but carrying an arbitrary amount of 
entropy are beamed toward a black hole. In order to analyze 
these situations, 
\textcite{Marolfetal04} considered a large 
enough black hole to reduce the problem again to the corresponding one 
with a Rindler horizon. They concluded that, although inertial 
observers assign an entropy equal to the logarithm of 
the number of internal states $n$ to an arbitrary object,
this would not be the case for Rindler observers. For bodies with 
a large number of internal states, $n \gg 1$, Rindler observers 
would assign an entropy  of only $S_R \approx E_R \beta$, 
where $E_R$ is the Killing energy associated with the Rindler 
observers  and $\beta^{-1} = \kappa/2 \pi$ is the corresponding 
temperature associated with the surface gravity $\kappa$. 
As a result, a falling object which crosses the horizon would 
respect the GSL according to the Rindler observers no matter how many 
internal states (i.e.~how large entropy is according
 to the inertial observers)
it might carry. The inertial observer, at the same time, should 
raise no doubt about the GSL since he/she would never lose 
sight of the object. This illustrates how subtle the entropy concept
can be in General Relativity.

\subsection{Einstein equations as an equation of state?}
\label{subsection:spacetimeentropy}

The four laws of black hole mechanics, which are closely connected
with the four laws of black hole thermodynamics were derived by assuming
the Einstein equations.  \textcite{Jacobson95} has put forward the 
intriguing idea of turning the logic around and deriving the Einstein 
equations by assuming  
(i)~the proportionality of entropy and the horizon area 
and
(ii)~the fundamental relation $\delta Q = T dS$,
where $\delta Q$ and $T$ would be interpreted as the energy flux
and Unruh temperature, respectively, seen by an accelerated observer 
just outside the horizon. In this sense, Einstein equations could 
be seen as an ``equation of state" of spacetime. 
Because of its importance to Thermodynamics, Relativity, Information 
Theory and Quantum Gravity, black hole thermodynamics will undoubtedly
continue 
attracting a lot of attention in the near future, and the Unruh effect 
should keep being a useful tool in the investigation of these issues.

\subsection{Miscellaneous topics}
\label{subsection:Miscellaneous}

Several other issues connected with the Unruh effect have attracted 
attention recently. In parallel to the investigation of the 
decoherence of the internal state of single accelerated detectors 
as commented in Sec.~\ref{subsection:Decoherence}, studies of the 
entanglement between independent accelerated detectors coupled 
to a background field can  be found in the literature 
[see, e.g., \textcite{Pringle89,Benattital04,Rezniktal05, Massaretal06}]. 
Recently \textcite{Alsingetal03}, 
\textcite{Alsingetal04} and \textcite{Alsingetal06} 
have considered the teleportation of a state between an inertial and a 
Rindler observer. 
Although the authors' conclusion that the fidelity of the 
teleportation will be in general reduced due to the Unruh effect 
may be correct in the end, 
the details will probably depend 
on the particular experimental set-up. For instance, ideal uniformly 
accelerated {\em rigid} cavities prepared in the Rindler vacuum  would keep 
thermal fluctuations out \cite{Levinetal92} and the Unruh effect 
would not be responsible, in principle, for fidelity loss 
[see \textcite{Schutzholdetal05} for further considerations]. 
More detailed investigations are expected in the near future when
the Unruh effect should begin to be studied in connection with 
quantum communication \cite{Bradler07}. A different sort of question 
which one may pose is whether or not sufficiently accelerated Rindler 
observers would see broken symmetries being restored because of the 
high temperature of the Unruh thermal bath \cite{Hill85,Ohsaku04,Ebertetal07,Kharzeevetal05}. 
Finally, the Unruh effect has also gained importance in
quantum gravity theories [see, e.g., 
\textcite{Susskindetal94}]\footnote{See also, e.g., 
\textcite{Parentanietal89}, who have considered uniformly 
accelerated observers in the vacuum of free strings.}
and condensed matter physics \cite{Unruh81} because of 
its close relation with the Hawking effect.

\section{Concluding remarks}
\label{section:Summary}

The Unruh effect has played a crucial role in our 
understanding that the particle content of a field 
theory is observer dependent. It expresses the fact that 
{\em uniformly accelerated observers} in Minkowski 
spacetime associate a thermal bath of {\em Rindler 
particles} to the {\em no-particle state} of {\em inertial 
observers}. As a Quantum Field Theory effect, it does {\em not} 
depend on extra structures such as particle detectors 
or other measuring apparatus.
By the same token, the Unruh effect does {\em not} require 
experimental confirmation any more than free quantum 
field theory does, although some observables can be more 
easily computed and interpreted from the point of view 
of uniformly accelerated observers using the Unruh effect.
This is a matter of convenience and not
of principle. We have dedicated Sec.~\ref{section:Applications}
to discuss in detail some physical phenomena using plain 
quantum field theory adapted to inertial observers and  
shown how the same observables can be recalculated 
from the point of view of Rindler observers with the help 
of the Unruh effect.

The Unruh effect is also useful as a {\em theoretical laboratory} 
to investigate phenomena such 
as the thermal emission of particles from 
black holes and cosmological horizons because it retains many essential 
features of these phenomena while reducing their technical 
complexity. Because of the importance of the Hawking (and Hawking-like) 
effect(s) to Thermodynamics, Information Theory, Quantum Gravity and 
Cosmology, the Unruh effect should continue being a valuable tool 
in the future to those who intend to investigate these issues.


\begin{acknowledgments}

We would like 
to thank C.~Fewster, S.~Fulling, B.~Kay, J.~Louko, W.~Unruh, 
D.~Vanzella and R.~Wald 
for useful conversations. L.~C. and G.~M. are also thankful to Conselho 
Nacional de Desenvolvimento Cient\'\i fico e Tecnol\'ogico for partial 
support. G.~M. is also thankful to Funda\c c\~ao de Amparo \`a Pesquisa do 
Estado de S\~ao Paulo for partial support.

\end{acknowledgments} 

\appendix

\section{Derivation of the positive-frequency solutions in the right
Rindler wedge}
\label{Appendix A}

In this Appendix we present a derivation of the normalized
positive-frequency modes in the right Rindler wedge.  First let us show
that the normalization condition (\ref{sec2:normalization}) leads to
the $\delta$-function normalization (\ref{sec2:delta-normal}) of the
function $g_{\omega k_\perp}(\xi)$.
Define
\begin{equation}
S_A(\omega,\omega^\prime) 
\equiv 
\int_{-A}^\infty d\xi\, 
g_{\omega k_\perp}^\ast (\xi) g_{\omega^\prime k_\perp}(\xi).
\end{equation}
 By the differential equation (\ref{sec2:bessel}) satisfied by
$g_{\omega k_\perp}$ and the condition (\ref{sec2:normalization}) 
we find
\begin{eqnarray}
&& (\omega^2 - \omega^{\prime 2}) S_A(\omega,\omega^\prime)\nonumber \\
&& = \left[g_{\omega' k_\perp}(\xi)\frac{d\ }{d\xi}g_{\omega
k_\perp}^\ast(\xi)
- g_{\omega k_\perp}^\ast (\xi)\frac{d\ }{d\xi}g_{\omega' k_\perp}(\xi)
\right]_{\xi=-A} \nonumber \\
&&  \approx \frac{1}{\pi}
\left\{(\omega-\omega^\prime)
\sin\left[(\omega+\omega^\prime)A-\gamma(\omega)-\gamma(\omega')
\right]\right.\nonumber \\
&& \left.\ \  
+(\omega+\omega')\sin\left[(\omega-\omega')A
-\gamma(\omega)+\gamma(\omega')\right]\right\}
\end{eqnarray}
for $\xi <0, \, |\xi| \gg1$.
Then, using the formula
\begin{equation}
\lim_{A\to \infty} [\sin (xA)] / x = \pi \delta(x),
\end{equation}
we find
\begin{eqnarray}
\int_{-\infty}^{+\infty}d\xi\,
g_{\omega k_\perp}^\ast(\xi)
g_{\omega^\prime k_\perp}(\xi)
& = & \lim_{A\to\infty}S_A(\omega,\omega^\prime) \nonumber \\
& = & \delta(\omega-\omega^\prime),
\end{eqnarray}
identifying bounded terms oscillating with frequency $\sim A$ with zero.

Now, by changing the variable in the differential equation 
(\ref{sec2:bessel}) as
\begin{equation}
\chi \equiv \frac{\sqrt{k_\perp^2 + m^2}}{a}e^{a\xi},
\end{equation}
we find that this equation becomes
\begin{equation}
\left( \frac{d^2\ }{d\chi^2}
+\frac{1}{\chi}\frac{d\ }{d\chi} - 1 +
\frac{(\omega/a)^2}{\chi^2}\right) g_{\omega k_\perp} = 0.
\end{equation}
This is a modified Bessel equation with index $i\omega/a$ (or 
$-i\omega/a$).  Hence,
together with the requirement that 
$|g_{\omega{\bf k}_\perp}(\xi)|$ should not tend to infinity as 
$\xi \to \infty$, we find
\begin{equation}
g_{\omega{\bf k}_\perp}(\xi) = 
C_{\omega k_\perp}K_{i\omega/a}[ (\kappa/a) e^{a\xi}],
\end{equation}
where $\kappa \equiv \sqrt{k_\perp^2 + m^2}$ and 
$C_{\omega k_\perp}$ is a constant.  
Now, the modified Bessel function $K_\nu(x)$ is defined by
\begin{equation}
K_\nu(x) \equiv - \frac{\pi}{2}\frac{i^{-\nu}J_\nu(ix)
- i^\nu J_{-\nu}(ix)}{\sin\nu\pi},
\end{equation}
and the Bessel function $J_\nu(x)$ for small $|x|$ is approximated as
\begin{equation}
J_\nu(x)\approx [\Gamma(\nu+1)]^{-1} ({x}/{2})^\nu.
\end{equation}
[See \textcite{Gradshteynbook}.]  Hence,
\begin{equation}
K_{i\omega/a}(x) \approx \frac{i\pi}{2\sinh(\pi\omega/a)}\left\{
\frac{(x/2)^{i\omega/a}}{\Gamma(1+i\omega/a)} 
- \frac{(x/2)^{-i\omega/a}}{\Gamma(1-i\omega/a)}\right\}.
\label{appendixA:small-argumentK}
\end{equation}
Note also
\begin{eqnarray}
\left|\Gamma(1+i\omega/a)\right|^2 & = & 
\Gamma(1+i\omega/a)\Gamma(1-i\omega/a)\nonumber \\
& = & \frac{i\omega}{a}\Gamma(i\omega/a)\Gamma(1-i\omega/a)\nonumber \\
& = & \frac{\pi\omega}{a\sinh(\pi\omega/a)}.
\end{eqnarray}
Hence
\begin{equation}
K_{i\omega/a}(x) \approx \sqrt{\frac{\pi a}{\omega\sinh(\pi\omega/a)}}
\,\left[e^{i\alpha}(x/2)^{i\omega/a} + {\rm c.c.}\right],
\end{equation}
where $\alpha$ is a real constant.
By comparing this formula with Eq.~(\ref{sec2:normalization}), 
we find that the function $g_{\omega k_\perp}(\xi)$ satisfying the
differential equation (\ref{sec2:bessel}) and the normalization
condition (\ref{sec2:normalization}) can be chosen as in 
Eq.~(\ref{sec2:normalizedgomega}).

\bibliography{rmpbibfinalsub}

\end{document}